\documentclass[floatfix,amsmath,amssymb,superscriptaddress,nofootinbib,longbibliography,twocolumn, aps, prb]{revtex4-2}

\usepackage{lipsum}
\usepackage{verbatim}
\usepackage[colorlinks=true,linkcolor=blue,citecolor=blue]{hyperref}
\usepackage[utf8]{inputenc}
\usepackage[T1]{fontenc}

\usepackage{array}
\usepackage{mathtools}
\usepackage{bm}

\usepackage{physics}
\usepackage{algorithmic}
\usepackage{algorithm}
\usepackage{siunitx}

\usepackage{graphicx}
\usepackage{tikz}

\usepackage{tabularx}
\usepackage{booktabs}
\newcolumntype{L}[1]{>{\raggedright\let\newline\\\arraybackslash\hspace{0pt}}m{#1}}
\newcolumntype{C}[1]{>{\centering\let\newline\\\arraybackslash\hspace{0pt}}m{#1}}
\newcolumntype{R}[1]{>{\raggedleft\let\newline\\\arraybackslash\hspace{0pt}}m{#1}}
\newcolumntype{x}[1]{>{\centering\arraybackslash\hspace{0pt}}p{#1}}

\usepackage{soul}

\usepackage[caption=false]{subfig}
\newcommand{\phantomsubfloat}[1]{
    {%
        \captionsetup[subfigure]{labelformat=empty}
        \subfloat[][]{#1}
    }%
}

\usepackage{cleveref}
\Crefname{equation}{Equation}{Equations}
\Crefname{section}{Section}{Sections}
\Crefname{appendix}{Appendix}{Appendices}
\Crefname{figure}{Figure}{Figures}
\Crefname{table}{Table}{Tables}
\crefname{equation}{Eq.}{Eqs.}
\crefname{section}{Sec.}{Secs.}
\crefname{appendix}{Appendix}{Appendices}
\crefname{figure}{Fig.}{Figs.}
\crefname{table}{Table}{Tables}

\newcommand{\normord}[1]{:\mathrel{\mkern2mu #1 \mkern2mu}:}
\newcommand{\trans}[1]{#1^{\mathsf{T}}}

\usepackage{orcidlink}

\usepackage{pifont}
\newcommand{\cmark}{\ding{51}}
\newcommand{\xmark}{\ding{55}}

\newcommand\ucdot{\ensuremath{{}\cdot{}}}
\DeclareSIUnit\angstrom{\text{Å}}
\DeclareSIUnit\unitcellarea{A_{uc}}
\DeclareSIUnit\carrier{carrier}

\usepackage{xspace}
\newcommand{\hBN}{\textit{h}-BN\xspace}
\newcommand{\ie}{\textit{i}.\textit{e}.}
\newcommand{\eg}{\textit{e}.\textit{g}.}

\definecolor{orange}{rgb}{1,0.5,0}

\newcommand{\icontext}[1]{\smash{\mbox{\hspace{-1.8pt}\smash{\raisebox{-4.2pt}{\includegraphics[height = 13pt]{#1}}}\hspace{-1.5pt}}}}

\hypersetup{pdftitle=BLG can cant}

\begin{document}

\title{Symmetry-broken metallic orders in spin-orbit-coupled Bernal bilayer graphene}

\author{Jin Ming Koh\,\orcidlink{0000-0002-6130-5591}}
\affiliation{Department of Physics, California Institute of Technology, Pasadena, California 91125, USA}
\affiliation{Department of Physics, Harvard University, Cambridge, Massachusetts 02138, USA}

\author{Alex Thomson\,\orcidlink{0000-0002-9938-5048}}
\affiliation{Department of Physics, University of California, Davis, California 95616, USA}

\author{Jason Alicea\,\orcidlink{0000-0001-9979-3423}}
\affiliation{Department of Physics, California Institute of Technology, Pasadena, California 91125, USA}
\affiliation{Institute for Quantum Information and Matter, California Institute of Technology, Pasadena, California 91125, USA}

\author{\'Etienne Lantagne-Hurtubise\,\orcidlink{0000-0003-0417-6452}}
\affiliation{Department of Physics, California Institute of Technology, Pasadena, California 91125, USA}
\affiliation{Institute for Quantum Information and Matter, California Institute of Technology, Pasadena, California 91125, USA}

\begin{abstract}
We explore Bernal bilayer graphene in the presence of long-range Coulomb interactions, short-range Hund's coupling, and proximity-induced Ising spin-orbit coupling using self-consistent Hartree-Fock simulations. We show that the interplay between these three ingredients produces an intricate phase diagram comprising a multitude of symmetry-broken metallic states tunable via doping and applied displacement field. In particular, we find intervalley coherent and spin-canted ground states that may hold the key to understanding spin-orbit-enabled superconductivity observed in this platform.  We also investigate various phase transitions where a continuous $\mathrm{U}(1)$ symmetry is broken to ascertain the possible role of critical fluctuations on pairing. 
\end{abstract}

\maketitle
\date{\today}

\section{Introduction}

\begin{figure}[!t]
    \centering
    \includegraphics[width = 1\linewidth]{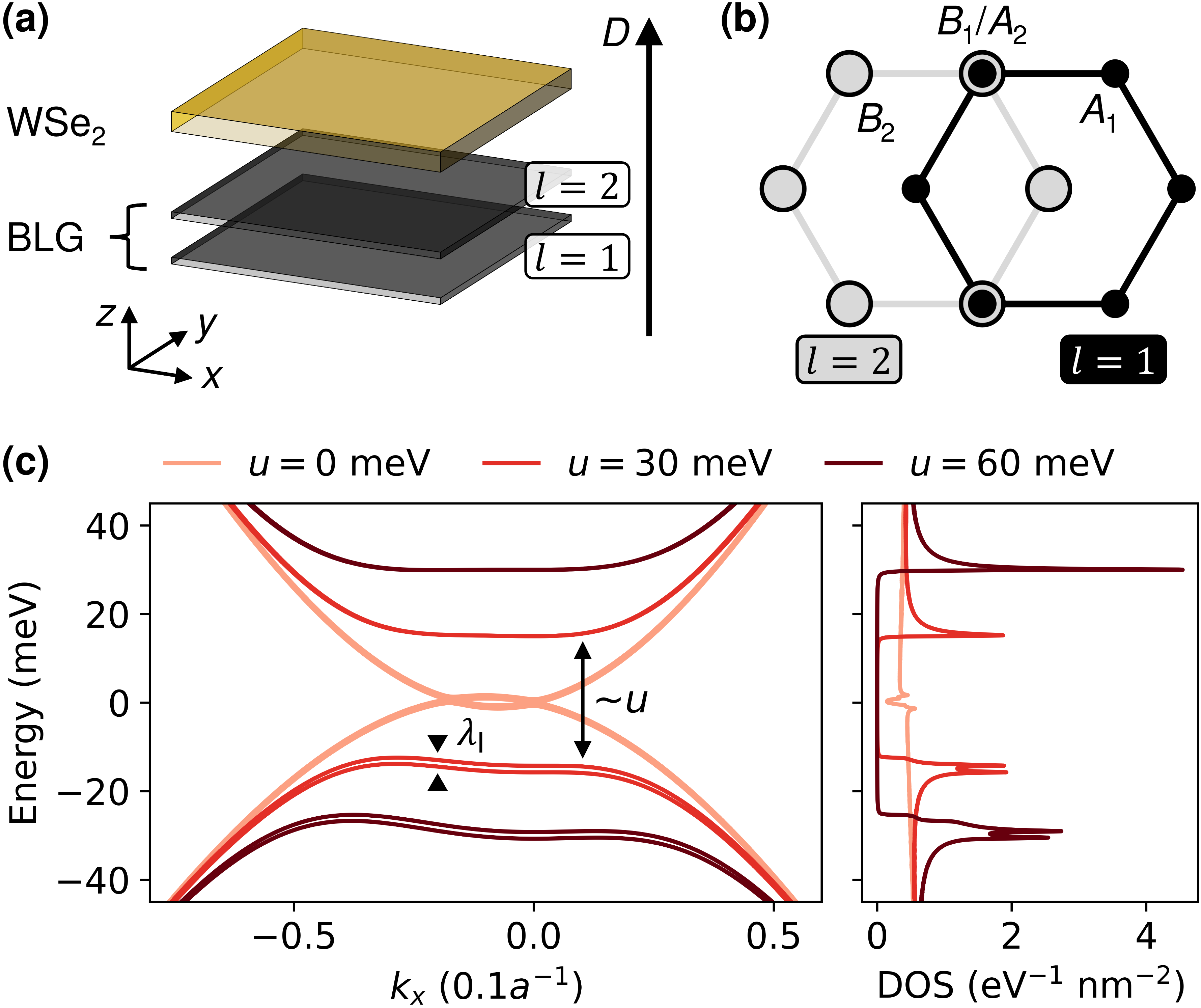}
    \phantomsubfloat{\label{fig:schematics-stack}}
    \phantomsubfloat{\label{fig:schematics-stacking}}
    \phantomsubfloat{\label{fig:schematics-band-structure}}
    \vspace{-1.6\baselineskip}
    \caption{\textbf{Schematic of spin-orbit proximitized BLG.} \textbf{(a)} A transition-metal dichalcogenide (such as WSe$_2$ or WS$_2$) placed in proximity to BLG induces $\si{\milli\electronvolt}$-scale spin-orbit coupling to the top graphene layer. Applying a perpendicular displacement field $D$ generates an interlayer potential difference $u$ through \cref{eq:definition-u}. \textbf{(b)} Bilayer graphene stacking configuration, with sublattices $A_l$, $B_l$ for layer $l \in \{1, 2\}$. \textbf{(c)} Low-energy band structure in valley $\vb{K}^+$ for $k_y = 0$ (left panel) and density of states (right panel) for different $u$ assuming Ising spin-orbit coupling of strength $\lambda_{\mathrm{I}} = \SI{1.5}{\milli\electronvolt}$. The $D$ field opens a gap at charge neutrality and enhances van Hove singularities that drive strong interaction effects. Valley $\vb{K}^-$ exhibits a band structure related by time reversal symmetry.}
    \label{fig:schematic}
\end{figure}

Rhombohedral graphene multilayers host an exceptionally rich and tunable set of quantum phases of matter~\cite{Shi2020, zhou2021half, zhou2021superconductivity, zhou2022isospin, Seiler2022, delaBarrera2022, Kerelsky2021, Han2023, liu2023interactiondriven, Han2023a, Seiler2024}. Their exquisite tunability originates from the ability to electrically address, in a dual-gated device, both the doping level (by varying the chemical potential) and the effective interaction strength (through the band-flattening effect of a perpendicular displacement field $D$), as illustrated in \cref{fig:schematic}. While the first member of the rhombohedral family, Bernal bilayer graphene (BLG), has a long history dating back to the early days of graphene research~\cite{Novoselov2006, McCann2006, Zhang2009, Weitz2010, mccann2013electronic}, its complex low-temperature phase diagram
presents an enduring challenge to the quantum matter community.

Of particular interest is the recent observation of superconductivity when BLG is perturbed by an in-plane magnetic field~\cite{zhou2022isospin} or by inducing spin-orbit coupling (SOC) via proximity to an adjacent WSe$_2$ layer~\cite{Zhang2023}. 
In the former case, the in-plane field nucleates a narrow sliver of superconductivity near the location of a Stoner-type phase transition in the normal state; this unconventional superconductor exhibits a low critical temperature $T_{\mathrm{c}} \sim \SI{30}{\milli\kelvin}$ and an extreme Pauli limit violation consistent with spin-triplet pairing. The phenomenology of the latter case is very different: Up to four new superconducting regions appear with proximity-induced SOC~\cite{Zhang2023, Holleis2023, li2024tunable, Yiran2024}. 
The most prominent dome features a critical temperature $T_{\mathrm{c}}$ enhanced up to $\sim \SI{500}{\milli\kelvin}$ and spans a large density window within a symmetry-broken normal state. Furthermore, the Pauli limit violation evolves nontrivially across different superconducting regions, emphasizing the important role of SOC on superconductivity.

An important step towards understanding the origin and mechanism~\cite{Pantaleon2023review,Yangzhi2022,Curtis2023,Jimeno-Pozo2022,Dong2023,dong2023signatures,li2023charge, Wagner2023,Shavit2023,son2024,Dong2024} of these spin-orbit-enabled superconductors consists of constraining the nature of their \emph{normal state}. This endeavor is critical for two reasons: First, candidate 
metallic states differ by their fermiology and the symmetries they preserve, which can have an outsize influence on their potential superconducting instabilities.  And second, the normal state can host collective modes (dependent on details of the underlying order) that potentially serve as a pairing glue~\cite{Kozii2022, Dong2024}. 

Our key objective in this work is to shed light on the interacting normal-state phase diagram of spin-orbit coupled BLG~\cite{Zhang2023, Holleis2023, li2024tunable, Seiler2024layerselective, Yiran2024} and to leverage these
insights to further our understanding of superconductivity in this system. We use self-consistent numerical Hartree-Fock simulations, following a recent implementation for rhombohedral trilayers~\cite{Koh2024}, and contrast our results to other Hartree-Fock studies on proximitized BLG~\cite{Zhang2023, Ming2023, Wang2024, Zhumagulov2024}. In particular, Ref.~\onlinecite{Ming2023} reported a thorough investigation of the role of Ising SOC on the phase diagram of BLG in the presence of long-range Coulomb repulsion. While we recover their results in the appropriate limit, our main contribution here is to explore the crucial role of additional short-range interactions, also known in the literature as Hund's coupling. As we show below the inclusion of this term produces ferromagnetic instabilities that can either coexist or compete with the effects of Ising SOC, leading to a rich landscape of symmetry-broken phases. Among the new metallic states we uncover, we note two instances of \emph{spin-canting order}, within which spins in each valley tilt away from their respective Ising axes toward the graphene planes. The first instance is a spin-canted generalized half metal\footnote{Throughout this work, we refer to metallic states with a reduced Fermi surface degeneracy as \emph{generalized half-metals} (two-fold degeneracy) or \emph{generalized quarter-metals} (one-fold degeneracy), irrespective of the presence of minority Fermi surfaces.}, with a two-fold Fermi surface degeneracy, that occupies most of the experimentally relevant region for the strongest superconducting state in BLG/WSe$_2$. The second instance possesses spin-canted intervalley coherent (IVC) order, hosts non-degenerate Fermi surfaces (generalized quarter metal), and may be relevant for one of the new superconductors (dubbed SC3) discovered in Ref.~\onlinecite{Yiran2024}.

\begin{figure*}
    \centering
    \includegraphics[width = \linewidth]{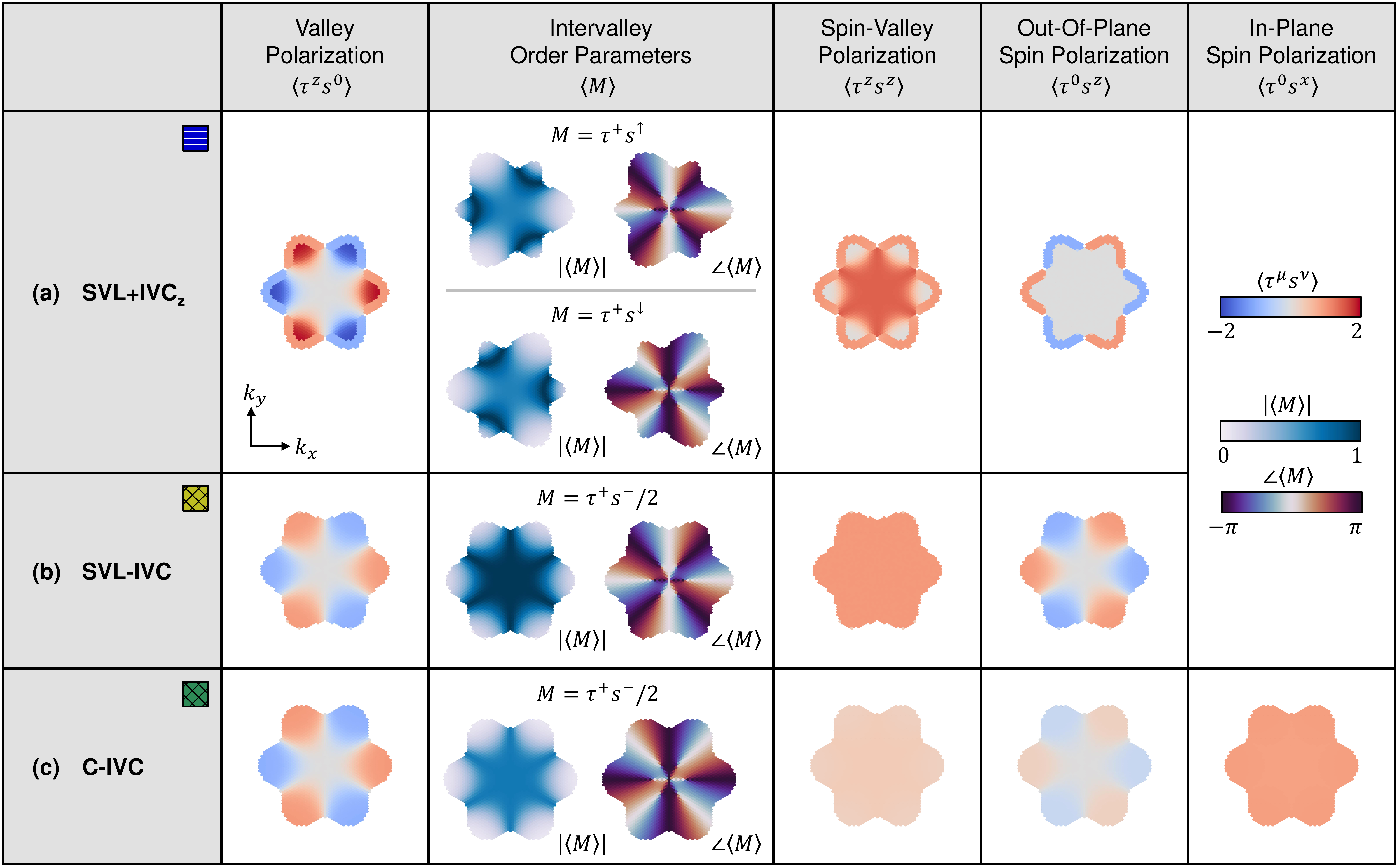}
    \phantomsubfloat{\label{fig:ivc-states-SVL-IVC-Z}}
    \phantomsubfloat{\label{fig:ivc-states-SVL-IVC}}
    \phantomsubfloat{\label{fig:ivc-states-SC-IVC}}
    \vspace{-1.6\baselineskip}
    \caption{\textbf{Momentum-resolved textures of intervalley-coherent ground states stabilized by Ising SOC.} Rows \textbf{(a)--(c)} illustrate the three IVC ground states obtained in the presence of Ising SOC in \cref{fig:results-w-soc}, projected to the occupied (hole) bands. The first column shows the valley polarization, which follows the trigonal-warping-induced energy difference between the two valleys. The second column depicts in-plane components $\tau^+ = \tau^x + i \tau^y$ of the valley pseudospin, accompanied by an appropriate spin operator. The third column shows the spin-valley-locked polarization, which couples directly to Ising SOC. The fourth and fifth columns depict the out-of-plane and in-plane spin polarization, respectively. \textbf{(a)} The doubly degenerate ($g = 2$) SVL+IVC$_{\text{z}}$ state exhibits a spin-valley-locked polarization ($\tau^z s^z$) that coexists with spin-triplet IVC order ($\tau^x s^z$), whereby intervalley coherence develops with an opposite sign for the two spin projections. The \textbf{(b)} SVL-IVC and \textbf{(c)} C-IVC states exhibit a single Fermi surface that arises by developing intervalley coherence within their respective (spin-valley locked or spin-canted) generalized half-metal state. The C-IVC state thus possesses a non-zero in-plane spin polarization, which distinguishes it from its SVL-IVC counterpart.}
    \label{fig:ivc-states}
\end{figure*}

Comparing with recent experiments~\cite{Yiran2024} that systematically vary the induced Ising SOC strength $\lambda_{\mathrm{I}}$ by engineering the twist angle at the BLG/WSe$_2$ interface~\cite{Li2019, David2019, Naimer2021, Yangzhi2022}, we identify two promising candidate normal states that can host the largest superconducting region. Namely, experiments reveal that while $T_{\mathrm{c}}$ is enhanced with increasing $\lambda_{\mathrm{I}}$, its extent \emph{shrinks} within the accessible displacement field and density regime. This observation suggests that the normal state that gives way to superconductivity (which exhibits two-fold Fermi surface degeneracy as per quantum oscillation measurements) is one of the symmetry-broken generalized half metals that \emph{competes} against the symmetric spin-valley-locked state that is trivially favored by $\lambda_{\mathrm{I}}$. The spin-canting order described above is one of the normal states we find that satisfies these properties; the other option is an IVC spin-triplet state that coexists with spin-valley locking (see \cref{fig:ivc-states-SVL-IVC-Z} for an illustration). While both candidates are consistent with currently available data, we propose future experiments to distinguish between them. We also connect these observations with a recent proposal wherein spin-canting fluctuations
provide a glue for superconductivity~\cite{Dong2024}, and contrast with other pairing mechanisms including Kohn-Luttinger~\cite{Jimeno-Pozo2022, Wagner2023, son2024}, acoustic phonons~\cite{Chou2022, Rubio2024} and IVC fluctuations~\cite{Chatterjee2022, dong2023signatures}.

Finally, we hone in on $\mathrm{U}(1)$-breaking transitions observed at mean-field level, of either IVC or spin-canting type. While we find IVC transitions to be generically first order, spin-canting transitions appear 
continuous (within the precision of our Hartree-Fock implementation). We speculate on the role of critical fluctuations associated with these putative continuous transitions.

The rest of this article is organized as follows. In \cref{sec:model-methods} we introduce the model Hamiltonian describing spin-orbit-proximitized BLG, discuss different (long-range vs.~short-range) interaction terms, and present the self-consistent Hartree-Fock technique used throughout this study. In \cref{sec:phase-diagrams} we investigate the phase diagram of the system. \Cref{sec:phase-diagrams/no-ising} first sets the stage by considering the situation without SOC, emphasizing the crucial role of Hund's coupling. In \cref{sec:phase-diagrams/ising} we then introduce Ising SOC, explore its effect on the delicate phase competition between symmetry-broken phases, and connect with recent experimental studies of BLG/WSe$_2$. We explore in more detail $\mathrm{U}(1)$-breaking phase transitions, towards either IVC or spin-canting order, in \cref{sec:phase-transitions}. We highlight how different symmetry-broken ground states respond to changes in screening in ~\cref{sec:screening}. Finally, \cref{sec:discussion} offers insights into spin-orbit-enabled superconductivity in BLG and suggests future research directions.

\section{Model and methods}
\label{sec:model-methods}

\begin{table*}
    \centering
    \newcommand{\legendicon}[1]{\smash{\raisebox{-4pt}{\includegraphics[height = 14pt]{#1}}}}
    \begin{tabular}{p{7.6cm} p{1.8cm} p{2.4cm} C{0.8cm} C{0.8cm} C{0.8cm} | C{0.8cm} | C{0.5cm} p{0.6cm}}
        \toprule 
        Order Description & Symbol & Order Operators & $\mathcal{T}$ & \hspace{-0.2cm} $\text{U}(1)_{\text{v}}$ & $\text{U}(1)_{\mathrm{s}}$ & $\mathcal{T}_{\mathrm{eff}}$  & $g$ & Leg. \\
        \midrule 
        Fully symmetric & FS
            & - 
            & \cmark & \cmark & \cmark & --- 
            & $4$
            & \legendicon{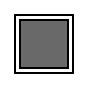}
            \\
        Valley-polarized & VP 
            & $\tau^z s^0$ 
            & \xmark & \cmark & \cmark & --- 
            & $2$ 
            & \legendicon{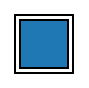}
            \\
        Spin-polarized & SP 
            & $\tau^0 s^z$ 
            & \xmark & \cmark & \cmark & $\mathcal{T}_0$  
            & $2$
            & \legendicon{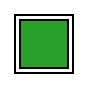}
            \\
        Spin-valley-locked & SVL 
            & $\tau^z s^z$ 
            & \cmark & \cmark & \cmark & --- 
            & $2$
            & \legendicon{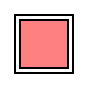}
            \\
        Spin-canted & C 
            & $\tau^z s^z, \tau^0 s^x$ 
            & \xmark & \cmark  & \xmark & $\mathcal{T}_{\mathrm{s}}$ 
            & $2$
            & \legendicon{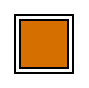}
            \\
        Intervalley-coherent spin-singlet & IVC$_{\text{0}}$ 
            & $\tau^x s^0$
            & \cmark & \xmark & \cmark & --- 
            & $2$
            & \legendicon{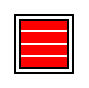}
            \\
        Intervalley-coherent spin-triplet & IVC$_{\text{z}}$ 
            & $\tau^x s^z$ 
            & \xmark & \xmark & \cmark  & $\mathcal{T}_{\mathrm{v}}$ 
            & $2$
            & \legendicon{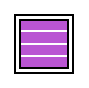}
            \\
        Intervalley-coherent spin-singlet + spin-valley-locked & SVL+IVC$_{\text{0}}$
            & $\tau^x s^0, \tau^z s^z$ 
            & \cmark & \xmark & \cmark & --- 
            & $2$
            & \legendicon{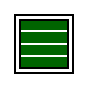}
            \\
        Intervalley-coherent spin-triplet + spin-valley locked & SVL+IVC$_{\text{z}}$
            & $\tau^x s^z, \tau^z s^z$ 
            & \xmark & \xmark & \cmark & $\mathcal{T}_{\mathrm{v}}$ 
            & $2$
            & \legendicon{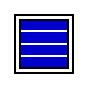}
            \\
       Spin-valley-polarized & SVP
            & $\tau^z s^0, \tau^0 s^z$ 
            & \xmark & \cmark & \cmark & ---
            & $1$
            & \legendicon{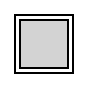}
            \\
        Spin-polarized intervalley-coherent & SP-IVC
            & $\tau^x s^0, \tau^0 s^z$ 
            & \xmark & \xmark & \cmark & $\mathcal{T}_0$  
            & $1$
            & \legendicon{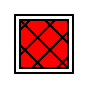}
            \\
        Spin-valley-locked intervalley-coherent & SVL-IVC
            & $\tau^x s^x, \tau^z s^z$ 
            & \xmark & \xmark & \xmark &  $\mathcal{T}_{\mathrm{v}/\mathrm{s}}$ 
            & $1$
            & \legendicon{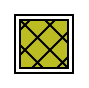}
            \\
        Spin-canted intervalley-coherent & C-IVC 
            & $\tau^x s^x, \tau^z s^z, \tau^0 s^x$
            & \xmark & \xmark & \xmark & $\mathcal{T}_{\mathrm{s}}$  
            & $1$
            & \legendicon{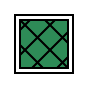}
            \\
        \bottomrule
    \end{tabular}
    \caption{\textbf{Symmetry classification of Hartree-Fock ground states.} A minimal set of spin-valley operators characterizing each class of ground states is listed along with their transformation properties under electronic time-reversal $\mathcal{T} = \tau^x s^y \mathcal{K}$, $\mathrm{U}(1)_{\mathrm{v}}$ valley conservation, and $\mathrm{U}(1)_{\mathrm{s}}$ spin rotations around the $z$ axis. Among states that break $\mathcal{T}$, we highlight cases that preserve an effective antiunitary symmetry $\mathcal{T}_{\mathrm{eff}}$ that enforces a nesting condition for zero-momentum pairing: either spinless time-reversal symmetry $\mathcal{T}_0 = \tau^x \mathcal{K}$, or a product of $\mathcal{T}$ with a valley ($\mathcal{T}_{\mathrm{v}}$) or spin ($\mathcal{T}_{\mathrm{s}}$) rotation. The integer $g$ denotes the spin-valley degeneracy of the Fermi surfaces, while the last column shows the color and hatching scheme used in phase diagrams throughout this work.}
    \label{tab:symmetry-orders-legends}
\end{table*}

\Cref{fig:schematics-stack} sketches a spin-orbit-proximitized BLG device featuring a displacement field $D$ applied normal to the graphene planes; see \cref{fig:schematics-stacking} for the BLG atomic structure.  Let us first discuss the physics of `pure' BLG without the proximitizing transition-metal dichalcogenide (TMD) layer. Relevant symmetries here include three-fold rotations $\mathrm{C}_3$, 
translations, time reversal $\mathcal{T}$, and ${\mathrm{SU}}(2)_{\mathrm{s}}$ spin rotations (neglecting weak native spin-orbit coupling). The system also exhibits an approximate $\mathrm{U}(1)_{\mathrm{v}}$ valley conservation at low energies.  

The tight-binding Hamiltonian describing BLG can be expanded near the two graphene valleys $\tau \in \{\pm 1\}$ as
\begin{equation}\begin{split}
    \hat{H}_{0} = \sum_{\vb{k}} \sum_{\tau s \sigma \sigma'} 
        h(\vb{K}^\tau + \vb{k})_{\sigma \sigma'} 
        c_{\tau s \sigma \vb{k}}^\dag 
        c_{\tau s \sigma' \vb{k}},
\end{split}\end{equation}
where $c_{\tau s \sigma \vb{k}}$ annihilates an electron with momentum $\vb{k}$ in valley $\tau$, with spin $s \in \smash{\{\uparrow, \downarrow\}}$ and sublattice $\sigma \in \{A_1,B_1, A_2, B_2\}$. The matrix $h$ contains the leading intra- and inter-layer tunneling matrix elements as well as on-site potentials~\cite{Jung2014}; see \cref{app-sec:hamiltonian}. The $D$ field in particular generates an interlayer potential difference $u$ that enters $\hat{H}_{0}$ as
\begin{equation}
     u = q_e d^\perp D / \epsilon_{\mathrm{r}}^\perp ,
     \label{eq:definition-u}
\end{equation}
with $q_{\text{e}}$ the electron charge, $d^\perp \approx \SI{3.3}{\angstrom}$ the interlayer distance 
and $\epsilon_{\mathrm{r}}^\perp = 4.4$ a dielectric constant describing the screening of perpendicular fields. The low-energy valence and conduction bands predicted by $\hat{H}_{0}$ host van Hove singularities that are amplified by nonzero $D$ (see \cref{fig:schematics-band-structure}), in turn dramatically enhancing interaction effects. 

Coulomb repulsion between electrons is included using a convenient decomposition into long- and short-range components. The long-range component
\begin{equation}\begin{split}
    \hat{H}_{\mathrm{C}} &= \frac{1}{2A} \sum_{\vb{q}} V_{\mathrm{C}}(\vb{q}) 
        \normord{\rho(\vb{q}) \rho(-\vb{q})}
\end{split}\end{equation}
involves the long-wavelength part of the electronic density, $\smash{\rho(\vb{q})} = \smash{\sum_{\vb{k} \alpha} c_{\alpha \vb{k}}^\dag c_{\alpha (\vb{k} + \vb{q})}}$, where 
$\alpha = (\tau, s, \sigma)$ is a combined flavor index.  The prefactor $A$ is the sample area, and we consider the dual-gated 
Coulomb potential $V_{\mathrm{C}}(\vb{q}) = (q_{\mathrm{e}}^2 / 2 \epsilon_{\mathrm{r}} \epsilon_0 q) \tanh{(q d)}$, with $d$ the distance from BLG to the gates, $\epsilon_{\mathrm{r}}$ the relative permittivity, and $\epsilon_0$ the permittivity of free space. Electronic screening (beyond that provided by the gates and the \hBN spacer layers) is accounted for by treating $\epsilon_{\mathrm{r}}$ as a tunable parameter fixed by comparing to experiments---see below. 

The interacting model $\smash{\hat{H}_{0}} + \smash{\hat{H}_{\mathrm{C}}}$ defined above preserves a non-generic $\mathrm{SU}(2) \times \mathrm{SU}(2)$ symmetry corresponding to independent spin rotations in each valley. This symmetry is reduced to global $\mathrm{SU}(2)_{\mathrm{s}}$  spin rotations by introducing the short-range component $\smash{\hat{H}_{\mathrm{V}}}$ (see \cref{app-sec:hamiltonian} for the explicit form), with coupling strength $J_{\mathrm{H}}$,
that captures exchange of electrons between valleys. Also known as the Hund's coupling, such a term (for $J_{\mathrm{H}}>0$) favors aligning the electron spins in the two valleys, consistent with experimental observations of metallic ferromagnetism in rhombohedral graphene multilayers~\cite{zhou2022isospin, Seiler2022, delaBarrera2022, zhou2021half}.

Proximitizing BLG with an adjacent TMD as sketched in \cref{fig:schematics-stack} breaks $\mathrm{SU}(2)_{\mathrm{s}}$ spin rotation symmetry by inducing $\si{\milli\electronvolt}$-scale SOC in the top graphene layer~\cite{Gmitra2016, Wang2016, Yang2017, Zihlmann2018, Island2019, Wang2019, Amann2022, Sun2022determining}. The two main contributions are Ising- and Rasbha-type SOC, whose strength can be tuned through the relative twist angle between the TMD and BLG~\cite{Li2019, David2019, Naimer2021, Yangzhi2022}, as demonstrated experimentally~\cite{Yiran2024}. The effective Rashba scale in the low-energy BLG bands is suppressed by the $D$-field-induced sublattice polarization of the corresponding wavefunctions~\cite{Zaletel2019}. In this work we therefore focus on Ising SOC, which takes the form
\begin{equation}\begin{split}
    \hat{H}_{\mathrm{I}} &= \frac{\lambda_{\mathrm{I}}}{2} \sum_{\vb{k}} 
        \vb{c}^\dag_{\vb{k}} \left( \tau^z s^z \mathbb{P}_2 \right) \vb{c}_{\vb{k}}.
\end{split}\end{equation}
Here $\trans{\vb{c}}_{\vb{k}} = \mqty[ c_{+ \uparrow A_1 \vb{k}} & \ldots & c_{- \downarrow B_2 \vb{k}} ]$ combines the relevant fermion operators, $\lambda_{\mathrm{I}}$ denotes the Ising energy scale, and $\mathbb{P}_2$ projects onto the top graphene layer. Throughout we respectively use $\tau^\mu$, $s^\mu$, and $\sigma^\mu$ to label valley, spin, and sublattice Pauli matrices. As shown in \cref{fig:schematics-band-structure}, Ising SOC leads to a noticeable spin splitting in the valence band for $D > 0$ (and in the conduction band for $D < 0$), due to the induced layer polarization of the low-energy wavefunctions~\cite{Gmitra2017, Khoo2017}. 

To explore the system's phase diagram in the presence of long-range Coulomb interactions, short-range Hund's coupling, and proximity-induced Ising SOC, we implement a self-consistent Hartree-Fock procedure described in detail in Ref.~\onlinecite{Koh2024} (see in particular Appendix B therein). We consider 
all symmetry-breaking orders in the four-dimensional spin and valley subspace spanned by $\tau^\mu s^\nu$ with $\mu, \nu \in \{0, x, y, z\}$, and compare the lowest-energy instance of each class of ground states obtained using a symmetry-restricted algorithm. Throughout this work we assume that $\vb{k}$ remains a good quantum number (\ie~we do not allow for translation symmetry breaking, except for commensurate $\smash{\sqrt{3}} \times \smash{\sqrt{3}}$ charge density waves associated with intervalley coherence). We further impose orbital C$_3$ symmetry (\ie~we do not attempt to capture nematic or momentum polarization instabilities
~\cite{Jung2015, Dong2021, Huang2023}, which are difficult to resolve numerically~\cite{Koh2024}).

The results presented in this work were computed on momentum grids comprising ${\sim} 1700$ points (except for higher-resolution linecuts shown in \cref{fig:results-linecuts}), keeping all four bands (per spin and valley) of $\smash{\hat{H}_0}$ and imposing a momentum cutoff $\Lambda = 0.12 a^{-1}$, where $a = \SI{2.46}{\angstrom}$ is the graphene lattice constant. In \cref{tab:symmetry-orders-legends} we list all the obtained symmetry-broken ground states (along with abbreviations and color schemes used throughout), their transformation properties under various symmetries, and their Fermi surface degeneracies labeled by the integer $g$.

\section{Hartree-Fock phase diagrams}
\label{sec:phase-diagrams}

\begin{figure*}
    \centering
    \includegraphics[width = 1\linewidth]{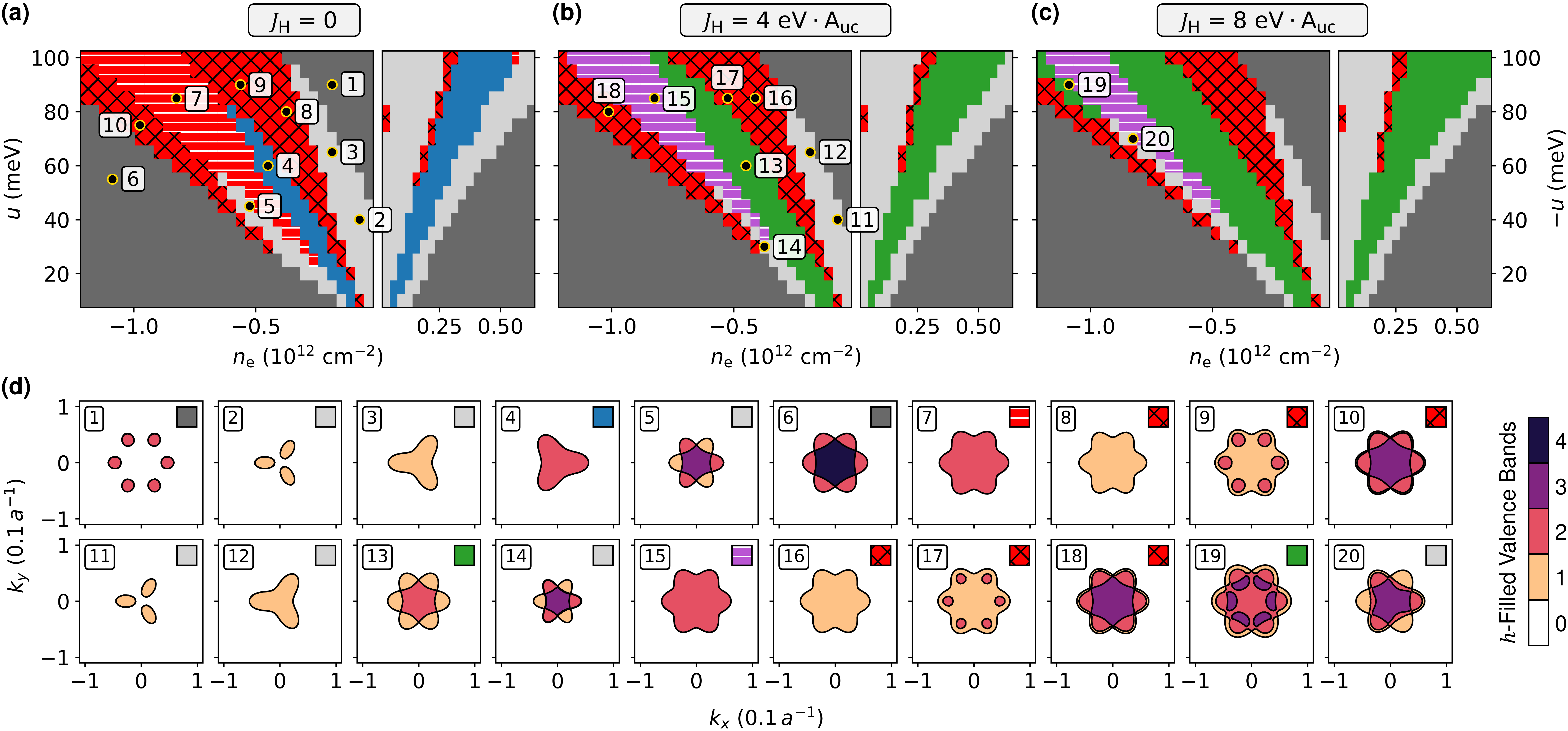}
    \phantomsubfloat{\label{fig:results-wo-soc-Jh-00}}
    \phantomsubfloat{\label{fig:results-wo-soc-Jh-04}}
    \phantomsubfloat{\label{fig:results-wo-soc-Jh-08}}
    \phantomsubfloat{\label{fig:results-wo-soc-fermi}}
    \vspace{-1.6\baselineskip}
    \caption{\textbf{Phase diagrams without SOC.} 
    Hole- and electron-doped phase diagrams of BLG as a function of charge density $n_{\text{e}}$ and interlayer potential $u$, at moderate Coulomb strength $\epsilon_{\mathrm{r}} = 20$. We consider cases \textbf{(a)} without Hund's coupling ($J_{\mathrm{H}} = 0$), and with ferromagnetic Hund's coupling \textbf{(b)} $J_{\mathrm{H}} = \SI{4}{\electronvolt\ucdot\unitcellarea}$ and \textbf{(c)} $J_{\mathrm{H}} = \SI{8}{\electronvolt\ucdot\unitcellarea}$. 
    Different phases are denoted by their color and hatching (see \cref{tab:symmetry-orders-legends} for legends). In panel \textbf{a}, unphysical degeneracies between different ground states arise due to the enlarged $\mathrm{SU}(2) \times \mathrm{SU}(2)$ symmetry of our model in the $J_{\mathrm{H}} = 0$ limit (see Appendix \ref{app:no-Hunds}). In panels \textbf{b} and \textbf{c}, Hund’s coupling breaks the degeneracy between the $g = 2$ Stoner ferromagnets in favor of the spin-polarized (\icontext{icon-HM-SP-nom}) phase, and similarly promotes IVC$_{\text{z}}$ (\icontext{icon-HM-IVC-Z-nom}) and SP-IVC (\icontext{icon-QM-SP-IVC-nom}) phases. \textbf{(d)} Fermi surface structure at numbered points in panel \textbf{a}. Colors denote the number of mean-field valence bands occupied by holes. Fermi surfaces in the electron-doped regime are similar.}
    \label{fig:results-wo-soc}
\end{figure*}

References~\onlinecite{Ming2023, Wang2024, Zhumagulov2024} previously studied the phase diagram of BLG in the presence of long-range Coulomb interactions and induced Ising SOC. Here we consider a crucial experimentally motivated addition to their treatment:  ferromagnetic Hund's coupling with $J_{\mathrm{H}} > 0$. As we show below, inclusion of this term qualitatively changes the nature of the stabilized orders and their fermiology, with potentially important implications for the nature of superconductivity arising in proximitized BLG devices. Throughout we mostly focus our attention on the hole-doped region, which yields richer physics both numerically and experimentally (although superconductivity was also recently observed on the electron-doped side at very high displacement fields~\cite{li2024tunable}). The relevant regimes for interaction parameters $\epsilon_{\mathrm{r}}$ and $J_{\mathrm{H}}$ can be estimated by benchmarking to experimental results~\cite{zhou2021half, zhou2022isospin, Zhang2023, Arp2024}. In this work, following Ref.~\onlinecite{Koh2024} we take $\epsilon_{\mathrm{r}} = 20$ and $J_{\mathrm{H}} = 4 - \SI{8}{\electronvolt\ucdot\unitcellarea} \approx 200 - \SI{400}{\milli\electronvolt\nano\meter\squared}$ (where $\si{\unitcellarea} \approx \SI{0.052}{\nano\meter\squared}$ denotes the unit cell area of graphene). Other values of $\epsilon_{\mathrm{r}}$ are explored in \cref{sec:screening}.

\begin{figure*}
    \centering
    \includegraphics[width = 1\linewidth]{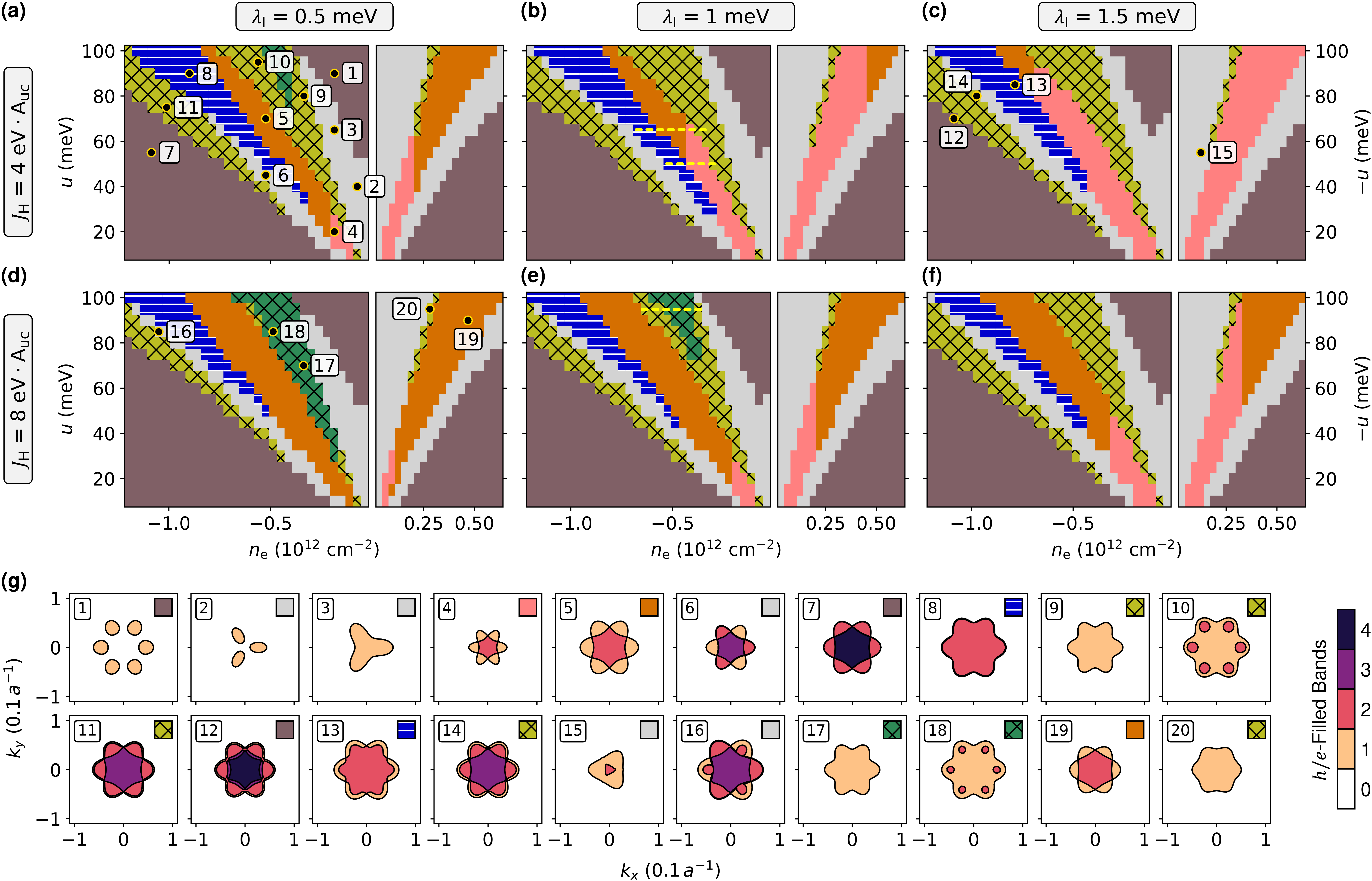}
    \phantomsubfloat{\label{fig:results-w-soc-Jh-04-lambdaI-05}}
    \phantomsubfloat{\label{fig:results-w-soc-Jh-04-lambdaI-10}}
    \phantomsubfloat{\label{fig:results-w-soc-Jh-04-lambdaI-15}}
    \phantomsubfloat{\label{fig:results-w-soc-Jh-08-lambdaI-05}}
    \phantomsubfloat{\label{fig:results-w-soc-Jh-08-lambdaI-10}}
    \phantomsubfloat{\label{fig:results-w-soc-Jh-08-lambdaI-15}}
    \phantomsubfloat{\label{fig:results-w-soc-fermi}}
    \vspace{-1.6\baselineskip}
    \caption{\textbf{Phase diagrams with SOC.} Hole- and electron-doped phase diagrams of BLG as a function of charge density $n_{\text{e}}$ and interlayer potential $u$, at moderate Coulomb strength $\epsilon_{\mathrm{r}} = 20$. In \textbf{(a)--(f)} we consider cases with ferromagnetic Hund's coupling $J_{\mathrm{H}} = 4, \SI{8}{\electronvolt\ucdot\unitcellarea}$ and induced Ising SOC strengths $\lambda_{\mathrm{I}} = 0.5, 1, \SI{1.5}{\milli\electronvolt}$.
    Different phases are denoted by their color and hatching (see \cref{tab:symmetry-orders-legends} for legends). The fully degenerate ($g = 4$) phases without SOC now acquire a small spin-valley-locked polarization (\icontext{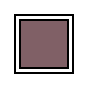}) due to the explicit symmetry breaking by $\lambda_{\mathrm{I}}$. Yellow dashed lines denote line-cuts further explored in \cref{fig:results-linecuts}. \textbf{(g)} Fermi surface structure at numbered points in panel \textbf{a}. Colors denote the number of mean-field valence bands and conduction bands occupied by carriers in the hole- and electron-doped regimes, respectively.}
    \label{fig:results-w-soc}
\end{figure*}

\subsection{No spin-orbit coupling}
\label{sec:phase-diagrams/no-ising}

To set the stage, we present in \cref{fig:results-wo-soc} the ground state phase diagram of BLG without SOC. \Cref{fig:results-wo-soc-Jh-00} showcases the situation without Hund's coupling, where only long-range Coulomb interactions are included. The structure of the phase diagram consists of a cascade of phase transitions as either electron or hole density is tuned across the van Hove singularity~\cite{Szabo2022}; generically, doping away from charge neutrality induces a series of transitions from the symmetry-unbroken phase (with full spin/valley degeneracy, $g = 4$) to generalized quarter metals ($g = 1$), half metals ($g = 2$), three-quarter metals $(g = 1$), and back to the symmetric phase. 
The different symmetry-broken phases can be further classified according to their transformation properties under $\mathrm{U}(1)_{\mathrm{v}}$ valley rotations (see \cref{tab:symmetry-orders-legends}). States that preserve $\mathrm{U}(1)_{\mathrm{v}}$ are generalized Stoner ferromagnets, in which a subset of the four spin/valley flavors is preferentially occupied, while states that break that symmetry exhibit IVC order that spontaneously hybridizes the two graphene valleys. We note that IVC phases are more prevalent in the phase diagram of bilayer graphene compared to its rhombohedral trilayer cousin~\cite{Koh2024}---this trend is particularly evident in the sector of $g = 2$ states\footnote{This trend can be ascribed to two tendencies. First, the van Hove singularities are generally more pronounced in rhombohedral trilayer graphene than BLG; the correspondingly enhanced interaction effects favor Stoner ferromagnets over IVC states (see also \cref{fig:results-interaction-strength}). Second, the Fermi surfaces of BLG tend to be more trigonally warped than their rhombohedral trilayer counterparts (compare for example \cref{fig:results-wo-soc} in this work to Figs.~4 and 5 in Ref.~\onlinecite{Koh2024}), which favors the development of intervalley coherence.}. In addition to the spin-degenerate, generalized half-metal ($g = 2$) IVC states, we find their spin-polarized, non-degenerate ($g = 1$) counterparts within the regions that host quarter-metal phases~\cite{Das2024}. Overall, the results we obtain in this limit are qualitatively consistent with those of Ref.~\onlinecite{Ming2023}.

Importantly, this phase diagram possesses numerous artificial degeneracies between different ground state orders as a consequence of
the enlarged $\mathrm{SU}(2) \times \mathrm{SU}(2)$ symmetry 
in the $J_{\mathrm{H}} = 0$ limit; see Appendix \ref{app:no-Hunds} for details.
As discussed above, the Hund's interaction reduces the symmetry of the model 
down to physical global $\mathrm{SU}(2)_{\mathrm{s}}$ rotations, and its inclusion is thus critical in order to capture the full richness of the problem. 
This reduction qualitatively affects the phase competition, resulting in the phase diagrams of \cref{fig:results-wo-soc-Jh-04,fig:results-wo-soc-Jh-08}.
First, $J_{\mathrm{H}} > 0$ breaks the degeneracy of the various $g = 2$ Stoner phases in favor of spin-polarized ferromagnets (\icontext{icon-HM-SP-nom}), consistent with experimental observations~\cite{zhou2022isospin, Seiler2022, delaBarrera2022}. Ferromagnetic Hund's coupling additionally breaks the degeneracy between the different $g = 2$ IVC states in favor of the spin-triplet IVC$_{\text{z}}$ (\icontext{icon-HM-IVC-Z-nom}). Likewise, the spin-polarized SP-IVC phase (\icontext{icon-QM-SP-IVC-nom}) is selected from the various $g = 1$ IVC states. More quantitatively, phase boundaries move with increasing $J_{\mathrm{H}}$ to prefer states that host a large magnetic moment (\eg~the fully spin-polarized half-metal phase), which enjoy a larger energy advantage from the Hund's coupling. Representative Fermi surfaces of the various metallic ground states are presented in \cref{fig:results-wo-soc-fermi}.

\begin{figure*}
    \centering
    \includegraphics[width = 1\linewidth]{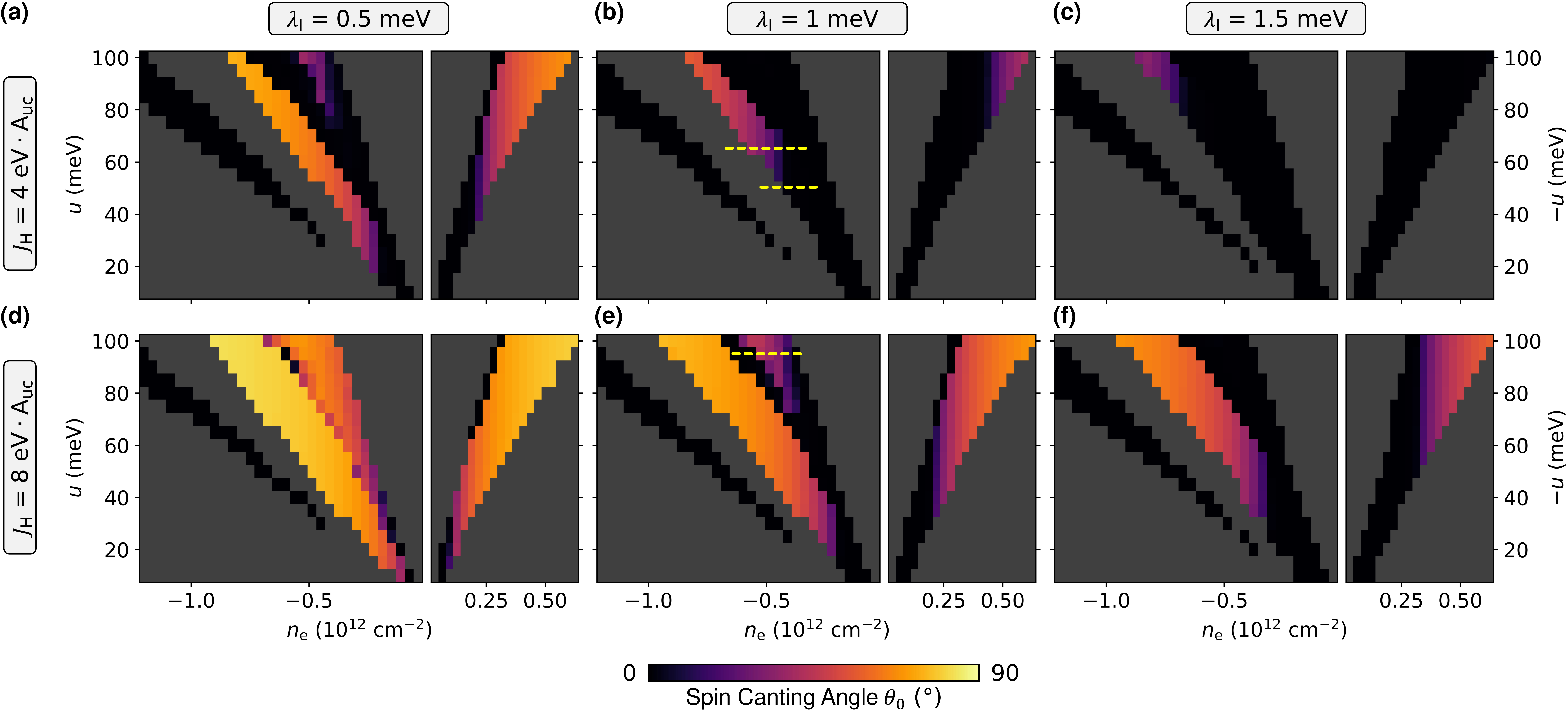}
    \phantomsubfloat{\label{fig:results-canting-Jh-04-lambdaI-05}}
    \phantomsubfloat{\label{fig:results-canting-Jh-04-lambdaI-10}}
    \phantomsubfloat{\label{fig:results-canting-Jh-04-lambdaI-15}}
    \phantomsubfloat{\label{fig:results-canting-Jh-08-lambdaI-05}}
    \phantomsubfloat{\label{fig:results-canting-Jh-08-lambdaI-10}}
    \phantomsubfloat{\label{fig:results-canting-Jh-08-lambdaI-15}}
    \vspace{-1.6\baselineskip}
    \caption{\textbf{Ferromagnetism and spin canting.} Spin canting angle $\theta_0$ extracted from our Hartree-Fock simulations using \cref{eq:canting_angle}, as a function of charge density $n_{\text{e}}$ and interlayer potential $u$. Here $\theta_0 = 0$ corresponds to a purely spin-valley-locked state (with zero net spin magnetization) while the limiting case $\theta_0 = 90^\circ$ denotes the ferromagnetic state expected without Ising SOC. We focus on $g = 2$ Stoner phases and $g = 1, 2$ IVC regions of the phase diagrams, which may exhibit spin-canting; remaining areas are masked in dark gray. Parameters used for the various panels follow those in \crefrange{fig:results-w-soc-Jh-04-lambdaI-05}{fig:results-w-soc-Jh-08-lambdaI-15}, respectively.}
    \label{fig:results-canting-angle}
\end{figure*}

\subsection{Effects of Ising SOC}
\label{sec:phase-diagrams/ising}

The case with Ising SOC but \emph{without} Hund's coupling---which also features a pathological symmetry group, $\mathrm{U}(1) \times \mathrm{U}(1)$---is described in \cref{app:no-Hunds} for completeness and comparison with the results of Ref.~\onlinecite{Ming2023}. In what follows we focus on simulations that include both Hund's coupling and $\si{\milli\electronvolt}$-scale Ising SOC; addition of the latter qualitatively changes the stabilized ground-state orders as shown in \cref{fig:results-w-soc}. 

First, and most intuitively, for sufficiently large $\lambda_{\mathrm{I}}$ or low $D$ field (corresponding to regimes where single-particle physics dominate over interaction effects), the preferred $g = 2$ Stoner ferromagnet becomes the spin-valley-locked (SVL) state (\icontext{icon-HM-SVL-nom}). 
Here, the electron spins minimize energy from Ising SOC by aligning in opposite out-of-plane directions in the two valleys, thus preserving all symmetries of the Hamiltonian. In contrast, for smaller $\lambda_{\mathrm{I}}$ and/or at larger $D$ fields, the system finds a compromise between the competing requirements of Hund's coupling and Ising SOC by tilting the electron spins in each valley by an angle $\theta_0$ away from their respective Ising axes~\cite{Koh2024, Arp2024, Dong2024}. In-plane spin polarization is thus generated along
a spontaneously chosen direction (parametrized \eg~by $\tau^0 s^x$), which coexists with spin-valley locking (described by $\tau^z s^z$). As these two order parameters anti-commute, the resulting phase (\icontext{icon-HM-SC-nom})---which we refer to in the following as the spin-canted Stoner ferromagnet (C)---remains doubly degenerate ($g = 2$). While time reversal $\mathcal{T} = \tau^x s^y \mathcal{K}$ (where $\mathcal{K}$ denotes complex conjugation) is broken by the net spin magnetization of the spin-canted state C, composing $\mathcal{T}$ with a global spin rotation around the $z$ axis yields an antiunitary symmetry $\mathcal{T}_{\mathrm{s}} = e^{i\pi s^z/2}\mathcal{T}$ that remains unbroken.

Ising SOC also alters the IVC ground states of \cref{fig:results-wo-soc-Jh-04,fig:results-wo-soc-Jh-08}.
The IVC$_{\text{z}}$ phase (\icontext{icon-HM-IVC-Z-nom}) 
now accommodates an underlying SVL polarization; we denote this state by SVL+IVC$_{\text{z}}$ (\icontext{icon-HM-SVL-IVC-Z-nom}).
Again, because spin-valley locking described by $\tau^z s^z$ anticommutes with the IVC$_{\text{z}}$ order parameter $\tau^x s^z$, the resulting state remains two-fold degenerate ($g = 2$). Although this phase breaks microscopic time reversal $\mathcal{T}$, it preserves an 
antiunitary symmetry $\mathcal{T}_{\mathrm{v}} = \tau^y s^y \mathcal{K}$ that combines $\mathcal{T}$ and a valley rotation~\cite{Koh2024}.

For moderate to large values of $\lambda_{\mathrm{I}} / J_{\mathrm{H}}$, the $g = 1$ IVC states align their spin quantization axes with the out-of-plane Ising axes, leading to a spin-valley-locked IVC phase (SVL-IVC/\icontext{icon-QM-SVC-nom}) whose valley and spin textures along the Fermi surfaces are intertwined~\cite{Koh2024, Zhumagulov2023} (see \cref{fig:ivc-states-SVL-IVC} for an illustration). This state was recently evidenced in a closely related system, rhombohedral trilayer graphene proximitized by WSe$_2$~\cite{Caitlin2024}. 
The SVL-IVC phase breaks time reversal $\mathcal{T}$ 
but preserves both the valley- and spin-rotated antiunitaries $\mathcal{T}_{\mathrm{v}}$ and $\mathcal{T}_{\mathrm{s}}$---a consequence of the underlying spin-valley locking. 

Finally, at low $\lambda_{\mathrm{I}}/J_{\mathrm{H}}$ (see \cref{fig:results-w-soc-Jh-04-lambdaI-05,fig:results-w-soc-Jh-08-lambdaI-05,fig:results-w-soc-Jh-08-lambdaI-10}), the competing requirements of Hund's coupling and Ising SOC generate a new type of intervalley coherent order: the spin-canted IVC phase (C-IVC/\icontext{icon-QM-SCC-nom}), where intervalley coherence sets in between the two Fermi surfaces of the corresponding $g = 2$ spin-canted state. The momentum-resolved texture of this state is very similar to its SVL-IVC cousin, except for the finite in-plane spin moment generated by spin-canting (see \cref{fig:ivc-states-SC-IVC}). 
Like the spin-canted C state, the C-IVC phase preserves the antiunitary symmetry $\mathcal{T}_{\mathrm{s}}$. Representative Fermi surfaces of various ground states obtained in the presence of Hund's coupling and Ising SOC are shown in \cref{fig:results-w-soc-fermi}.

Importantly, all of the IVC states we obtain
possess an antiunitary symmetry that enforces Fermi surface degeneracies, making them well-nested for zero-momentum pairing instabilities. This symmetry protection is operative provided the relevant $\mathrm{U}(1)$ symmetry [either $\mathrm{U}(1)_{\mathrm{v}}$ or $\mathrm{U}(1)_{\mathrm{s}}$] is preserved microscopically (\ie~broken spontaneously rather than explicitly).
In particular, short-range disorder potentials will be detrimental to phases relying on $\mathcal{T}_{\mathrm{v}}$ by inducing intervalley scatterings, whereas Rashba SOC 
will counteract the protection offered by $\mathcal{T}_{\mathrm{s}}$.

We note that the spin-canting transition between the two $g = 1$ IVC phases (SVL-IVC/\icontext{icon-QM-SVC-nom} and C-IVC/\icontext{icon-QM-SCC-nom}) occurs at a much \emph{smaller} value of $\lambda_{\mathrm{I}} / J_{\mathrm{H}}$ as compared to the transition between their $g = 2$ valley-symmetric counterparts (SVL/\icontext{icon-HM-SVL-nom} and C/\icontext{icon-HM-SC-nom});
see \eg~the evolution in \cref{fig:results-w-soc-Jh-08-lambdaI-05} through \cref{fig:results-w-soc-Jh-08-lambdaI-15}.
This trend occurs because the competition between Ising SOC (a single-particle effect) and Hund's coupling (an interaction effect) is density dependent~\cite{Koh2024, Arp2024, Dong2024}. For example, in a simple Ginzburg-Landau description of the spin-canting transition~\cite{Dong2024, Caitlin2024}, canting order disappears in favor of spin-valley locking when $\lambda_{\mathrm{I}} \geq \lambda_{\mathrm{I}}^{\mathrm{c}}$, with the critical SOC strength $\lambda_{\mathrm{I}}^{\mathrm{c}} = 2 J_{\mathrm{H}} n_0$. Here $n_0$ is the polarization density of the phase---\ie~the density difference between the majority and minority spin components, which equals the total carrier density for a fully polarized state. For low polarization densities (as in the $g = 1$ IVC phases mentioned above)
Ising SOC can therefore more efficiently combat Hund's coupling tendencies, as compared to larger polarization densities in the corresponding $g = 2$ states.

\Cref{fig:results-w-soc} also reveals interesting trends with increasing $\lambda_{\mathrm{I}}$ that can be compared to recent experiments systematically inducing Ising SOC through twist-angle engineering of the BLG/WSe$_2$ interface~\cite{Yiran2024}. In particular, the experimental data exhibit a counter-intuitive trend in the properties of the most prominent superconducting region of BLG/WSe$_2$ when $\lambda_{\mathrm{I}}$ increases: the area populated by superconductivity \emph{shrinks} within the regime of accessible $D$ field and density, while the optimal critical temperature is \emph{enhanced}. The first property can be naturally tied to the competition between different phases evidenced by our calculations. Namely, the symmetric spin-valley-locked (SVL) phase (\icontext{icon-HM-SVL-nom}) is preferred by Ising SOC, and correspondingly expands with increasing $\lambda_{\mathrm{I}}$ at the expense of the two competing symmetry-broken $g = 2$ states: the intervalley coherent SVL+IVC$_{\text{z}}$ state (\icontext{icon-HM-SVL-IVC-Z-nom}) and the spin-canted C state (\icontext{icon-HM-SC-nom}). These two states are therefore natural candidates for the normal state hosting superconductivity in these devices, as their phase space also shrinks upon increasing $\lambda_{\mathrm{I}}$
---see \cref{sec:screening} and \cref{sec:discussion} for more discussion on this important point. 

\Cref{fig:results-canting-angle} shows how the spin canting angle $\theta_0$ evolves throughout the phase diagram, with different parameters $J_{\mathrm{H}}$ and $\lambda_{\mathrm{I}}$. We extract $\theta_0$ from our simulations as
\begin{equation}
\label{eq:canting_angle}
    \theta_0 = \arctan \abs{ \frac{ \langle \tau^0 s^x \rangle}{ \langle \tau^z s^z \rangle} },
\end{equation}
where $\langle \hdots \rangle$ denotes an expectation value taken in the Hartree-Fock ground state. Throughout the spin-canted region of the phase diagram $\theta_0$ evolves continuously as shown in \cref{fig:results-canting-angle}, with larger canting angles (\ie~closer to the ferromagnetic limit $\theta_0 = 90^\circ$) for larger $D$ fields and doping. Both of these trends can be understood from the free-energy picture described above~\cite{Dong2024}: stronger $D$ fields and increased doping level generally lead to a larger polarization density $n_0$, which favors Hund's coupling's effectiveness over Ising SOC. Increasing $J_{\mathrm{H}}$ leads to an expansion of the spin-canting phase and canting angles closer to $\theta_0 = 90^\circ$. In contrast, enhancing $\lambda_{\mathrm{I}}$ leads to a reduction of the phase space occupied by canting order, as well as smaller canting angles.  In the pairing scenario of Ref.~\onlinecite{Dong2024}, which relies on the magnon modes associated with the development of canting order, small canting angles (closer to the spin-valley-locked limit $\theta_0 = 0$) yield stronger pairing interaction. Hence, together with our Hartree-Fock analysis this scenario provides a plausible explanation for the experimentally observed tunable enhancement of $T_{\mathrm{c}}$ with $\lambda_{\mathrm{I}}$ accompanied by a phase space reduction for superconductivity~\cite{Yiran2024}.

\section{Intervalley coherent and spin-canting phase transitions}
\label{sec:phase-transitions}

\begin{figure}
    \centering
    \includegraphics[width = 1\linewidth]{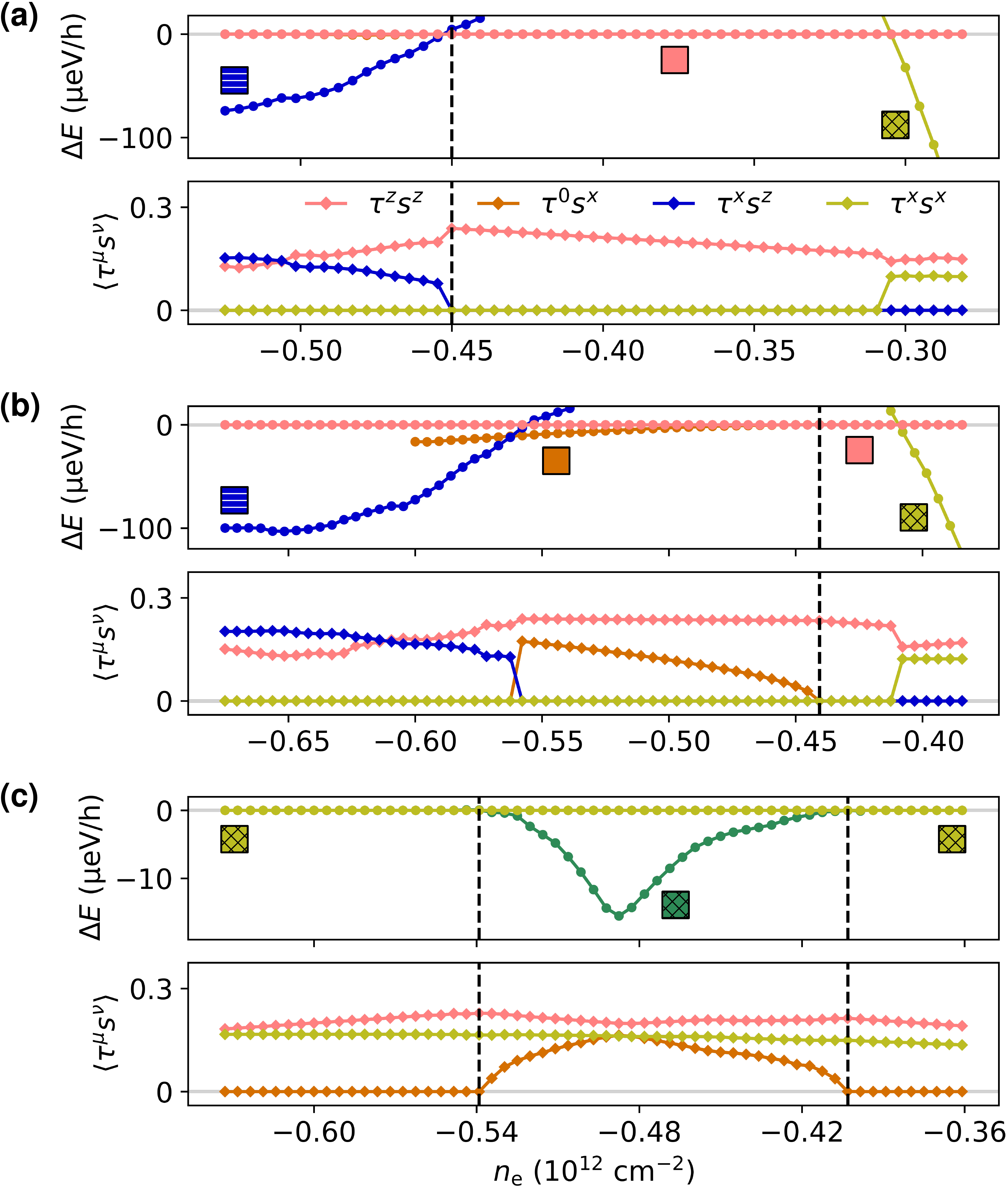}
    \phantomsubfloat{\label{fig:results-linecuts-Jh-04-lambdaI-10-u-050}}
    \phantomsubfloat{\label{fig:results-linecuts-Jh-04-lambdaI-10-u-065}}
    \phantomsubfloat{\label{fig:results-linecuts-Jh-08-lambdaI-05-u-095}}
    \vspace{-1.6\baselineskip}
    \caption{ \textbf{Nature of $\mathrm{U}(1)$-breaking phase transitions.} Energy difference between relevant ground states, normalized by the total hole density (top panels), and expectation value of order parameters $\tau^\mu s^\nu$ of interest (bottom panels), along three different cuts shown by yellow dashed lines in \cref{fig:results-w-soc,fig:results-canting-angle} for $\lambda_{\mathrm{I}} = \SI{1}{\milli\electronvolt}$. $\mathrm{U}(1)$-breaking phase transitions of interest are identified by vertical black dashed lines. \textbf{(a)} Focusing on $J_{\mathrm{H}} = \SI{4}{\electronvolt\ucdot\unitcellarea}$ and $u = \SI{50}{\milli\electronvolt}$, we characterize the $\mathrm{U}(1)_{\mathrm{v}}$-breaking transition between the symmetric SVL state (\icontext{icon-HM-SVL-nom}) and the SVL+IVC$_{\text{z}}$ state (\icontext{icon-HM-SVL-IVC-Z-nom}). The transition appears weakly first order, as evidenced by the level crossing between the respective ground state energies and the small jump in the $\tau^x s^z$ order parameter. \textbf{(b)} Taking a higher $u = \SI{65}{\milli\electronvolt}$, we see the appearance of the spin-canting (C) solution (\icontext{icon-HM-SC-nom}). While the C to SVL+IVC$_{\text{z}}$ transition is first order due to symmetry considerations, the onset of spin canting 
    (\ie~the SVL to C transition) appears continuous within the numerical accuracy of our simulations.
    \textbf{(c)} A similar scenario plays out for both spin-canting transitions between SVL-IVC (\icontext{icon-QM-SVC-nom}) and C-IVC (\icontext{icon-QM-SCC-nom}) states, shown here for $u = \SI{95}{\milli\electronvolt}$ and larger $J_{\mathrm{H}} = \SI{8}{\electronvolt\ucdot\unitcellarea}$. These results were computed on a finer momentum grid with ${\sim}3300$ points.
    }
    \label{fig:results-linecuts}
\end{figure}

As shown above, the mean-field phase diagram of spin-orbit-coupled BLG hosts multiple phase transitions where a $\mathrm{U}(1)$ symmetry is spontaneously broken. These come in two broad categories: transitions associated with the onset of IVC order, where $\mathrm{U}(1)_{\mathrm{v}}$ is broken, and those where spin canting sets in, which break $\mathrm{U}(1)_{\mathrm{s}}$ spin rotations about the $z$ axis.

Such transitions can occur between states with different Fermi surface degeneracies---\eg~from a SVL-IVC  (\icontext{icon-QM-SVC-nom}) to a SVL (\icontext{icon-HM-SVL-nom}) state (\cref{fig:results-linecuts-Jh-04-lambdaI-10-u-050,fig:results-linecuts-Jh-04-lambdaI-10-u-065})---or that break different symmetries of the underlying Hamiltonian---\eg~from a C (\icontext{icon-HM-SC-nom}) state to a SVL+IVC$_{\text{z}}$ (\icontext{icon-HM-SVL-IVC-Z-nom}) state (\cref{fig:results-linecuts-Jh-04-lambdaI-10-u-065}). In these cases one expects first-order transitions, as confirmed in our Hartree-Fock numerics.  On the other hand, continuous transitions would be of interest to exotic proposals for pairing in BLG\footnote{First-order phase transitions could also be relevant for critical pairing scenarios if they are associated with a (nearly) divergent susceptibility~\cite{Arp2024, Raines2024unconventional}.}. At mean-field level, a promising setting to look for continuous transitions arises when only the relevant $\mathrm{U}(1)$ symmetry is spontaneously broken at a transition between states with the same $g$. This situation occurs for `pure' IVC or spin-canting transitions detailed below. Here, symmetry considerations alone are insufficient to predict the order of the transition, which is ultimately determined by Stoner~\cite{Coleman2015} or IVC energetics.

\Cref{fig:results-linecuts} hones in on three representative examples of $\mathrm{U}(1)$-breaking phase transitions, following line cuts identified by yellow dashed lines in \cref{fig:results-w-soc,fig:results-canting-angle}.
\Cref{fig:results-linecuts-Jh-04-lambdaI-10-u-050} explores the $\mathrm{U}(1)_{\mathrm{v}}$ breaking transition where the symmetric SVL state (\icontext{icon-HM-SVL-nom}) develops intervalley coherence and turns into the SVL+IVC$_{\text{z}}$ state (\icontext{icon-HM-SVL-IVC-Z-nom}) with increased hole-doping. Even though both states are generalized half metals with $g = 2$, the transition appears first order, with a visible level crossing of the two (symmetry-broken and symmetric) ground state energies and a jump in the $\tau^x s^z$ order parameter. In \cref{fig:results-linecuts-Jh-04-lambdaI-10-u-065} we show the spin-canting transition, where in-plane magnetization is spontaneously generated starting from the symmetric SVL phase (\icontext{icon-HM-SVL-nom}), leading to the C phase (\icontext{icon-HM-SC-nom}). This transition appears continuous within the precision of our Hartree-Fock algorithm: 
the $\tau^0 s^x$ order parameter smoothly vanishes, and the energy of the symmetry-breaking solution merges with the symmetry-preserving one. A similar situation occurs for spin-canting transitions\footnote{Note that the energy advantage from developing spin canting is much lower than for other symmetry-breaking phases. This occurs because only Ising SOC and Hund's coupling are involved in the canting transitions---both long-range Coulomb repulsion and kinetic energy are agnostic to the presence of canting order.}  between the two $g = 1$ IVC states, SVL-IVC (\icontext{icon-QM-SVC-nom}) and C-IVC (\icontext{icon-QM-SCC-nom}), as shown in \cref{fig:results-linecuts-Jh-08-lambdaI-05-u-095}. The presence of these two putative continuous phase transitions 
calls for an examination of pairing mechanisms based on critical spin canting fluctuations.

\section{Distinguishing normal-state orders through screening}
\label{sec:screening}

\begin{figure*}
    \centering
    \includegraphics[width = 1\linewidth]{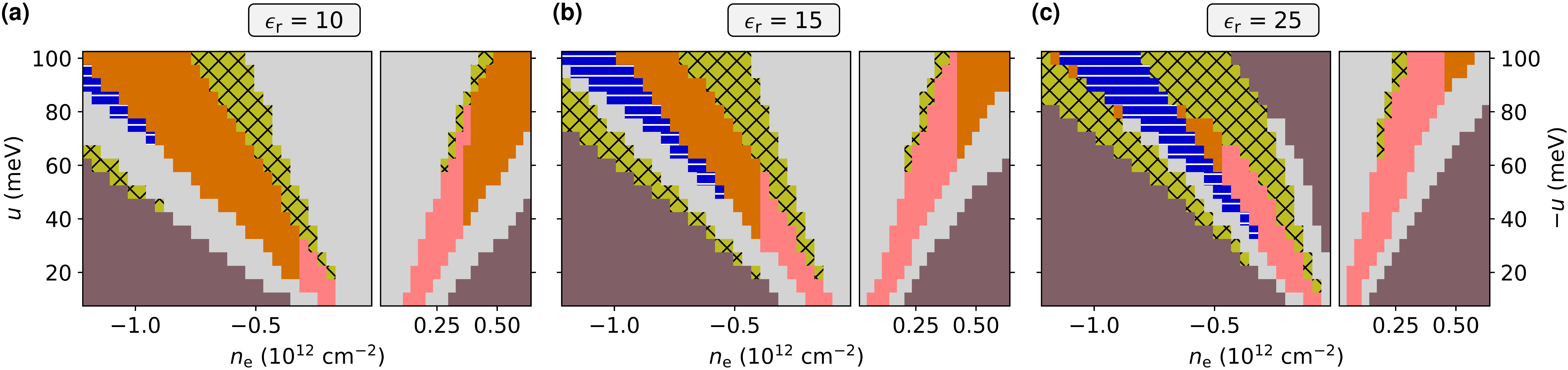}
    \phantomsubfloat{\label{fig:results-interaction-eps-10}}
    \phantomsubfloat{\label{fig:results-interaction-eps-15}}
    \phantomsubfloat{\label{fig:results-interaction-eps-25}}
    \vspace{-1.6\baselineskip}
    \caption{\textbf{Effects of screening on the phase competition.} Hole- and electron-doped phase diagrams of BLG as a function of charge density $n_{\text{e}}$ and interlayer potential $u$, at Coulomb strengths \textbf{(a)} $\epsilon_{\mathrm{r}} = 10$, \textbf{(b)} $\epsilon_{\mathrm{r}} = 15$, and \textbf{(c)} $\epsilon_{\mathrm{r}} = 25$. All panels correspond to Hund's coupling $J_{\mathrm{H}} = \SI{4}{\electronvolt\ucdot\unitcellarea}$ and  Ising SOC strength $\lambda_{\mathrm{I}} = \SI{1}{\milli\electronvolt}$. The $\epsilon_{\mathrm{r}} = 20$ case is presented in \cref{fig:results-w-soc-Jh-04-lambdaI-10}.}
    \label{fig:results-interaction-strength}
\end{figure*}

Based on their trends with increasing $\lambda_{\mathrm{I}}$ and the presence of an antiunitary symmetry ($\mathcal{T}_{\mathrm{s}}$ or $\mathcal{T}_{\mathrm{v}}$), we identified in \cref{sec:phase-diagrams/ising} two natural candidates for the normal state of the dominant superconducting region in BLG/WSe$_2$: either the spin-canted C (\icontext{icon-HM-SC-nom}) or intervalley coherent SVL+IVC$_{\text{z}}$ (\icontext{icon-HM-SVL-IVC-Z-nom}) generalized half metals ($g = 2$).  Experimentally distinguishing these orders---or ruling them out---poses an important challenge.  Direct probes that search for the in-plane magnetic moment emerging in the spin-canted phase or lattice-scale symmetry-breaking order stemming from intervalley coherence provide one possible detection avenue (see next section for further discussion).  Here, we propose an alternative approach that exploits the different response these orders exhibit to changes in screening.  

To this end, \cref{fig:results-interaction-strength} investigates the role of long-range Coulomb interaction strength, controlled by the screening parameter $\epsilon_{\mathrm{r}}$, on the phase diagram of spin-orbit-coupled BLG. For concreteness we fix $\lambda_{\mathrm{I}} = \SI{1}{\milli\electronvolt}$ and $J_{\mathrm{H}} = \SI{4}{\electronvolt\ucdot\unitcellarea}$ and present results for $\epsilon_{\mathrm{r}} = 10$, $15$ and $25$; compare to $\epsilon_{\mathrm{r}} = 20$ shown in \cref{fig:results-w-soc-Jh-04-lambdaI-10}. When interaction strength is increased ($\epsilon_{\mathrm{r}}$ is decreased), the symmetry-breaking regions of the phase diagram expand, as expected. The spin-canted phase in particular \emph{grows} at the expense of the neighboring symmetry-preserving spin-valley-locked phase. The competition between IVC phases and their Stoner ferromagnet counterparts is also influenced by $\epsilon_{\mathrm{r}}$---but crucially with an opposite trend.  Generalized Stoner ferromagnets, which exhibit roughly uniform polarization across their Fermi sea, are favored at large interaction strength.  By contrast, IVC phases are more prominent at weaker interaction strength, as they minimize their energy from a combination of polarization energy and their ability to rotate the valley pseudospin as a function of momentum $\vb{k}$ to adjust to the trigonally warped band structure of BLG (see \cref{fig:ivc-states}).  In particular, the hole-doped SVL+IVC$_{\text{z}}$ half-metal state \emph{shrinks} as the interaction strength increases.  

These opposing, screening-dependent trends are expected to be mirrored in superconducting phases whose emergence is controlled by either candidate normal state, \eg~via Goldstone~\cite{Dong2024} or critical fluctuations~\cite{Chatterjee2022,Dong2023,dong2023signatures}.
In this case, superconductivity descending from the spin-canted C (intervalley coherent SVL+IVC$_{\text{z}}$) phase is expected to occupy a smaller (larger) phase space upon an increase in screening. 
Conversely, other 
pairing mechanisms that are less reliant on the underlying normal state---such as acoustic phonons~\cite{Chou2022, Rubio2024} or Kohn-Luttinger~\cite{Jimeno-Pozo2022, Wagner2023, son2024}---are also sensitive to increased screening: respectively, an enhancement and a reduction of $T_c$ is then predicted.
A superconducting region's response to screening strength could therefore be used to help infer the pairing mechanism when considered in conjunction with normal-state evolution.

Experimentally, screening can be controlled in a variety of ways. For example, the gate distance could be varied by encapsulating with different thicknesses of \hBN~\cite{Saito2020, Stepanov2020}; the \hBN substrate could also be supplemented by a stronger dielectric~\cite{Veyrat2020, Coissard2022} or another 2D material with tunable charge density~\cite{Liu2021, Liu2022b}.

\section{Outlook and discussion}
\label{sec:discussion}

We have provided a detailed Hartree-Fock investigation of the delicate phase competition in BLG with proximity-induced spin-orbit interaction. In particular, inclusion of Hund's coupling leads to qualitative differences as compared to previous studies~\cite{Ming2023, Wang2024, Zhumagulov2024}. The most striking addition comes in the form of \emph{spin-canting} order, 
in which the system spontaneously generates an in-plane spin magnetization on top of the spin-valley locking mandated by Ising SOC. 
Such an order can exist both on its own (giving the generalized half-metal, $g = 2$ state, C/\icontext{icon-HM-SC-nom}) or in combination with intervalley coherence (leading to a generalized quarter-metal, $g = 1$ state, C-IVC/\icontext{icon-QM-SCC-nom}).
Moreover, we showed that spin-triplet intervalley coherent order and spin-valley locking coexist in the preferred $g = 2$ IVC state (SVL+IVC$_{\text{z}}$/\icontext{icon-HM-SVL-IVC-Z-nom}) while maintaining a two-fold Fermi surface degeneracy.

These results may prove crucial to understanding the phenomenology of SOC-proximitized BLG devices. First, the area occupied by both symmetry-broken generalized half metals (SVL+IVC$_{\text{z}}$ and the spin-canted phase C) is reduced by increasing $\lambda_{\mathrm{I}}$---with a more pronounced effect in the spin-canted case. These two states are thus promising candidates to host the strongest superconducting dome observed in BLG/WSe$_2$, whose area is also markedly reduced with increasing $\lambda_{\mathrm{I}}$, and furnish a promising starting point to investigate superconducting instabilities. In particular, both of these states spontaneously break a continuous $\mathrm{U}(1)$ symmetry and therefore host low-energy collective (Goldstone) modes. A recent work~\cite{Dong2024} focusing on the spin-canted phase described a pairing mechanism that crucially relies on its soft magnon modes; a similar analysis for low-lying fluctuations around the IVC orders identified here would be an interesting direction to pursue~\cite{Chatterjee2022, You2022, dong2023signatures}. Furthermore, the onset of canting order appears continuous within our Hartree-Fock treatment, which motivates further study of critical spin-canting fluctuations. 

One of the new superconducting regions reported in Ref.~\onlinecite{Yiran2024} arises out of a singly degenerate, $g = 1$ normal state. Because valley polarization is highly detrimental to zero-momentum Cooper pairing (see however Ref.~\onlinecite{han2024signatureschiralsuperconductivityrhombohedral} for a possible finite-momentum pairing state arising from a valley-polarized quarter metal in tetralayer graphene),
the corresponding normal state likely exhibits intervalley coherence.
Whether this putative IVC state is spin-valley locked (SVL-IVC/\icontext{icon-QM-SVC-nom}) or also possesses spin-canting order (C-IVC/\icontext{icon-QM-SCC-nom}), and whether the (almost) continuous transition between the two, explored in \cref{fig:results-linecuts-Jh-08-lambdaI-05-u-095}, can be tied to superconductivity in this system, constitute interesting directions for future research. 
Moreover, superconductivity was recently reported on the \emph{electron-doped} side of proximitized BLG/WSe$_2$ devices. This superconductor exhibits a fairly large $T_{\mathrm{c}} \sim \SI{200}{\milli\kelvin}$ and arises from a generalized half-metal ($g = 2$) state at markedly higher displacement fields compared to the hole-doped superconductors~\cite{li2024tunable}. Interestingly, spin-canting instabilities in our Hartree-Fock simulations (see \cref{fig:results-w-soc,fig:results-canting-angle}) also appear at higher $D$ fields on the electron-doped side, due to the particle-hole asymmetry in the low-energy band structure of BLG (see also \cref{fig:schematic}).
Intervalley coherence with $g = 2$, in contrast, is not observed on the electron-doped side in our calculations.

Future experiments will be necessary to distinguish between the candidate normal states we propose (or different competing states outside the scope of the present study), and to determine whether superconductivity on the hole and electron-doped side of BLG/WSe$_2$ has a common origin. Nano-SQUID measurements~\cite{Vasyukov2013} could image fringe magnetic fields associated with magnetic ordering and thus track the onset of spin canting~\cite{Arp2024, Caitlin2024}. The behavior of phase boundaries with applied (in-plane and out-of-plane) magnetic fields, as extracted from compressibility and/or transport, can help constrain the spin and orbital moment of the phase of interest~\cite{zhou2022isospin, Seiler2022, delaBarrera2022, Arp2024, Caitlin2024}.
Scanning tunneling probes can directly image the atomic-scale charge-density wave characteristic of an IVC state. In rhombohedral graphene devices, the need for dual gating presents a challenge to conventional STM measurements (as compared to \eg~studies of intervalley coherence in monolayer graphene~\cite{Coissard2022, Liu2022} or moir\'e  systems~\cite{Nuckolls2023, Kim2023}). This challenge could perhaps be alleviated in a gate-defined junction geometry, with the characteristic $\smash{\sqrt{3}} \times \smash{\sqrt{3}}$ modulation accessible in the exposed junction area~\cite{Thomson2022, Xie2023}. Magnetoresistance measurements were also recently proposed as a transport signature of intervalley coherence in rhombohedral graphene~\cite{Wei2023weaklocalizationprobeintervalley}.

A complementary approach to resolve the nature of superconductivity in spin-orbit-proximitized BLG consists of transport experiments with variable screening.  
In \cref{sec:screening}
we showed that our two candidate normal states with $g = 2$ have opposing behaviors as screening is enhanced (recall \cref{fig:results-interaction-strength}): the SVL+IVC$_{\text{z}}$ state expands, while the spin-canted solution shrinks by moving towards higher $D$ fields and (electron or hole) doping levels. 
Combining different experimental probes may prove crucial to constrain the nature of the normal states hosting spin-orbit-enabled superconductivity in BLG and other rhombohedral graphene systems and ultimately shed light on their pairing mechanism.

\section*{Acknowledgments}

We are grateful to Zhiyu Dong, Cyprian Lewandowski, Stevan Nadj-Perge, Gal Shavit, Andrea Young, and Yiran Zhang for insightful discussions and collaborations on related projects. 
J.~M.~K.~is grateful for support from the Agency for Science, Technology and Research (A*STAR) Graduate Academy, Singapore. A.~T.~acknowledges support from the National Science Foundation under Grant No.~2341066. \'E.~L.-H.~was supported by the Gordon and Betty Moore Foundation’s EPiQS Initiative, Grant GBMF8682. The U.S. Department of Energy, Office of Science, National Quantum Information Science Research Centers, Quantum Science Center supported the high-performance computing as well as the symmetry analysis component of this work. Additional support was provided by the Caltech Institute for Quantum Information and Matter, an NSF Physics Frontiers Center (Grant No. PHY-2317110), and the Walter Burke Institute for Theoretical Physics at Caltech. This work was performed in part at the Aspen Center for Physics, which is supported by National Science Foundation grant PHY-2210452.

\bibliography{references}

\begin{thebibliography}{81}%
\makeatletter
\providecommand \@ifxundefined [1]{%
 \@ifx{#1\undefined}
}%
\providecommand \@ifnum [1]{%
 \ifnum #1\expandafter \@firstoftwo
 \else \expandafter \@secondoftwo
 \fi
}%
\providecommand \@ifx [1]{%
 \ifx #1\expandafter \@firstoftwo
 \else \expandafter \@secondoftwo
 \fi
}%
\providecommand \natexlab [1]{#1}%
\providecommand \enquote  [1]{``#1''}%
\providecommand \bibnamefont  [1]{#1}%
\providecommand \bibfnamefont [1]{#1}%
\providecommand \citenamefont [1]{#1}%
\providecommand \href@noop [0]{\@secondoftwo}%
\providecommand \href [0]{\begingroup \@sanitize@url \@href}%
\providecommand \@href[1]{\@@startlink{#1}\@@href}%
\providecommand \@@href[1]{\endgroup#1\@@endlink}%
\providecommand \@sanitize@url [0]{\catcode `\\12\catcode `\$12\catcode `\&12\catcode `\#12\catcode `\^12\catcode `\_12\catcode `\%12\relax}%
\providecommand \@@startlink[1]{}%
\providecommand \@@endlink[0]{}%
\providecommand \url  [0]{\begingroup\@sanitize@url \@url }%
\providecommand \@url [1]{\endgroup\@href {#1}{\urlprefix }}%
\providecommand \urlprefix  [0]{URL }%
\providecommand \Eprint [0]{\href }%
\providecommand \doibase [0]{https://doi.org/}%
\providecommand \selectlanguage [0]{\@gobble}%
\providecommand \bibinfo  [0]{\@secondoftwo}%
\providecommand \bibfield  [0]{\@secondoftwo}%
\providecommand \translation [1]{[#1]}%
\providecommand \BibitemOpen [0]{}%
\providecommand \bibitemStop [0]{}%
\providecommand \bibitemNoStop [0]{.\EOS\space}%
\providecommand \EOS [0]{\spacefactor3000\relax}%
\providecommand \BibitemShut  [1]{\csname bibitem#1\endcsname}%
\let\auto@bib@innerbib\@empty
\bibitem [{\citenamefont {Shi}\ \emph {et~al.}(2020)\citenamefont {Shi}, \citenamefont {Xu}, \citenamefont {Yang}, \citenamefont {Slizovskiy}, \citenamefont {Morozov}, \citenamefont {Son}, \citenamefont {Ozdemir}, \citenamefont {Mullan}, \citenamefont {Barrier}, \citenamefont {Yin}, \citenamefont {Berdyugin}, \citenamefont {Piot}, \citenamefont {Taniguchi}, \citenamefont {Watanabe}, \citenamefont {Fal'ko}, \citenamefont {Novoselov}, \citenamefont {Geim},\ and\ \citenamefont {Mishchenko}}]{Shi2020}%
  \BibitemOpen
  \bibfield  {author} {\bibinfo {author} {\bibfnamefont {Y.}~\bibnamefont {Shi}}, \bibinfo {author} {\bibfnamefont {S.}~\bibnamefont {Xu}}, \bibinfo {author} {\bibfnamefont {Y.}~\bibnamefont {Yang}}, \bibinfo {author} {\bibfnamefont {S.}~\bibnamefont {Slizovskiy}}, \bibinfo {author} {\bibfnamefont {S.~V.}\ \bibnamefont {Morozov}}, \bibinfo {author} {\bibfnamefont {S.-K.}\ \bibnamefont {Son}}, \bibinfo {author} {\bibfnamefont {S.}~\bibnamefont {Ozdemir}}, \bibinfo {author} {\bibfnamefont {C.}~\bibnamefont {Mullan}}, \bibinfo {author} {\bibfnamefont {J.}~\bibnamefont {Barrier}}, \bibinfo {author} {\bibfnamefont {J.}~\bibnamefont {Yin}}, \bibinfo {author} {\bibfnamefont {A.~I.}\ \bibnamefont {Berdyugin}}, \bibinfo {author} {\bibfnamefont {B.~A.}\ \bibnamefont {Piot}}, \bibinfo {author} {\bibfnamefont {T.}~\bibnamefont {Taniguchi}}, \bibinfo {author} {\bibfnamefont {K.}~\bibnamefont {Watanabe}}, \bibinfo {author} {\bibfnamefont {V.~I.}\ \bibnamefont {Fal'ko}}, \bibinfo {author} {\bibfnamefont {K.~S.}\
  \bibnamefont {Novoselov}}, \bibinfo {author} {\bibfnamefont {A.~K.}\ \bibnamefont {Geim}},\ and\ \bibinfo {author} {\bibfnamefont {A.}~\bibnamefont {Mishchenko}},\ }\bibfield  {title} {\bibinfo {title} {Electronic phase separation in multilayer rhombohedral graphite},\ }\href {https://doi.org/10.1038/s41586-020-2568-2} {\bibfield  {journal} {\bibinfo  {journal} {Nature}\ }\textbf {\bibinfo {volume} {584}},\ \bibinfo {pages} {210} (\bibinfo {year} {2020})}\BibitemShut {NoStop}%
\bibitem [{\citenamefont {Zhou}\ \emph {et~al.}(2021{\natexlab{a}})\citenamefont {Zhou}, \citenamefont {Xie}, \citenamefont {Ghazaryan}, \citenamefont {Holder}, \citenamefont {Ehrets}, \citenamefont {Spanton}, \citenamefont {Taniguchi}, \citenamefont {Watanabe}, \citenamefont {Berg}, \citenamefont {Serbyn},\ and\ \citenamefont {Young}}]{zhou2021half}%
  \BibitemOpen
  \bibfield  {author} {\bibinfo {author} {\bibfnamefont {H.}~\bibnamefont {Zhou}}, \bibinfo {author} {\bibfnamefont {T.}~\bibnamefont {Xie}}, \bibinfo {author} {\bibfnamefont {A.}~\bibnamefont {Ghazaryan}}, \bibinfo {author} {\bibfnamefont {T.}~\bibnamefont {Holder}}, \bibinfo {author} {\bibfnamefont {J.~R.}\ \bibnamefont {Ehrets}}, \bibinfo {author} {\bibfnamefont {E.~M.}\ \bibnamefont {Spanton}}, \bibinfo {author} {\bibfnamefont {T.}~\bibnamefont {Taniguchi}}, \bibinfo {author} {\bibfnamefont {K.}~\bibnamefont {Watanabe}}, \bibinfo {author} {\bibfnamefont {E.}~\bibnamefont {Berg}}, \bibinfo {author} {\bibfnamefont {M.}~\bibnamefont {Serbyn}},\ and\ \bibinfo {author} {\bibfnamefont {A.~F.}\ \bibnamefont {Young}},\ }\bibfield  {title} {\bibinfo {title} {Half- and quarter-metals in rhombohedral trilayer graphene},\ }\href {https://doi.org/10.1038/s41586-021-03938-w} {\bibfield  {journal} {\bibinfo  {journal} {Nature}\ }\textbf {\bibinfo {volume} {598}},\ \bibinfo {pages} {429} (\bibinfo {year}
  {2021}{\natexlab{a}})}\BibitemShut {NoStop}%
\bibitem [{\citenamefont {Zhou}\ \emph {et~al.}(2021{\natexlab{b}})\citenamefont {Zhou}, \citenamefont {Xie}, \citenamefont {Taniguchi}, \citenamefont {Watanabe},\ and\ \citenamefont {Young}}]{zhou2021superconductivity}%
  \BibitemOpen
  \bibfield  {author} {\bibinfo {author} {\bibfnamefont {H.}~\bibnamefont {Zhou}}, \bibinfo {author} {\bibfnamefont {T.}~\bibnamefont {Xie}}, \bibinfo {author} {\bibfnamefont {T.}~\bibnamefont {Taniguchi}}, \bibinfo {author} {\bibfnamefont {K.}~\bibnamefont {Watanabe}},\ and\ \bibinfo {author} {\bibfnamefont {A.~F.}\ \bibnamefont {Young}},\ }\bibfield  {title} {\bibinfo {title} {Superconductivity in rhombohedral trilayer graphene},\ }\href {https://doi.org/10.1038/s41586-021-03926-0} {\bibfield  {journal} {\bibinfo  {journal} {Nature}\ }\textbf {\bibinfo {volume} {598}},\ \bibinfo {pages} {434} (\bibinfo {year} {2021}{\natexlab{b}})}\BibitemShut {NoStop}%
\bibitem [{\citenamefont {Zhou}\ \emph {et~al.}(2022)\citenamefont {Zhou}, \citenamefont {Holleis}, \citenamefont {Saito}, \citenamefont {Cohen}, \citenamefont {Huynh}, \citenamefont {Patterson}, \citenamefont {Yang}, \citenamefont {Taniguchi}, \citenamefont {Watanabe},\ and\ \citenamefont {Young}}]{zhou2022isospin}%
  \BibitemOpen
  \bibfield  {author} {\bibinfo {author} {\bibfnamefont {H.}~\bibnamefont {Zhou}}, \bibinfo {author} {\bibfnamefont {L.}~\bibnamefont {Holleis}}, \bibinfo {author} {\bibfnamefont {Y.}~\bibnamefont {Saito}}, \bibinfo {author} {\bibfnamefont {L.}~\bibnamefont {Cohen}}, \bibinfo {author} {\bibfnamefont {W.}~\bibnamefont {Huynh}}, \bibinfo {author} {\bibfnamefont {C.~L.}\ \bibnamefont {Patterson}}, \bibinfo {author} {\bibfnamefont {F.}~\bibnamefont {Yang}}, \bibinfo {author} {\bibfnamefont {T.}~\bibnamefont {Taniguchi}}, \bibinfo {author} {\bibfnamefont {K.}~\bibnamefont {Watanabe}},\ and\ \bibinfo {author} {\bibfnamefont {A.~F.}\ \bibnamefont {Young}},\ }\bibfield  {title} {\bibinfo {title} {Isospin magnetism and spin-polarized superconductivity in {B}ernal bilayer graphene},\ }\href {https://doi.org/10.1126/science.abm8386} {\bibfield  {journal} {\bibinfo  {journal} {Science}\ }\textbf {\bibinfo {volume} {375}},\ \bibinfo {pages} {774} (\bibinfo {year} {2022})}\BibitemShut {NoStop}%
\bibitem [{\citenamefont {Seiler}\ \emph {et~al.}(2022)\citenamefont {Seiler}, \citenamefont {Geisenhof}, \citenamefont {Winterer}, \citenamefont {Watanabe}, \citenamefont {Taniguchi}, \citenamefont {Xu}, \citenamefont {Zhang},\ and\ \citenamefont {Weitz}}]{Seiler2022}%
  \BibitemOpen
  \bibfield  {author} {\bibinfo {author} {\bibfnamefont {A.~M.}\ \bibnamefont {Seiler}}, \bibinfo {author} {\bibfnamefont {F.~R.}\ \bibnamefont {Geisenhof}}, \bibinfo {author} {\bibfnamefont {F.}~\bibnamefont {Winterer}}, \bibinfo {author} {\bibfnamefont {K.}~\bibnamefont {Watanabe}}, \bibinfo {author} {\bibfnamefont {T.}~\bibnamefont {Taniguchi}}, \bibinfo {author} {\bibfnamefont {T.}~\bibnamefont {Xu}}, \bibinfo {author} {\bibfnamefont {F.}~\bibnamefont {Zhang}},\ and\ \bibinfo {author} {\bibfnamefont {R.~T.}\ \bibnamefont {Weitz}},\ }\bibfield  {title} {\bibinfo {title} {Quantum cascade of correlated phases in trigonally warped bilayer graphene},\ }\href {https://doi.org/10.1038/s41586-022-04937-1} {\bibfield  {journal} {\bibinfo  {journal} {Nature}\ }\textbf {\bibinfo {volume} {608}},\ \bibinfo {pages} {298} (\bibinfo {year} {2022})}\BibitemShut {NoStop}%
\bibitem [{\citenamefont {de~la Barrera}\ \emph {et~al.}(2022)\citenamefont {de~la Barrera}, \citenamefont {Aronson}, \citenamefont {Zheng}, \citenamefont {Watanabe}, \citenamefont {Taniguchi}, \citenamefont {Ma}, \citenamefont {Jarillo-Herrero},\ and\ \citenamefont {Ashoori}}]{delaBarrera2022}%
  \BibitemOpen
  \bibfield  {author} {\bibinfo {author} {\bibfnamefont {S.~C.}\ \bibnamefont {de~la Barrera}}, \bibinfo {author} {\bibfnamefont {S.}~\bibnamefont {Aronson}}, \bibinfo {author} {\bibfnamefont {Z.}~\bibnamefont {Zheng}}, \bibinfo {author} {\bibfnamefont {K.}~\bibnamefont {Watanabe}}, \bibinfo {author} {\bibfnamefont {T.}~\bibnamefont {Taniguchi}}, \bibinfo {author} {\bibfnamefont {Q.}~\bibnamefont {Ma}}, \bibinfo {author} {\bibfnamefont {P.}~\bibnamefont {Jarillo-Herrero}},\ and\ \bibinfo {author} {\bibfnamefont {R.}~\bibnamefont {Ashoori}},\ }\bibfield  {title} {\bibinfo {title} {Cascade of isospin phase transitions in {B}ernal-stacked bilayer graphene at zero magnetic field},\ }\href {https://doi.org/10.1038/s41567-022-01616-w} {\bibfield  {journal} {\bibinfo  {journal} {Nat. Phys.}\ }\textbf {\bibinfo {volume} {18}},\ \bibinfo {pages} {771} (\bibinfo {year} {2022})}\BibitemShut {NoStop}%
\bibitem [{\citenamefont {Kerelsky}\ \emph {et~al.}(2021)\citenamefont {Kerelsky}, \citenamefont {Rubio-Verd{\'{u}}}, \citenamefont {Xian}, \citenamefont {Kennes}, \citenamefont {Halbertal}, \citenamefont {Finney}, \citenamefont {Song}, \citenamefont {Turkel}, \citenamefont {Wang}, \citenamefont {Watanabe}, \citenamefont {Taniguchi}, \citenamefont {Hone}, \citenamefont {Dean}, \citenamefont {Basov}, \citenamefont {Rubio},\ and\ \citenamefont {Pasupathy}}]{Kerelsky2021}%
  \BibitemOpen
  \bibfield  {author} {\bibinfo {author} {\bibfnamefont {A.}~\bibnamefont {Kerelsky}}, \bibinfo {author} {\bibfnamefont {C.}~\bibnamefont {Rubio-Verd{\'{u}}}}, \bibinfo {author} {\bibfnamefont {L.}~\bibnamefont {Xian}}, \bibinfo {author} {\bibfnamefont {D.~M.}\ \bibnamefont {Kennes}}, \bibinfo {author} {\bibfnamefont {D.}~\bibnamefont {Halbertal}}, \bibinfo {author} {\bibfnamefont {N.}~\bibnamefont {Finney}}, \bibinfo {author} {\bibfnamefont {L.}~\bibnamefont {Song}}, \bibinfo {author} {\bibfnamefont {S.}~\bibnamefont {Turkel}}, \bibinfo {author} {\bibfnamefont {L.}~\bibnamefont {Wang}}, \bibinfo {author} {\bibfnamefont {K.}~\bibnamefont {Watanabe}}, \bibinfo {author} {\bibfnamefont {T.}~\bibnamefont {Taniguchi}}, \bibinfo {author} {\bibfnamefont {J.}~\bibnamefont {Hone}}, \bibinfo {author} {\bibfnamefont {C.}~\bibnamefont {Dean}}, \bibinfo {author} {\bibfnamefont {D.~N.}\ \bibnamefont {Basov}}, \bibinfo {author} {\bibfnamefont {A.}~\bibnamefont {Rubio}},\ and\ \bibinfo {author} {\bibfnamefont {A.~N.}\
  \bibnamefont {Pasupathy}},\ }\bibfield  {title} {\bibinfo {title} {Moir{\'{e}}less correlations in {ABCA} graphene},\ }\href {https://doi.org/10.1073/pnas.2017366118} {\bibfield  {journal} {\bibinfo  {journal} {PNAS}\ }\textbf {\bibinfo {volume} {118}} (\bibinfo {year} {2021})}\BibitemShut {NoStop}%
\bibitem [{\citenamefont {Han}\ \emph {et~al.}(2023{\natexlab{a}})\citenamefont {Han}, \citenamefont {Lu}, \citenamefont {Scuri}, \citenamefont {Sung}, \citenamefont {Wang}, \citenamefont {Han}, \citenamefont {Watanabe}, \citenamefont {Taniguchi}, \citenamefont {Park},\ and\ \citenamefont {Ju}}]{Han2023}%
  \BibitemOpen
  \bibfield  {author} {\bibinfo {author} {\bibfnamefont {T.}~\bibnamefont {Han}}, \bibinfo {author} {\bibfnamefont {Z.}~\bibnamefont {Lu}}, \bibinfo {author} {\bibfnamefont {G.}~\bibnamefont {Scuri}}, \bibinfo {author} {\bibfnamefont {J.}~\bibnamefont {Sung}}, \bibinfo {author} {\bibfnamefont {J.}~\bibnamefont {Wang}}, \bibinfo {author} {\bibfnamefont {T.}~\bibnamefont {Han}}, \bibinfo {author} {\bibfnamefont {K.}~\bibnamefont {Watanabe}}, \bibinfo {author} {\bibfnamefont {T.}~\bibnamefont {Taniguchi}}, \bibinfo {author} {\bibfnamefont {H.}~\bibnamefont {Park}},\ and\ \bibinfo {author} {\bibfnamefont {L.}~\bibnamefont {Ju}},\ }\bibfield  {title} {\bibinfo {title} {Correlated insulator and chern insulators in pentalayer rhombohedral-stacked graphene},\ }\href {https://doi.org/10.1038/s41565-023-01520-1} {\bibfield  {journal} {\bibinfo  {journal} {Nat. Nanotechnology}\ }\textbf {\bibinfo {volume} {19}},\ \bibinfo {pages} {181–187} (\bibinfo {year} {2023}{\natexlab{a}})}\BibitemShut {NoStop}%
\bibitem [{\citenamefont {Liu}\ \emph {et~al.}(2023)\citenamefont {Liu}, \citenamefont {Zheng}, \citenamefont {Sha}, \citenamefont {Lyu}, \citenamefont {Li}, \citenamefont {Park}, \citenamefont {Ren}, \citenamefont {Watanabe}, \citenamefont {Taniguchi}, \citenamefont {Jia}, \citenamefont {Luo}, \citenamefont {Shi}, \citenamefont {Jung},\ and\ \citenamefont {Chen}}]{liu2023interactiondriven}%
  \BibitemOpen
  \bibfield  {author} {\bibinfo {author} {\bibfnamefont {K.}~\bibnamefont {Liu}}, \bibinfo {author} {\bibfnamefont {J.}~\bibnamefont {Zheng}}, \bibinfo {author} {\bibfnamefont {Y.}~\bibnamefont {Sha}}, \bibinfo {author} {\bibfnamefont {B.}~\bibnamefont {Lyu}}, \bibinfo {author} {\bibfnamefont {F.}~\bibnamefont {Li}}, \bibinfo {author} {\bibfnamefont {Y.}~\bibnamefont {Park}}, \bibinfo {author} {\bibfnamefont {Y.}~\bibnamefont {Ren}}, \bibinfo {author} {\bibfnamefont {K.}~\bibnamefont {Watanabe}}, \bibinfo {author} {\bibfnamefont {T.}~\bibnamefont {Taniguchi}}, \bibinfo {author} {\bibfnamefont {J.}~\bibnamefont {Jia}}, \bibinfo {author} {\bibfnamefont {W.}~\bibnamefont {Luo}}, \bibinfo {author} {\bibfnamefont {Z.}~\bibnamefont {Shi}}, \bibinfo {author} {\bibfnamefont {J.}~\bibnamefont {Jung}},\ and\ \bibinfo {author} {\bibfnamefont {G.}~\bibnamefont {Chen}},\ }\bibfield  {title} {\bibinfo {title} {Spontaneous broken-symmetry insulator and metals in tetralayer rhombohedral graphene},\ }\href
  {https://doi.org/10.1038/s41565-023-01558-1} {\bibfield  {journal} {\bibinfo  {journal} {Nature Nanotechnology}\ }\textbf {\bibinfo {volume} {19}},\ \bibinfo {pages} {188–195} (\bibinfo {year} {2023})}\BibitemShut {NoStop}%
\bibitem [{\citenamefont {Han}\ \emph {et~al.}(2023{\natexlab{b}})\citenamefont {Han}, \citenamefont {Lu}, \citenamefont {Scuri}, \citenamefont {Sung}, \citenamefont {Wang}, \citenamefont {Han}, \citenamefont {Watanabe}, \citenamefont {Taniguchi}, \citenamefont {Fu}, \citenamefont {Park},\ and\ \citenamefont {Ju}}]{Han2023a}%
  \BibitemOpen
  \bibfield  {author} {\bibinfo {author} {\bibfnamefont {T.}~\bibnamefont {Han}}, \bibinfo {author} {\bibfnamefont {Z.}~\bibnamefont {Lu}}, \bibinfo {author} {\bibfnamefont {G.}~\bibnamefont {Scuri}}, \bibinfo {author} {\bibfnamefont {J.}~\bibnamefont {Sung}}, \bibinfo {author} {\bibfnamefont {J.}~\bibnamefont {Wang}}, \bibinfo {author} {\bibfnamefont {T.}~\bibnamefont {Han}}, \bibinfo {author} {\bibfnamefont {K.}~\bibnamefont {Watanabe}}, \bibinfo {author} {\bibfnamefont {T.}~\bibnamefont {Taniguchi}}, \bibinfo {author} {\bibfnamefont {L.}~\bibnamefont {Fu}}, \bibinfo {author} {\bibfnamefont {H.}~\bibnamefont {Park}},\ and\ \bibinfo {author} {\bibfnamefont {L.}~\bibnamefont {Ju}},\ }\bibfield  {title} {\bibinfo {title} {Orbital multiferroicity in pentalayer rhombohedral graphene},\ }\href {https://doi.org/10.1038/s41586-023-06572-w} {\bibfield  {journal} {\bibinfo  {journal} {Nature}\ }\textbf {\bibinfo {volume} {623}},\ \bibinfo {pages} {41–47} (\bibinfo {year} {2023}{\natexlab{b}})}\BibitemShut {NoStop}%
\bibitem [{\citenamefont {Seiler}\ \emph {et~al.}(2024{\natexlab{a}})\citenamefont {Seiler}, \citenamefont {Statz}, \citenamefont {Weimer}, \citenamefont {Jacobsen}, \citenamefont {Watanabe}, \citenamefont {Taniguchi}, \citenamefont {Dong}, \citenamefont {Levitov},\ and\ \citenamefont {Weitz}}]{Seiler2024}%
  \BibitemOpen
  \bibfield  {author} {\bibinfo {author} {\bibfnamefont {A.~M.}\ \bibnamefont {Seiler}}, \bibinfo {author} {\bibfnamefont {M.}~\bibnamefont {Statz}}, \bibinfo {author} {\bibfnamefont {I.}~\bibnamefont {Weimer}}, \bibinfo {author} {\bibfnamefont {N.}~\bibnamefont {Jacobsen}}, \bibinfo {author} {\bibfnamefont {K.}~\bibnamefont {Watanabe}}, \bibinfo {author} {\bibfnamefont {T.}~\bibnamefont {Taniguchi}}, \bibinfo {author} {\bibfnamefont {Z.}~\bibnamefont {Dong}}, \bibinfo {author} {\bibfnamefont {L.~S.}\ \bibnamefont {Levitov}},\ and\ \bibinfo {author} {\bibfnamefont {R.~T.}\ \bibnamefont {Weitz}},\ }\bibfield  {title} {\bibinfo {title} {Interaction-driven quasi-insulating ground states of gapped electron-doped bilayer graphene},\ }\href {https://doi.org/10.1103/PhysRevLett.133.066301} {\bibfield  {journal} {\bibinfo  {journal} {Phys. Rev. Lett.}\ }\textbf {\bibinfo {volume} {133}},\ \bibinfo {pages} {066301} (\bibinfo {year} {2024}{\natexlab{a}})}\BibitemShut {NoStop}%
\bibitem [{\citenamefont {Novoselov}\ \emph {et~al.}(2006)\citenamefont {Novoselov}, \citenamefont {McCann}, \citenamefont {Morozov}, \citenamefont {Fal’ko}, \citenamefont {Katsnelson}, \citenamefont {Zeitler}, \citenamefont {Jiang}, \citenamefont {Schedin},\ and\ \citenamefont {Geim}}]{Novoselov2006}%
  \BibitemOpen
  \bibfield  {author} {\bibinfo {author} {\bibfnamefont {K.~S.}\ \bibnamefont {Novoselov}}, \bibinfo {author} {\bibfnamefont {E.}~\bibnamefont {McCann}}, \bibinfo {author} {\bibfnamefont {S.~V.}\ \bibnamefont {Morozov}}, \bibinfo {author} {\bibfnamefont {V.~I.}\ \bibnamefont {Fal’ko}}, \bibinfo {author} {\bibfnamefont {M.~I.}\ \bibnamefont {Katsnelson}}, \bibinfo {author} {\bibfnamefont {U.}~\bibnamefont {Zeitler}}, \bibinfo {author} {\bibfnamefont {D.}~\bibnamefont {Jiang}}, \bibinfo {author} {\bibfnamefont {F.}~\bibnamefont {Schedin}},\ and\ \bibinfo {author} {\bibfnamefont {A.~K.}\ \bibnamefont {Geim}},\ }\bibfield  {title} {\bibinfo {title} {Unconventional quantum {H}all effect and {B}erry’s phase of 2\ensuremath{\pi} in bilayer graphene},\ }\href {https://doi.org/10.1038/nphys245} {\bibfield  {journal} {\bibinfo  {journal} {Nat. Phys.}\ }\textbf {\bibinfo {volume} {2}},\ \bibinfo {pages} {177–180} (\bibinfo {year} {2006})}\BibitemShut {NoStop}%
\bibitem [{\citenamefont {McCann}\ and\ \citenamefont {Fal'ko}(2006)}]{McCann2006}%
  \BibitemOpen
  \bibfield  {author} {\bibinfo {author} {\bibfnamefont {E.}~\bibnamefont {McCann}}\ and\ \bibinfo {author} {\bibfnamefont {V.~I.}\ \bibnamefont {Fal'ko}},\ }\bibfield  {title} {\bibinfo {title} {Landau-level degeneracy and quantum {H}all effect in a graphite bilayer},\ }\href {https://doi.org/10.1103/PhysRevLett.96.086805} {\bibfield  {journal} {\bibinfo  {journal} {Phys. Rev. Lett.}\ }\textbf {\bibinfo {volume} {96}},\ \bibinfo {pages} {086805} (\bibinfo {year} {2006})}\BibitemShut {NoStop}%
\bibitem [{\citenamefont {Zhang}\ \emph {et~al.}(2009)\citenamefont {Zhang}, \citenamefont {Tang}, \citenamefont {Girit}, \citenamefont {Hao}, \citenamefont {Martin}, \citenamefont {Zettl}, \citenamefont {Crommie}, \citenamefont {Shen},\ and\ \citenamefont {Wang}}]{Zhang2009}%
  \BibitemOpen
  \bibfield  {author} {\bibinfo {author} {\bibfnamefont {Y.}~\bibnamefont {Zhang}}, \bibinfo {author} {\bibfnamefont {T.-T.}\ \bibnamefont {Tang}}, \bibinfo {author} {\bibfnamefont {C.}~\bibnamefont {Girit}}, \bibinfo {author} {\bibfnamefont {Z.}~\bibnamefont {Hao}}, \bibinfo {author} {\bibfnamefont {M.~C.}\ \bibnamefont {Martin}}, \bibinfo {author} {\bibfnamefont {A.}~\bibnamefont {Zettl}}, \bibinfo {author} {\bibfnamefont {M.~F.}\ \bibnamefont {Crommie}}, \bibinfo {author} {\bibfnamefont {Y.~R.}\ \bibnamefont {Shen}},\ and\ \bibinfo {author} {\bibfnamefont {F.}~\bibnamefont {Wang}},\ }\bibfield  {title} {\bibinfo {title} {Direct observation of a widely tunable bandgap in bilayer graphene},\ }\href {https://doi.org/10.1038/nature08105} {\bibfield  {journal} {\bibinfo  {journal} {Nature}\ }\textbf {\bibinfo {volume} {459}},\ \bibinfo {pages} {820–823} (\bibinfo {year} {2009})}\BibitemShut {NoStop}%
\bibitem [{\citenamefont {Weitz}\ \emph {et~al.}(2010)\citenamefont {Weitz}, \citenamefont {Allen}, \citenamefont {Feldman}, \citenamefont {Martin},\ and\ \citenamefont {Yacoby}}]{Weitz2010}%
  \BibitemOpen
  \bibfield  {author} {\bibinfo {author} {\bibfnamefont {R.~T.}\ \bibnamefont {Weitz}}, \bibinfo {author} {\bibfnamefont {M.~T.}\ \bibnamefont {Allen}}, \bibinfo {author} {\bibfnamefont {B.~E.}\ \bibnamefont {Feldman}}, \bibinfo {author} {\bibfnamefont {J.}~\bibnamefont {Martin}},\ and\ \bibinfo {author} {\bibfnamefont {A.}~\bibnamefont {Yacoby}},\ }\bibfield  {title} {\bibinfo {title} {Broken-symmetry states in doubly gated suspended bilayer graphene},\ }\href {https://doi.org/10.1126/science.1194988} {\bibfield  {journal} {\bibinfo  {journal} {Science}\ }\textbf {\bibinfo {volume} {330}},\ \bibinfo {pages} {812} (\bibinfo {year} {2010})}\BibitemShut {NoStop}%
\bibitem [{\citenamefont {McCann}\ and\ \citenamefont {Koshino}(2013)}]{mccann2013electronic}%
  \BibitemOpen
  \bibfield  {author} {\bibinfo {author} {\bibfnamefont {E.}~\bibnamefont {McCann}}\ and\ \bibinfo {author} {\bibfnamefont {M.}~\bibnamefont {Koshino}},\ }\bibfield  {title} {\bibinfo {title} {The electronic properties of bilayer graphene},\ }\href {https://doi.org/10.1088/0034-4885/76/5/056503} {\bibfield  {journal} {\bibinfo  {journal} {Rep. Prog. Phys.}\ }\textbf {\bibinfo {volume} {76}},\ \bibinfo {pages} {056503} (\bibinfo {year} {2013})}\BibitemShut {NoStop}%
\bibitem [{\citenamefont {Zhang}\ \emph {et~al.}(2023)\citenamefont {Zhang}, \citenamefont {Polski}, \citenamefont {Thomson}, \citenamefont {Lantagne-Hurtubise}, \citenamefont {Lewandowski}, \citenamefont {Zhou}, \citenamefont {Watanabe}, \citenamefont {Taniguchi}, \citenamefont {Alicea},\ and\ \citenamefont {Nadj-Perge}}]{Zhang2023}%
  \BibitemOpen
  \bibfield  {author} {\bibinfo {author} {\bibfnamefont {Y.}~\bibnamefont {Zhang}}, \bibinfo {author} {\bibfnamefont {R.}~\bibnamefont {Polski}}, \bibinfo {author} {\bibfnamefont {A.}~\bibnamefont {Thomson}}, \bibinfo {author} {\bibfnamefont {{\'E}.}~\bibnamefont {Lantagne-Hurtubise}}, \bibinfo {author} {\bibfnamefont {C.}~\bibnamefont {Lewandowski}}, \bibinfo {author} {\bibfnamefont {H.}~\bibnamefont {Zhou}}, \bibinfo {author} {\bibfnamefont {K.}~\bibnamefont {Watanabe}}, \bibinfo {author} {\bibfnamefont {T.}~\bibnamefont {Taniguchi}}, \bibinfo {author} {\bibfnamefont {J.}~\bibnamefont {Alicea}},\ and\ \bibinfo {author} {\bibfnamefont {S.}~\bibnamefont {Nadj-Perge}},\ }\bibfield  {title} {\bibinfo {title} {Enhanced superconductivity in spin--orbit proximitized bilayer graphene},\ }\href {https://doi.org/10.1038/s41586-022-05446-x} {\bibfield  {journal} {\bibinfo  {journal} {Nature}\ }\textbf {\bibinfo {volume} {613}},\ \bibinfo {pages} {268} (\bibinfo {year} {2023})}\BibitemShut {NoStop}%
\bibitem [{\citenamefont {Holleis}\ \emph {et~al.}(2024)\citenamefont {Holleis}, \citenamefont {Patterson}, \citenamefont {Zhang}, \citenamefont {Vituri}, \citenamefont {Yoo}, \citenamefont {Zhou}, \citenamefont {Taniguchi}, \citenamefont {Watanabe}, \citenamefont {Berg}, \citenamefont {Nadj-Perge},\ and\ \citenamefont {Young}}]{Holleis2023}%
  \BibitemOpen
  \bibfield  {author} {\bibinfo {author} {\bibfnamefont {L.}~\bibnamefont {Holleis}}, \bibinfo {author} {\bibfnamefont {C.~L.}\ \bibnamefont {Patterson}}, \bibinfo {author} {\bibfnamefont {Y.}~\bibnamefont {Zhang}}, \bibinfo {author} {\bibfnamefont {Y.}~\bibnamefont {Vituri}}, \bibinfo {author} {\bibfnamefont {H.~M.}\ \bibnamefont {Yoo}}, \bibinfo {author} {\bibfnamefont {H.}~\bibnamefont {Zhou}}, \bibinfo {author} {\bibfnamefont {T.}~\bibnamefont {Taniguchi}}, \bibinfo {author} {\bibfnamefont {K.}~\bibnamefont {Watanabe}}, \bibinfo {author} {\bibfnamefont {E.}~\bibnamefont {Berg}}, \bibinfo {author} {\bibfnamefont {S.}~\bibnamefont {Nadj-Perge}},\ and\ \bibinfo {author} {\bibfnamefont {A.~F.}\ \bibnamefont {Young}},\ }\href {https://arxiv.org/abs/2303.00742} {\bibinfo {title} {Nematicity and orbital depairing in superconducting {B}ernal bilayer graphene with strong spin orbit coupling}} (\bibinfo {year} {2024}),\ \Eprint {https://arxiv.org/abs/2303.00742} {arXiv:2303.00742 [cond-mat.supr-con]} \BibitemShut
  {NoStop}%
\bibitem [{\citenamefont {Li}\ \emph {et~al.}(2024)\citenamefont {Li}, \citenamefont {Xu}, \citenamefont {Li}, \citenamefont {Li}, \citenamefont {Li}, \citenamefont {Watanabe}, \citenamefont {Taniguchi}, \citenamefont {Tong}, \citenamefont {Shen}, \citenamefont {Lu}, \citenamefont {Jia}, \citenamefont {Wu}, \citenamefont {Liu},\ and\ \citenamefont {Li}}]{li2024tunable}%
  \BibitemOpen
  \bibfield  {author} {\bibinfo {author} {\bibfnamefont {C.}~\bibnamefont {Li}}, \bibinfo {author} {\bibfnamefont {F.}~\bibnamefont {Xu}}, \bibinfo {author} {\bibfnamefont {B.}~\bibnamefont {Li}}, \bibinfo {author} {\bibfnamefont {J.}~\bibnamefont {Li}}, \bibinfo {author} {\bibfnamefont {G.}~\bibnamefont {Li}}, \bibinfo {author} {\bibfnamefont {K.}~\bibnamefont {Watanabe}}, \bibinfo {author} {\bibfnamefont {T.}~\bibnamefont {Taniguchi}}, \bibinfo {author} {\bibfnamefont {B.}~\bibnamefont {Tong}}, \bibinfo {author} {\bibfnamefont {J.}~\bibnamefont {Shen}}, \bibinfo {author} {\bibfnamefont {L.}~\bibnamefont {Lu}}, \bibinfo {author} {\bibfnamefont {J.}~\bibnamefont {Jia}}, \bibinfo {author} {\bibfnamefont {F.}~\bibnamefont {Wu}}, \bibinfo {author} {\bibfnamefont {X.}~\bibnamefont {Liu}},\ and\ \bibinfo {author} {\bibfnamefont {T.}~\bibnamefont {Li}},\ }\bibfield  {title} {\bibinfo {title} {Tunable superconductivity in electron- and hole-doped bernal bilayer graphene},\ }\href
  {https://doi.org/10.1038/s41586-024-07584-w} {\bibfield  {journal} {\bibinfo  {journal} {Nature}\ }\textbf {\bibinfo {volume} {631}},\ \bibinfo {pages} {300–306} (\bibinfo {year} {2024})}\BibitemShut {NoStop}%
\bibitem [{\citenamefont {Zhang}\ \emph {et~al.}(2024)\citenamefont {Zhang}, \citenamefont {Shavit}, \citenamefont {Ma}, \citenamefont {Han}, \citenamefont {Watanabe}, \citenamefont {Taniguchi}, \citenamefont {Hsieh}, \citenamefont {Lewandowski}, \citenamefont {von Oppen}, \citenamefont {Oreg},\ and\ \citenamefont {Nadj-Perge}}]{Yiran2024}%
  \BibitemOpen
  \bibfield  {author} {\bibinfo {author} {\bibfnamefont {Y.}~\bibnamefont {Zhang}}, \bibinfo {author} {\bibfnamefont {G.}~\bibnamefont {Shavit}}, \bibinfo {author} {\bibfnamefont {H.}~\bibnamefont {Ma}}, \bibinfo {author} {\bibfnamefont {Y.}~\bibnamefont {Han}}, \bibinfo {author} {\bibfnamefont {K.}~\bibnamefont {Watanabe}}, \bibinfo {author} {\bibfnamefont {T.}~\bibnamefont {Taniguchi}}, \bibinfo {author} {\bibfnamefont {D.}~\bibnamefont {Hsieh}}, \bibinfo {author} {\bibfnamefont {C.}~\bibnamefont {Lewandowski}}, \bibinfo {author} {\bibfnamefont {F.}~\bibnamefont {von Oppen}}, \bibinfo {author} {\bibfnamefont {Y.}~\bibnamefont {Oreg}},\ and\ \bibinfo {author} {\bibfnamefont {S.}~\bibnamefont {Nadj-Perge}},\ }\href {https://arxiv.org/abs/2408.10335} {\bibinfo {title} {Twist-programmable superconductivity in spin-orbit coupled bilayer graphene}} (\bibinfo {year} {2024}),\ \Eprint {https://arxiv.org/abs/2408.10335} {arXiv:2408.10335 [cond-mat.supr-con]} \BibitemShut {NoStop}%
\bibitem [{\citenamefont {Pantaleón}\ \emph {et~al.}(2023)\citenamefont {Pantaleón}, \citenamefont {Jimeno-Pozo}, \citenamefont {Sainz-Cruz}, \citenamefont {Phong}, \citenamefont {Cea},\ and\ \citenamefont {Guinea}}]{Pantaleon2023review}%
  \BibitemOpen
  \bibfield  {author} {\bibinfo {author} {\bibfnamefont {P.~A.}\ \bibnamefont {Pantaleón}}, \bibinfo {author} {\bibfnamefont {A.}~\bibnamefont {Jimeno-Pozo}}, \bibinfo {author} {\bibfnamefont {H.}~\bibnamefont {Sainz-Cruz}}, \bibinfo {author} {\bibfnamefont {V.~T.}\ \bibnamefont {Phong}}, \bibinfo {author} {\bibfnamefont {T.}~\bibnamefont {Cea}},\ and\ \bibinfo {author} {\bibfnamefont {F.}~\bibnamefont {Guinea}},\ }\bibfield  {title} {\bibinfo {title} {Superconductivity and correlated phases in non-twisted bilayer and trilayer graphene},\ }\href {https://doi.org/10.1038/s42254-023-00575-2} {\bibfield  {journal} {\bibinfo  {journal} {Nat. Rev. Phys.}\ }\textbf {\bibinfo {volume} {5}},\ \bibinfo {pages} {304–315} (\bibinfo {year} {2023})}\BibitemShut {NoStop}%
\bibitem [{\citenamefont {Chou}\ \emph {et~al.}(2022{\natexlab{a}})\citenamefont {Chou}, \citenamefont {Wu},\ and\ \citenamefont {Das~Sarma}}]{Yangzhi2022}%
  \BibitemOpen
  \bibfield  {author} {\bibinfo {author} {\bibfnamefont {Y.-Z.}\ \bibnamefont {Chou}}, \bibinfo {author} {\bibfnamefont {F.}~\bibnamefont {Wu}},\ and\ \bibinfo {author} {\bibfnamefont {S.}~\bibnamefont {Das~Sarma}},\ }\bibfield  {title} {\bibinfo {title} {Enhanced superconductivity through virtual tunneling in {B}ernal bilayer graphene coupled to ${\mathrm{wse}}_{2}$},\ }\href {https://doi.org/10.1103/PhysRevB.106.L180502} {\bibfield  {journal} {\bibinfo  {journal} {Phys. Rev. B}\ }\textbf {\bibinfo {volume} {106}},\ \bibinfo {pages} {L180502} (\bibinfo {year} {2022}{\natexlab{a}})}\BibitemShut {NoStop}%
\bibitem [{\citenamefont {Curtis}\ \emph {et~al.}(2023)\citenamefont {Curtis}, \citenamefont {Poniatowski}, \citenamefont {Xie}, \citenamefont {Yacoby}, \citenamefont {Demler},\ and\ \citenamefont {Narang}}]{Curtis2023}%
  \BibitemOpen
  \bibfield  {author} {\bibinfo {author} {\bibfnamefont {J.~B.}\ \bibnamefont {Curtis}}, \bibinfo {author} {\bibfnamefont {N.~R.}\ \bibnamefont {Poniatowski}}, \bibinfo {author} {\bibfnamefont {Y.}~\bibnamefont {Xie}}, \bibinfo {author} {\bibfnamefont {A.}~\bibnamefont {Yacoby}}, \bibinfo {author} {\bibfnamefont {E.}~\bibnamefont {Demler}},\ and\ \bibinfo {author} {\bibfnamefont {P.}~\bibnamefont {Narang}},\ }\bibfield  {title} {\bibinfo {title} {Stabilizing fluctuating spin-triplet superconductivity in graphene via induced spin-orbit coupling},\ }\href {https://doi.org/10.1103/PhysRevLett.130.196001} {\bibfield  {journal} {\bibinfo  {journal} {Phys. Rev. Lett.}\ }\textbf {\bibinfo {volume} {130}},\ \bibinfo {pages} {196001} (\bibinfo {year} {2023})}\BibitemShut {NoStop}%
\bibitem [{\citenamefont {Jimeno-Pozo}\ \emph {et~al.}(2023)\citenamefont {Jimeno-Pozo}, \citenamefont {Sainz-Cruz}, \citenamefont {Cea}, \citenamefont {Pantale\'on},\ and\ \citenamefont {Guinea}}]{Jimeno-Pozo2022}%
  \BibitemOpen
  \bibfield  {author} {\bibinfo {author} {\bibfnamefont {A.}~\bibnamefont {Jimeno-Pozo}}, \bibinfo {author} {\bibfnamefont {H.}~\bibnamefont {Sainz-Cruz}}, \bibinfo {author} {\bibfnamefont {T.}~\bibnamefont {Cea}}, \bibinfo {author} {\bibfnamefont {P.~A.}\ \bibnamefont {Pantale\'on}},\ and\ \bibinfo {author} {\bibfnamefont {F.}~\bibnamefont {Guinea}},\ }\bibfield  {title} {\bibinfo {title} {Superconductivity from electronic interactions and spin-orbit enhancement in bilayer and trilayer graphene},\ }\href {https://doi.org/10.1103/PhysRevB.107.L161106} {\bibfield  {journal} {\bibinfo  {journal} {Phys. Rev. B}\ }\textbf {\bibinfo {volume} {107}},\ \bibinfo {pages} {L161106} (\bibinfo {year} {2023})}\BibitemShut {NoStop}%
\bibitem [{\citenamefont {Dong}\ \emph {et~al.}(2023{\natexlab{a}})\citenamefont {Dong}, \citenamefont {Chubukov},\ and\ \citenamefont {Levitov}}]{Dong2023}%
  \BibitemOpen
  \bibfield  {author} {\bibinfo {author} {\bibfnamefont {Z.}~\bibnamefont {Dong}}, \bibinfo {author} {\bibfnamefont {A.~V.}\ \bibnamefont {Chubukov}},\ and\ \bibinfo {author} {\bibfnamefont {L.}~\bibnamefont {Levitov}},\ }\bibfield  {title} {\bibinfo {title} {Transformer spin-triplet superconductivity at the onset of isospin order in bilayer graphene},\ }\href {https://doi.org/10.1103/PhysRevB.107.174512} {\bibfield  {journal} {\bibinfo  {journal} {Phys. Rev. B}\ }\textbf {\bibinfo {volume} {107}},\ \bibinfo {pages} {174512} (\bibinfo {year} {2023}{\natexlab{a}})}\BibitemShut {NoStop}%
\bibitem [{\citenamefont {Dong}\ \emph {et~al.}(2023{\natexlab{b}})\citenamefont {Dong}, \citenamefont {Lee},\ and\ \citenamefont {Levitov}}]{dong2023signatures}%
  \BibitemOpen
  \bibfield  {author} {\bibinfo {author} {\bibfnamefont {Z.}~\bibnamefont {Dong}}, \bibinfo {author} {\bibfnamefont {P.~A.}\ \bibnamefont {Lee}},\ and\ \bibinfo {author} {\bibfnamefont {L.~S.}\ \bibnamefont {Levitov}},\ }\bibfield  {title} {\bibinfo {title} {Signatures of {C}ooper pair dynamics and quantum-critical superconductivity in tunable carrier bands},\ }\href {http://dx.doi.org/10.1073/pnas.2305943120} {\bibfield  {journal} {\bibinfo  {journal} {PNAS}\ }\textbf {\bibinfo {volume} {120}} (\bibinfo {year} {2023}{\natexlab{b}})}\BibitemShut {NoStop}%
\bibitem [{\citenamefont {Li}\ \emph {et~al.}(2023)\citenamefont {Li}, \citenamefont {Kuang}, \citenamefont {Jimeno-Pozo}, \citenamefont {Sainz-Cruz}, \citenamefont {Zhan}, \citenamefont {Yuan},\ and\ \citenamefont {Guinea}}]{li2023charge}%
  \BibitemOpen
  \bibfield  {author} {\bibinfo {author} {\bibfnamefont {Z.}~\bibnamefont {Li}}, \bibinfo {author} {\bibfnamefont {X.}~\bibnamefont {Kuang}}, \bibinfo {author} {\bibfnamefont {A.}~\bibnamefont {Jimeno-Pozo}}, \bibinfo {author} {\bibfnamefont {H.}~\bibnamefont {Sainz-Cruz}}, \bibinfo {author} {\bibfnamefont {Z.}~\bibnamefont {Zhan}}, \bibinfo {author} {\bibfnamefont {S.}~\bibnamefont {Yuan}},\ and\ \bibinfo {author} {\bibfnamefont {F.}~\bibnamefont {Guinea}},\ }\bibfield  {title} {\bibinfo {title} {Charge fluctuations, phonons, and superconductivity in multilayer graphene},\ }\href {https://doi.org/10.1103/PhysRevB.108.045404} {\bibfield  {journal} {\bibinfo  {journal} {Phys. Rev. B}\ }\textbf {\bibinfo {volume} {108}},\ \bibinfo {pages} {045404} (\bibinfo {year} {2023})}\BibitemShut {NoStop}%
\bibitem [{\citenamefont {Wagner}\ \emph {et~al.}(2023)\citenamefont {Wagner}, \citenamefont {Kwan}, \citenamefont {Bultinck}, \citenamefont {Simon},\ and\ \citenamefont {Parameswaran}}]{Wagner2023}%
  \BibitemOpen
  \bibfield  {author} {\bibinfo {author} {\bibfnamefont {G.}~\bibnamefont {Wagner}}, \bibinfo {author} {\bibfnamefont {Y.~H.}\ \bibnamefont {Kwan}}, \bibinfo {author} {\bibfnamefont {N.}~\bibnamefont {Bultinck}}, \bibinfo {author} {\bibfnamefont {S.~H.}\ \bibnamefont {Simon}},\ and\ \bibinfo {author} {\bibfnamefont {S.~A.}\ \bibnamefont {Parameswaran}},\ }\href {https://arxiv.org/abs/2302.00682} {\bibinfo {title} {Superconductivity from repulsive interactions in {B}ernal-stacked bilayer graphene}} (\bibinfo {year} {2023}),\ \Eprint {https://arxiv.org/abs/2302.00682} {arXiv:2302.00682 [cond-mat.supr-con]} \BibitemShut {NoStop}%
\bibitem [{\citenamefont {Shavit}\ and\ \citenamefont {Oreg}(2023)}]{Shavit2023}%
  \BibitemOpen
  \bibfield  {author} {\bibinfo {author} {\bibfnamefont {G.}~\bibnamefont {Shavit}}\ and\ \bibinfo {author} {\bibfnamefont {Y.}~\bibnamefont {Oreg}},\ }\bibfield  {title} {\bibinfo {title} {Inducing superconductivity in bilayer graphene by alleviation of the {S}toner blockade},\ }\href {https://doi.org/10.1103/PhysRevB.108.024510} {\bibfield  {journal} {\bibinfo  {journal} {Phys. Rev. B}\ }\textbf {\bibinfo {volume} {108}},\ \bibinfo {pages} {024510} (\bibinfo {year} {2023})}\BibitemShut {NoStop}%
\bibitem [{\citenamefont {Son}\ \emph {et~al.}(2024)\citenamefont {Son}, \citenamefont {Hsu},\ and\ \citenamefont {Kim}}]{son2024}%
  \BibitemOpen
  \bibfield  {author} {\bibinfo {author} {\bibfnamefont {J.~H.}\ \bibnamefont {Son}}, \bibinfo {author} {\bibfnamefont {Y.-T.}\ \bibnamefont {Hsu}},\ and\ \bibinfo {author} {\bibfnamefont {E.-A.}\ \bibnamefont {Kim}},\ }\href {https://arxiv.org/abs/2405.05442} {\bibinfo {title} {Switching between superconductivity and current density waves in {B}ernal bilayer graphene}} (\bibinfo {year} {2024}),\ \Eprint {https://arxiv.org/abs/2405.05442} {arXiv:2405.05442 [cond-mat.str-el]} \BibitemShut {NoStop}%
\bibitem [{\citenamefont {Dong}\ \emph {et~al.}(2024)\citenamefont {Dong}, \citenamefont {Étienne Lantagne-Hurtubise},\ and\ \citenamefont {Alicea}}]{Dong2024}%
  \BibitemOpen
  \bibfield  {author} {\bibinfo {author} {\bibfnamefont {Z.}~\bibnamefont {Dong}}, \bibinfo {author} {\bibnamefont {Étienne Lantagne-Hurtubise}},\ and\ \bibinfo {author} {\bibfnamefont {J.}~\bibnamefont {Alicea}},\ }\href {https://arxiv.org/abs/2406.17036} {\bibinfo {title} {Superconductivity from spin-canting fluctuations in rhombohedral graphene}} (\bibinfo {year} {2024}),\ \Eprint {https://arxiv.org/abs/2406.17036} {arXiv:2406.17036 [cond-mat.supr-con]} \BibitemShut {NoStop}%
\bibitem [{\citenamefont {Kozii}\ \emph {et~al.}(2022)\citenamefont {Kozii}, \citenamefont {Zaletel},\ and\ \citenamefont {Bultinck}}]{Kozii2022}%
  \BibitemOpen
  \bibfield  {author} {\bibinfo {author} {\bibfnamefont {V.}~\bibnamefont {Kozii}}, \bibinfo {author} {\bibfnamefont {M.~P.}\ \bibnamefont {Zaletel}},\ and\ \bibinfo {author} {\bibfnamefont {N.}~\bibnamefont {Bultinck}},\ }\bibfield  {title} {\bibinfo {title} {Spin-triplet superconductivity from intervalley {G}oldstone modes in magic-angle graphene},\ }\href {https://doi.org/10.1103/PhysRevB.106.235157} {\bibfield  {journal} {\bibinfo  {journal} {Phys. Rev. B}\ }\textbf {\bibinfo {volume} {106}},\ \bibinfo {pages} {235157} (\bibinfo {year} {2022})}\BibitemShut {NoStop}%
\bibitem [{\citenamefont {Seiler}\ \emph {et~al.}(2024{\natexlab{b}})\citenamefont {Seiler}, \citenamefont {Zhumagulov}, \citenamefont {Zollner}, \citenamefont {Yoon}, \citenamefont {Urbaniak}, \citenamefont {Geisenhof}, \citenamefont {Watanabe}, \citenamefont {Taniguchi}, \citenamefont {Fabian}, \citenamefont {Zhang},\ and\ \citenamefont {Weitz}}]{Seiler2024layerselective}%
  \BibitemOpen
  \bibfield  {author} {\bibinfo {author} {\bibfnamefont {A.~M.}\ \bibnamefont {Seiler}}, \bibinfo {author} {\bibfnamefont {Y.}~\bibnamefont {Zhumagulov}}, \bibinfo {author} {\bibfnamefont {K.}~\bibnamefont {Zollner}}, \bibinfo {author} {\bibfnamefont {C.}~\bibnamefont {Yoon}}, \bibinfo {author} {\bibfnamefont {D.}~\bibnamefont {Urbaniak}}, \bibinfo {author} {\bibfnamefont {F.~R.}\ \bibnamefont {Geisenhof}}, \bibinfo {author} {\bibfnamefont {K.}~\bibnamefont {Watanabe}}, \bibinfo {author} {\bibfnamefont {T.}~\bibnamefont {Taniguchi}}, \bibinfo {author} {\bibfnamefont {J.}~\bibnamefont {Fabian}}, \bibinfo {author} {\bibfnamefont {F.}~\bibnamefont {Zhang}},\ and\ \bibinfo {author} {\bibfnamefont {R.~T.}\ \bibnamefont {Weitz}},\ }\href {https://arxiv.org/abs/2403.17140} {\bibinfo {title} {Layer-selective spin-orbit coupling and strong correlation in bilayer graphene}} (\bibinfo {year} {2024}{\natexlab{b}}),\ \Eprint {https://arxiv.org/abs/2403.17140} {arXiv:2403.17140 [cond-mat.mes-hall]} \BibitemShut {NoStop}%
\bibitem [{\citenamefont {Koh}\ \emph {et~al.}(2024)\citenamefont {Koh}, \citenamefont {Alicea},\ and\ \citenamefont {Lantagne-Hurtubise}}]{Koh2024}%
  \BibitemOpen
  \bibfield  {author} {\bibinfo {author} {\bibfnamefont {J.~M.}\ \bibnamefont {Koh}}, \bibinfo {author} {\bibfnamefont {J.}~\bibnamefont {Alicea}},\ and\ \bibinfo {author} {\bibfnamefont {E.}~\bibnamefont {Lantagne-Hurtubise}},\ }\bibfield  {title} {\bibinfo {title} {Correlated phases in spin-orbit-coupled rhombohedral trilayer graphene},\ }\href {https://doi.org/10.1103/PhysRevB.109.035113} {\bibfield  {journal} {\bibinfo  {journal} {Phys. Rev. B}\ }\textbf {\bibinfo {volume} {109}},\ \bibinfo {pages} {035113} (\bibinfo {year} {2024})}\BibitemShut {NoStop}%
\bibitem [{\citenamefont {Xie}\ and\ \citenamefont {Das~Sarma}(2023)}]{Ming2023}%
  \BibitemOpen
  \bibfield  {author} {\bibinfo {author} {\bibfnamefont {M.}~\bibnamefont {Xie}}\ and\ \bibinfo {author} {\bibfnamefont {S.}~\bibnamefont {Das~Sarma}},\ }\bibfield  {title} {\bibinfo {title} {Flavor symmetry breaking in spin-orbit coupled bilayer graphene},\ }\href {https://doi.org/10.1103/PhysRevB.107.L201119} {\bibfield  {journal} {\bibinfo  {journal} {Phys. Rev. B}\ }\textbf {\bibinfo {volume} {107}},\ \bibinfo {pages} {L201119} (\bibinfo {year} {2023})}\BibitemShut {NoStop}%
\bibitem [{\citenamefont {Wang}\ \emph {et~al.}(2024)\citenamefont {Wang}, \citenamefont {Vila}, \citenamefont {Zaletel},\ and\ \citenamefont {Chatterjee}}]{Wang2024}%
  \BibitemOpen
  \bibfield  {author} {\bibinfo {author} {\bibfnamefont {T.}~\bibnamefont {Wang}}, \bibinfo {author} {\bibfnamefont {M.}~\bibnamefont {Vila}}, \bibinfo {author} {\bibfnamefont {M.~P.}\ \bibnamefont {Zaletel}},\ and\ \bibinfo {author} {\bibfnamefont {S.}~\bibnamefont {Chatterjee}},\ }\bibfield  {title} {\bibinfo {title} {Electrical control of spin and valley in spin-orbit coupled graphene multilayers},\ }\href {https://doi.org/10.1103/PhysRevLett.132.116504} {\bibfield  {journal} {\bibinfo  {journal} {Phys. Rev. Lett.}\ }\textbf {\bibinfo {volume} {132}},\ \bibinfo {pages} {116504} (\bibinfo {year} {2024})}\BibitemShut {NoStop}%
\bibitem [{\citenamefont {Zhumagulov}\ \emph {et~al.}(2024{\natexlab{a}})\citenamefont {Zhumagulov}, \citenamefont {Kochan},\ and\ \citenamefont {Fabian}}]{Zhumagulov2024}%
  \BibitemOpen
  \bibfield  {author} {\bibinfo {author} {\bibfnamefont {Y.}~\bibnamefont {Zhumagulov}}, \bibinfo {author} {\bibfnamefont {D.}~\bibnamefont {Kochan}},\ and\ \bibinfo {author} {\bibfnamefont {J.}~\bibnamefont {Fabian}},\ }\bibfield  {title} {\bibinfo {title} {Swapping exchange and spin-orbit induced correlated phases in proximitized bernal bilayer graphene},\ }\href {https://doi.org/10.1103/PhysRevB.110.045427} {\bibfield  {journal} {\bibinfo  {journal} {Phys. Rev. B}\ }\textbf {\bibinfo {volume} {110}},\ \bibinfo {pages} {045427} (\bibinfo {year} {2024}{\natexlab{a}})}\BibitemShut {NoStop}%
\bibitem [{\citenamefont {Li}\ and\ \citenamefont {Koshino}(2019)}]{Li2019}%
  \BibitemOpen
  \bibfield  {author} {\bibinfo {author} {\bibfnamefont {Y.}~\bibnamefont {Li}}\ and\ \bibinfo {author} {\bibfnamefont {M.}~\bibnamefont {Koshino}},\ }\bibfield  {title} {\bibinfo {title} {Twist-angle dependence of the proximity spin-orbit coupling in graphene on transition-metal dichalcogenides},\ }\href {https://doi.org/10.1103/PhysRevB.99.075438} {\bibfield  {journal} {\bibinfo  {journal} {Phys. Rev. B}\ }\textbf {\bibinfo {volume} {99}},\ \bibinfo {pages} {075438} (\bibinfo {year} {2019})}\BibitemShut {NoStop}%
\bibitem [{\citenamefont {David}\ \emph {et~al.}(2019)\citenamefont {David}, \citenamefont {Rakyta}, \citenamefont {Korm\'anyos},\ and\ \citenamefont {Burkard}}]{David2019}%
  \BibitemOpen
  \bibfield  {author} {\bibinfo {author} {\bibfnamefont {A.}~\bibnamefont {David}}, \bibinfo {author} {\bibfnamefont {P.}~\bibnamefont {Rakyta}}, \bibinfo {author} {\bibfnamefont {A.}~\bibnamefont {Korm\'anyos}},\ and\ \bibinfo {author} {\bibfnamefont {G.}~\bibnamefont {Burkard}},\ }\bibfield  {title} {\bibinfo {title} {Induced spin-orbit coupling in twisted graphene--transition metal dichalcogenide heterobilayers: Twistronics meets spintronics},\ }\href {https://doi.org/10.1103/PhysRevB.100.085412} {\bibfield  {journal} {\bibinfo  {journal} {Phys. Rev. B}\ }\textbf {\bibinfo {volume} {100}},\ \bibinfo {pages} {085412} (\bibinfo {year} {2019})}\BibitemShut {NoStop}%
\bibitem [{\citenamefont {Naimer}\ \emph {et~al.}(2021)\citenamefont {Naimer}, \citenamefont {Zollner}, \citenamefont {Gmitra},\ and\ \citenamefont {Fabian}}]{Naimer2021}%
  \BibitemOpen
  \bibfield  {author} {\bibinfo {author} {\bibfnamefont {T.}~\bibnamefont {Naimer}}, \bibinfo {author} {\bibfnamefont {K.}~\bibnamefont {Zollner}}, \bibinfo {author} {\bibfnamefont {M.}~\bibnamefont {Gmitra}},\ and\ \bibinfo {author} {\bibfnamefont {J.}~\bibnamefont {Fabian}},\ }\bibfield  {title} {\bibinfo {title} {Twist-angle dependent proximity induced spin-orbit coupling in graphene/transition metal dichalcogenide heterostructures},\ }\href {https://doi.org/10.1103/PhysRevB.104.195156} {\bibfield  {journal} {\bibinfo  {journal} {Phys. Rev. B}\ }\textbf {\bibinfo {volume} {104}},\ \bibinfo {pages} {195156} (\bibinfo {year} {2021})}\BibitemShut {NoStop}%
\bibitem [{\citenamefont {Chou}\ \emph {et~al.}(2022{\natexlab{b}})\citenamefont {Chou}, \citenamefont {Wu}, \citenamefont {Sau},\ and\ \citenamefont {Das~Sarma}}]{Chou2022}%
  \BibitemOpen
  \bibfield  {author} {\bibinfo {author} {\bibfnamefont {Y.-Z.}\ \bibnamefont {Chou}}, \bibinfo {author} {\bibfnamefont {F.}~\bibnamefont {Wu}}, \bibinfo {author} {\bibfnamefont {J.~D.}\ \bibnamefont {Sau}},\ and\ \bibinfo {author} {\bibfnamefont {S.}~\bibnamefont {Das~Sarma}},\ }\bibfield  {title} {\bibinfo {title} {Acoustic-phonon-mediated superconductivity in {B}ernal bilayer graphene},\ }\href {https://doi.org/10.1103/PhysRevB.105.L100503} {\bibfield  {journal} {\bibinfo  {journal} {Phys. Rev. B}\ }\textbf {\bibinfo {volume} {105}},\ \bibinfo {pages} {L100503} (\bibinfo {year} {2022}{\natexlab{b}})}\BibitemShut {NoStop}%
\bibitem [{\citenamefont {Viñas~Bostr\"{o}m}\ \emph {et~al.}(2024)\citenamefont {Viñas~Bostr\"{o}m}, \citenamefont {Fischer}, \citenamefont {Profe}, \citenamefont {Zhang}, \citenamefont {Kennes},\ and\ \citenamefont {Rubio}}]{Rubio2024}%
  \BibitemOpen
  \bibfield  {author} {\bibinfo {author} {\bibfnamefont {E.}~\bibnamefont {Viñas~Bostr\"{o}m}}, \bibinfo {author} {\bibfnamefont {A.}~\bibnamefont {Fischer}}, \bibinfo {author} {\bibfnamefont {J.~B.}\ \bibnamefont {Profe}}, \bibinfo {author} {\bibfnamefont {J.}~\bibnamefont {Zhang}}, \bibinfo {author} {\bibfnamefont {D.~M.}\ \bibnamefont {Kennes}},\ and\ \bibinfo {author} {\bibfnamefont {A.}~\bibnamefont {Rubio}},\ }\bibfield  {title} {\bibinfo {title} {Phonon-mediated unconventional superconductivity in rhombohedral stacked multilayer graphene},\ }\href {http://dx.doi.org/10.1038/s41524-024-01345-z} {\bibfield  {journal} {\bibinfo  {journal} {npj Computational Materials}\ }\textbf {\bibinfo {volume} {10}} (\bibinfo {year} {2024})}\BibitemShut {NoStop}%
\bibitem [{\citenamefont {Chatterjee}\ \emph {et~al.}(2022)\citenamefont {Chatterjee}, \citenamefont {Wang}, \citenamefont {Berg},\ and\ \citenamefont {Zaletel}}]{Chatterjee2022}%
  \BibitemOpen
  \bibfield  {author} {\bibinfo {author} {\bibfnamefont {S.}~\bibnamefont {Chatterjee}}, \bibinfo {author} {\bibfnamefont {T.}~\bibnamefont {Wang}}, \bibinfo {author} {\bibfnamefont {E.}~\bibnamefont {Berg}},\ and\ \bibinfo {author} {\bibfnamefont {M.~P.}\ \bibnamefont {Zaletel}},\ }\bibfield  {title} {\bibinfo {title} {Inter-valley coherent order and isospin fluctuation mediated superconductivity in rhombohedral trilayer graphene},\ }\href {https://doi.org/10.1038/s41467-022-33561-w} {\bibfield  {journal} {\bibinfo  {journal} {Nat. Commun.}\ }\textbf {\bibinfo {volume} {13}},\ \bibinfo {pages} {6013} (\bibinfo {year} {2022})}\BibitemShut {NoStop}%
\bibitem [{\citenamefont {Jung}\ and\ \citenamefont {MacDonald}(2014)}]{Jung2014}%
  \BibitemOpen
  \bibfield  {author} {\bibinfo {author} {\bibfnamefont {J.}~\bibnamefont {Jung}}\ and\ \bibinfo {author} {\bibfnamefont {A.~H.}\ \bibnamefont {MacDonald}},\ }\bibfield  {title} {\bibinfo {title} {Accurate tight-binding models for the $\ensuremath{\pi}$ bands of bilayer graphene},\ }\href {https://doi.org/10.1103/PhysRevB.89.035405} {\bibfield  {journal} {\bibinfo  {journal} {Phys. Rev. B}\ }\textbf {\bibinfo {volume} {89}},\ \bibinfo {pages} {035405} (\bibinfo {year} {2014})}\BibitemShut {NoStop}%
\bibitem [{\citenamefont {Gmitra}\ \emph {et~al.}(2016)\citenamefont {Gmitra}, \citenamefont {Kochan}, \citenamefont {H\"ogl},\ and\ \citenamefont {Fabian}}]{Gmitra2016}%
  \BibitemOpen
  \bibfield  {author} {\bibinfo {author} {\bibfnamefont {M.}~\bibnamefont {Gmitra}}, \bibinfo {author} {\bibfnamefont {D.}~\bibnamefont {Kochan}}, \bibinfo {author} {\bibfnamefont {P.}~\bibnamefont {H\"ogl}},\ and\ \bibinfo {author} {\bibfnamefont {J.}~\bibnamefont {Fabian}},\ }\bibfield  {title} {\bibinfo {title} {Trivial and inverted {D}irac bands and the emergence of quantum spin {H}all states in graphene on transition-metal dichalcogenides},\ }\href {https://doi.org/10.1103/PhysRevB.93.155104} {\bibfield  {journal} {\bibinfo  {journal} {Phys. Rev. B}\ }\textbf {\bibinfo {volume} {93}},\ \bibinfo {pages} {155104} (\bibinfo {year} {2016})}\BibitemShut {NoStop}%
\bibitem [{\citenamefont {Wang}\ \emph {et~al.}(2016)\citenamefont {Wang}, \citenamefont {Ki}, \citenamefont {Khoo}, \citenamefont {Mauro}, \citenamefont {Berger}, \citenamefont {Levitov},\ and\ \citenamefont {Morpurgo}}]{Wang2016}%
  \BibitemOpen
  \bibfield  {author} {\bibinfo {author} {\bibfnamefont {Z.}~\bibnamefont {Wang}}, \bibinfo {author} {\bibfnamefont {D.-K.}\ \bibnamefont {Ki}}, \bibinfo {author} {\bibfnamefont {J.~Y.}\ \bibnamefont {Khoo}}, \bibinfo {author} {\bibfnamefont {D.}~\bibnamefont {Mauro}}, \bibinfo {author} {\bibfnamefont {H.}~\bibnamefont {Berger}}, \bibinfo {author} {\bibfnamefont {L.~S.}\ \bibnamefont {Levitov}},\ and\ \bibinfo {author} {\bibfnamefont {A.~F.}\ \bibnamefont {Morpurgo}},\ }\bibfield  {title} {\bibinfo {title} {Origin and magnitude of `designer' spin-orbit interaction in graphene on semiconducting transition metal dichalcogenides},\ }\href {https://doi.org/10.1103/PhysRevX.6.041020} {\bibfield  {journal} {\bibinfo  {journal} {Phys. Rev. X}\ }\textbf {\bibinfo {volume} {6}},\ \bibinfo {pages} {041020} (\bibinfo {year} {2016})}\BibitemShut {NoStop}%
\bibitem [{\citenamefont {Yang}\ \emph {et~al.}(2017)\citenamefont {Yang}, \citenamefont {Lohmann}, \citenamefont {Barroso}, \citenamefont {Liao}, \citenamefont {Lin}, \citenamefont {Liu}, \citenamefont {Bartels}, \citenamefont {Watanabe}, \citenamefont {Taniguchi},\ and\ \citenamefont {Shi}}]{Yang2017}%
  \BibitemOpen
  \bibfield  {author} {\bibinfo {author} {\bibfnamefont {B.}~\bibnamefont {Yang}}, \bibinfo {author} {\bibfnamefont {M.}~\bibnamefont {Lohmann}}, \bibinfo {author} {\bibfnamefont {D.}~\bibnamefont {Barroso}}, \bibinfo {author} {\bibfnamefont {I.}~\bibnamefont {Liao}}, \bibinfo {author} {\bibfnamefont {Z.}~\bibnamefont {Lin}}, \bibinfo {author} {\bibfnamefont {Y.}~\bibnamefont {Liu}}, \bibinfo {author} {\bibfnamefont {L.}~\bibnamefont {Bartels}}, \bibinfo {author} {\bibfnamefont {K.}~\bibnamefont {Watanabe}}, \bibinfo {author} {\bibfnamefont {T.}~\bibnamefont {Taniguchi}},\ and\ \bibinfo {author} {\bibfnamefont {J.}~\bibnamefont {Shi}},\ }\bibfield  {title} {\bibinfo {title} {Strong electron-hole symmetric {R}ashba spin-orbit coupling in graphene/monolayer transition metal dichalcogenide heterostructures},\ }\href {https://doi.org/10.1103/PhysRevB.96.041409} {\bibfield  {journal} {\bibinfo  {journal} {Phys. Rev. B}\ }\textbf {\bibinfo {volume} {96}},\ \bibinfo {pages} {041409(R)} (\bibinfo {year}
  {2017})}\BibitemShut {NoStop}%
\bibitem [{\citenamefont {Zihlmann}\ \emph {et~al.}(2018)\citenamefont {Zihlmann}, \citenamefont {Cummings}, \citenamefont {Garcia}, \citenamefont {Kedves}, \citenamefont {Watanabe}, \citenamefont {Taniguchi}, \citenamefont {Sch\"onenberger},\ and\ \citenamefont {Makk}}]{Zihlmann2018}%
  \BibitemOpen
  \bibfield  {author} {\bibinfo {author} {\bibfnamefont {S.}~\bibnamefont {Zihlmann}}, \bibinfo {author} {\bibfnamefont {A.~W.}\ \bibnamefont {Cummings}}, \bibinfo {author} {\bibfnamefont {J.~H.}\ \bibnamefont {Garcia}}, \bibinfo {author} {\bibfnamefont {M.}~\bibnamefont {Kedves}}, \bibinfo {author} {\bibfnamefont {K.}~\bibnamefont {Watanabe}}, \bibinfo {author} {\bibfnamefont {T.}~\bibnamefont {Taniguchi}}, \bibinfo {author} {\bibfnamefont {C.}~\bibnamefont {Sch\"onenberger}},\ and\ \bibinfo {author} {\bibfnamefont {P.}~\bibnamefont {Makk}},\ }\bibfield  {title} {\bibinfo {title} {Large spin relaxation anisotropy and valley-{Z}eeman spin-orbit coupling in ${\mathrm{wse}}_{2}$/graphene/$h$-bn heterostructures},\ }\href {https://doi.org/10.1103/PhysRevB.97.075434} {\bibfield  {journal} {\bibinfo  {journal} {Phys. Rev. B}\ }\textbf {\bibinfo {volume} {97}},\ \bibinfo {pages} {075434} (\bibinfo {year} {2018})}\BibitemShut {NoStop}%
\bibitem [{\citenamefont {Island}\ \emph {et~al.}(2019)\citenamefont {Island}, \citenamefont {Cui}, \citenamefont {Lewandowski}, \citenamefont {Khoo}, \citenamefont {Spanton}, \citenamefont {Zhou}, \citenamefont {Rhodes}, \citenamefont {Hone}, \citenamefont {Taniguchi}, \citenamefont {Watanabe}, \citenamefont {Levitov}, \citenamefont {Zaletel},\ and\ \citenamefont {Young}}]{Island2019}%
  \BibitemOpen
  \bibfield  {author} {\bibinfo {author} {\bibfnamefont {J.~O.}\ \bibnamefont {Island}}, \bibinfo {author} {\bibfnamefont {X.}~\bibnamefont {Cui}}, \bibinfo {author} {\bibfnamefont {C.}~\bibnamefont {Lewandowski}}, \bibinfo {author} {\bibfnamefont {J.~Y.}\ \bibnamefont {Khoo}}, \bibinfo {author} {\bibfnamefont {E.~M.}\ \bibnamefont {Spanton}}, \bibinfo {author} {\bibfnamefont {H.}~\bibnamefont {Zhou}}, \bibinfo {author} {\bibfnamefont {D.}~\bibnamefont {Rhodes}}, \bibinfo {author} {\bibfnamefont {J.~C.}\ \bibnamefont {Hone}}, \bibinfo {author} {\bibfnamefont {T.}~\bibnamefont {Taniguchi}}, \bibinfo {author} {\bibfnamefont {K.}~\bibnamefont {Watanabe}}, \bibinfo {author} {\bibfnamefont {L.~S.}\ \bibnamefont {Levitov}}, \bibinfo {author} {\bibfnamefont {M.~P.}\ \bibnamefont {Zaletel}},\ and\ \bibinfo {author} {\bibfnamefont {A.~F.}\ \bibnamefont {Young}},\ }\bibfield  {title} {\bibinfo {title} {Spin{\textendash}orbit-driven band inversion in bilayer graphene by the van der {W}aals proximity effect},\ }\href
  {https://doi.org/10.1038/s41586-019-1304-2} {\bibfield  {journal} {\bibinfo  {journal} {Nature}\ }\textbf {\bibinfo {volume} {571}},\ \bibinfo {pages} {85} (\bibinfo {year} {2019})}\BibitemShut {NoStop}%
\bibitem [{\citenamefont {Wang}\ \emph {et~al.}(2019)\citenamefont {Wang}, \citenamefont {Che}, \citenamefont {Cao}, \citenamefont {Lyu}, \citenamefont {Watanabe}, \citenamefont {Taniguchi}, \citenamefont {Lau},\ and\ \citenamefont {Bockrath}}]{Wang2019}%
  \BibitemOpen
  \bibfield  {author} {\bibinfo {author} {\bibfnamefont {D.}~\bibnamefont {Wang}}, \bibinfo {author} {\bibfnamefont {S.}~\bibnamefont {Che}}, \bibinfo {author} {\bibfnamefont {G.}~\bibnamefont {Cao}}, \bibinfo {author} {\bibfnamefont {R.}~\bibnamefont {Lyu}}, \bibinfo {author} {\bibfnamefont {K.}~\bibnamefont {Watanabe}}, \bibinfo {author} {\bibfnamefont {T.}~\bibnamefont {Taniguchi}}, \bibinfo {author} {\bibfnamefont {C.~N.}\ \bibnamefont {Lau}},\ and\ \bibinfo {author} {\bibfnamefont {M.}~\bibnamefont {Bockrath}},\ }\bibfield  {title} {\bibinfo {title} {Quantum {H}all effect measurement of spin{\textendash}orbit coupling strengths in ultraclean bilayer graphene/${\mathrm{wse}}_{2}$ heterostructures},\ }\href {https://doi.org/10.1021/acs.nanolett.9b02445} {\bibfield  {journal} {\bibinfo  {journal} {Nano Lett.}\ }\textbf {\bibinfo {volume} {19}},\ \bibinfo {pages} {7028} (\bibinfo {year} {2019})}\BibitemShut {NoStop}%
\bibitem [{\citenamefont {Amann}\ \emph {et~al.}(2022)\citenamefont {Amann}, \citenamefont {V\"olkl}, \citenamefont {Rockinger}, \citenamefont {Kochan}, \citenamefont {Watanabe}, \citenamefont {Taniguchi}, \citenamefont {Fabian}, \citenamefont {Weiss},\ and\ \citenamefont {Eroms}}]{Amann2022}%
  \BibitemOpen
  \bibfield  {author} {\bibinfo {author} {\bibfnamefont {J.}~\bibnamefont {Amann}}, \bibinfo {author} {\bibfnamefont {T.}~\bibnamefont {V\"olkl}}, \bibinfo {author} {\bibfnamefont {T.}~\bibnamefont {Rockinger}}, \bibinfo {author} {\bibfnamefont {D.}~\bibnamefont {Kochan}}, \bibinfo {author} {\bibfnamefont {K.}~\bibnamefont {Watanabe}}, \bibinfo {author} {\bibfnamefont {T.}~\bibnamefont {Taniguchi}}, \bibinfo {author} {\bibfnamefont {J.}~\bibnamefont {Fabian}}, \bibinfo {author} {\bibfnamefont {D.}~\bibnamefont {Weiss}},\ and\ \bibinfo {author} {\bibfnamefont {J.}~\bibnamefont {Eroms}},\ }\bibfield  {title} {\bibinfo {title} {Counterintuitive gate dependence of weak antilocalization in bilayer $\mathrm{graphene}/{\mathrm{wse}}_{2}$ heterostructures},\ }\href {https://doi.org/10.1103/PhysRevB.105.115425} {\bibfield  {journal} {\bibinfo  {journal} {Phys. Rev. B}\ }\textbf {\bibinfo {volume} {105}},\ \bibinfo {pages} {115425} (\bibinfo {year} {2022})}\BibitemShut {NoStop}%
\bibitem [{\citenamefont {Sun}\ \emph {et~al.}(2023)\citenamefont {Sun}, \citenamefont {Rademaker}, \citenamefont {Mauro}, \citenamefont {Scarfato}, \citenamefont {P{\'a}sztor}, \citenamefont {Guti{\'e}rrez-Lezama}, \citenamefont {Wang}, \citenamefont {Martinez-Castro}, \citenamefont {Morpurgo},\ and\ \citenamefont {Renner}}]{Sun2022determining}%
  \BibitemOpen
  \bibfield  {author} {\bibinfo {author} {\bibfnamefont {L.}~\bibnamefont {Sun}}, \bibinfo {author} {\bibfnamefont {L.}~\bibnamefont {Rademaker}}, \bibinfo {author} {\bibfnamefont {D.}~\bibnamefont {Mauro}}, \bibinfo {author} {\bibfnamefont {A.}~\bibnamefont {Scarfato}}, \bibinfo {author} {\bibfnamefont {{\'A}.}~\bibnamefont {P{\'a}sztor}}, \bibinfo {author} {\bibfnamefont {I.}~\bibnamefont {Guti{\'e}rrez-Lezama}}, \bibinfo {author} {\bibfnamefont {Z.}~\bibnamefont {Wang}}, \bibinfo {author} {\bibfnamefont {J.}~\bibnamefont {Martinez-Castro}}, \bibinfo {author} {\bibfnamefont {A.~F.}\ \bibnamefont {Morpurgo}},\ and\ \bibinfo {author} {\bibfnamefont {C.}~\bibnamefont {Renner}},\ }\bibfield  {title} {\bibinfo {title} {Determining spin-orbit coupling in graphene by quasiparticle interference imaging},\ }\href {https://doi.org/10.1038/s41467-023-39453-x} {\bibfield  {journal} {\bibinfo  {journal} {Nat. Commun.}\ }\textbf {\bibinfo {volume} {14}},\ \bibinfo {pages} {3771} (\bibinfo {year} {2023})}\BibitemShut
  {NoStop}%
\bibitem [{\citenamefont {Zaletel}\ and\ \citenamefont {Khoo}(2019)}]{Zaletel2019}%
  \BibitemOpen
  \bibfield  {author} {\bibinfo {author} {\bibfnamefont {M.~P.}\ \bibnamefont {Zaletel}}\ and\ \bibinfo {author} {\bibfnamefont {J.~Y.}\ \bibnamefont {Khoo}},\ }\href {https://arxiv.org/abs/1901.01294} {\bibinfo {title} {The gate-tunable strong and fragile topology of multilayer-graphene on a transition metal dichalcogenide}} (\bibinfo {year} {2019}),\ \Eprint {https://arxiv.org/abs/1901.01294} {arXiv:1901.01294 [cond-mat.mes-hall]} \BibitemShut {NoStop}%
\bibitem [{\citenamefont {Gmitra}\ and\ \citenamefont {Fabian}(2017)}]{Gmitra2017}%
  \BibitemOpen
  \bibfield  {author} {\bibinfo {author} {\bibfnamefont {M.}~\bibnamefont {Gmitra}}\ and\ \bibinfo {author} {\bibfnamefont {J.}~\bibnamefont {Fabian}},\ }\bibfield  {title} {\bibinfo {title} {Proximity effects in bilayer graphene on monolayer ${\mathrm{wse}}_{2}$: Field-effect spin valley locking, spin-orbit valve, and spin transistor},\ }\href {https://doi.org/10.1103/PhysRevLett.119.146401} {\bibfield  {journal} {\bibinfo  {journal} {Phys. Rev. Lett.}\ }\textbf {\bibinfo {volume} {119}},\ \bibinfo {pages} {146401} (\bibinfo {year} {2017})}\BibitemShut {NoStop}%
\bibitem [{\citenamefont {Khoo}\ \emph {et~al.}(2017)\citenamefont {Khoo}, \citenamefont {Morpurgo},\ and\ \citenamefont {Levitov}}]{Khoo2017}%
  \BibitemOpen
  \bibfield  {author} {\bibinfo {author} {\bibfnamefont {J.~Y.}\ \bibnamefont {Khoo}}, \bibinfo {author} {\bibfnamefont {A.~F.}\ \bibnamefont {Morpurgo}},\ and\ \bibinfo {author} {\bibfnamefont {L.}~\bibnamefont {Levitov}},\ }\bibfield  {title} {\bibinfo {title} {On-demand spin{\textendash}orbit interaction from which-layer tunability in bilayer graphene},\ }\href {https://doi.org/10.1021/acs.nanolett.7b03604} {\bibfield  {journal} {\bibinfo  {journal} {Nano Lett.}\ }\textbf {\bibinfo {volume} {17}},\ \bibinfo {pages} {7003} (\bibinfo {year} {2017})}\BibitemShut {NoStop}%
\bibitem [{\citenamefont {Jung}\ \emph {et~al.}(2015)\citenamefont {Jung}, \citenamefont {Polini},\ and\ \citenamefont {MacDonald}}]{Jung2015}%
  \BibitemOpen
  \bibfield  {author} {\bibinfo {author} {\bibfnamefont {J.}~\bibnamefont {Jung}}, \bibinfo {author} {\bibfnamefont {M.}~\bibnamefont {Polini}},\ and\ \bibinfo {author} {\bibfnamefont {A.~H.}\ \bibnamefont {MacDonald}},\ }\bibfield  {title} {\bibinfo {title} {Persistent current states in bilayer graphene},\ }\href {https://doi.org/10.1103/PhysRevB.91.155423} {\bibfield  {journal} {\bibinfo  {journal} {Phys. Rev. B}\ }\textbf {\bibinfo {volume} {91}},\ \bibinfo {pages} {155423} (\bibinfo {year} {2015})}\BibitemShut {NoStop}%
\bibitem [{\citenamefont {Dong}\ \emph {et~al.}(2023{\natexlab{c}})\citenamefont {Dong}, \citenamefont {Davydova}, \citenamefont {Ogunnaike},\ and\ \citenamefont {Levitov}}]{Dong2021}%
  \BibitemOpen
  \bibfield  {author} {\bibinfo {author} {\bibfnamefont {Z.}~\bibnamefont {Dong}}, \bibinfo {author} {\bibfnamefont {M.}~\bibnamefont {Davydova}}, \bibinfo {author} {\bibfnamefont {O.}~\bibnamefont {Ogunnaike}},\ and\ \bibinfo {author} {\bibfnamefont {L.}~\bibnamefont {Levitov}},\ }\bibfield  {title} {\bibinfo {title} {Isospin- and momentum-polarized orders in bilayer graphene},\ }\href {https://doi.org/10.1103/PhysRevB.107.075108} {\bibfield  {journal} {\bibinfo  {journal} {Phys. Rev. B}\ }\textbf {\bibinfo {volume} {107}},\ \bibinfo {pages} {075108} (\bibinfo {year} {2023}{\natexlab{c}})}\BibitemShut {NoStop}%
\bibitem [{\citenamefont {Huang}\ \emph {et~al.}(2023)\citenamefont {Huang}, \citenamefont {Wolf}, \citenamefont {Qin}, \citenamefont {Wei}, \citenamefont {Blinov},\ and\ \citenamefont {MacDonald}}]{Huang2023}%
  \BibitemOpen
  \bibfield  {author} {\bibinfo {author} {\bibfnamefont {C.}~\bibnamefont {Huang}}, \bibinfo {author} {\bibfnamefont {T.~M.~R.}\ \bibnamefont {Wolf}}, \bibinfo {author} {\bibfnamefont {W.}~\bibnamefont {Qin}}, \bibinfo {author} {\bibfnamefont {N.}~\bibnamefont {Wei}}, \bibinfo {author} {\bibfnamefont {I.~V.}\ \bibnamefont {Blinov}},\ and\ \bibinfo {author} {\bibfnamefont {A.~H.}\ \bibnamefont {MacDonald}},\ }\bibfield  {title} {\bibinfo {title} {Spin and orbital metallic magnetism in rhombohedral trilayer graphene},\ }\href {https://doi.org/10.1103/PhysRevB.107.L121405} {\bibfield  {journal} {\bibinfo  {journal} {Phys. Rev. B}\ }\textbf {\bibinfo {volume} {107}},\ \bibinfo {pages} {L121405} (\bibinfo {year} {2023})}\BibitemShut {NoStop}%
\bibitem [{\citenamefont {Arp}\ \emph {et~al.}(2024)\citenamefont {Arp}, \citenamefont {Sheekey}, \citenamefont {Zhou}, \citenamefont {Tschirhart}, \citenamefont {Patterson}, \citenamefont {Yoo}, \citenamefont {Holleis}, \citenamefont {Redekop}, \citenamefont {Babikyan}, \citenamefont {Xie}, \citenamefont {Xiao}, \citenamefont {Vituri}, \citenamefont {Holder}, \citenamefont {Taniguchi}, \citenamefont {Watanabe}, \citenamefont {Huber}, \citenamefont {Berg},\ and\ \citenamefont {Young}}]{Arp2024}%
  \BibitemOpen
  \bibfield  {author} {\bibinfo {author} {\bibfnamefont {T.}~\bibnamefont {Arp}}, \bibinfo {author} {\bibfnamefont {O.}~\bibnamefont {Sheekey}}, \bibinfo {author} {\bibfnamefont {H.}~\bibnamefont {Zhou}}, \bibinfo {author} {\bibfnamefont {C.~L.}\ \bibnamefont {Tschirhart}}, \bibinfo {author} {\bibfnamefont {C.~L.}\ \bibnamefont {Patterson}}, \bibinfo {author} {\bibfnamefont {H.~M.}\ \bibnamefont {Yoo}}, \bibinfo {author} {\bibfnamefont {L.}~\bibnamefont {Holleis}}, \bibinfo {author} {\bibfnamefont {E.}~\bibnamefont {Redekop}}, \bibinfo {author} {\bibfnamefont {G.}~\bibnamefont {Babikyan}}, \bibinfo {author} {\bibfnamefont {T.}~\bibnamefont {Xie}}, \bibinfo {author} {\bibfnamefont {J.}~\bibnamefont {Xiao}}, \bibinfo {author} {\bibfnamefont {Y.}~\bibnamefont {Vituri}}, \bibinfo {author} {\bibfnamefont {T.}~\bibnamefont {Holder}}, \bibinfo {author} {\bibfnamefont {T.}~\bibnamefont {Taniguchi}}, \bibinfo {author} {\bibfnamefont {K.}~\bibnamefont {Watanabe}}, \bibinfo {author} {\bibfnamefont {M.~E.}\ \bibnamefont
  {Huber}}, \bibinfo {author} {\bibfnamefont {E.}~\bibnamefont {Berg}},\ and\ \bibinfo {author} {\bibfnamefont {A.~F.}\ \bibnamefont {Young}},\ }\bibfield  {title} {\bibinfo {title} {Intervalley coherence and intrinsic spin–orbit coupling in rhombohedral trilayer graphene},\ }\href {https://doi.org/10.1038/s41567-024-02560-7} {\bibfield  {journal} {\bibinfo  {journal} {Nat. Phys.}\ }\textbf {\bibinfo {volume} {20}},\ \bibinfo {pages} {1413–1420} (\bibinfo {year} {2024})}\BibitemShut {NoStop}%
\bibitem [{\citenamefont {Szab\'o}\ and\ \citenamefont {Roy}(2022)}]{Szabo2022}%
  \BibitemOpen
  \bibfield  {author} {\bibinfo {author} {\bibfnamefont {A.~L.}\ \bibnamefont {Szab\'o}}\ and\ \bibinfo {author} {\bibfnamefont {B.}~\bibnamefont {Roy}},\ }\bibfield  {title} {\bibinfo {title} {Competing orders and cascade of degeneracy lifting in doped {B}ernal bilayer graphene},\ }\href {https://doi.org/10.1103/PhysRevB.105.L201107} {\bibfield  {journal} {\bibinfo  {journal} {Phys. Rev. B}\ }\textbf {\bibinfo {volume} {105}},\ \bibinfo {pages} {L201107} (\bibinfo {year} {2022})}\BibitemShut {NoStop}%
\bibitem [{\citenamefont {Das}\ and\ \citenamefont {Huang}(2024)}]{Das2024}%
  \BibitemOpen
  \bibfield  {author} {\bibinfo {author} {\bibfnamefont {M.}~\bibnamefont {Das}}\ and\ \bibinfo {author} {\bibfnamefont {C.}~\bibnamefont {Huang}},\ }\bibfield  {title} {\bibinfo {title} {Quarter-metal phases in multilayer graphene: Ising-xy and annular lifshitz transitions},\ }\href {https://doi.org/10.1103/PhysRevB.110.035103} {\bibfield  {journal} {\bibinfo  {journal} {Phys. Rev. B}\ }\textbf {\bibinfo {volume} {110}},\ \bibinfo {pages} {035103} (\bibinfo {year} {2024})}\BibitemShut {NoStop}%
\bibitem [{\citenamefont {Zhumagulov}\ \emph {et~al.}(2024{\natexlab{b}})\citenamefont {Zhumagulov}, \citenamefont {Kochan},\ and\ \citenamefont {Fabian}}]{Zhumagulov2023}%
  \BibitemOpen
  \bibfield  {author} {\bibinfo {author} {\bibfnamefont {Y.}~\bibnamefont {Zhumagulov}}, \bibinfo {author} {\bibfnamefont {D.}~\bibnamefont {Kochan}},\ and\ \bibinfo {author} {\bibfnamefont {J.}~\bibnamefont {Fabian}},\ }\bibfield  {title} {\bibinfo {title} {Emergent correlated phases in rhombohedral trilayer graphene induced by proximity spin-orbit and exchange coupling},\ }\href {https://doi.org/10.1103/PhysRevLett.132.186401} {\bibfield  {journal} {\bibinfo  {journal} {Phys. Rev. Lett.}\ }\textbf {\bibinfo {volume} {132}},\ \bibinfo {pages} {186401} (\bibinfo {year} {2024}{\natexlab{b}})}\BibitemShut {NoStop}%
\bibitem [{\citenamefont {Patterson}\ \emph {et~al.}(2024)\citenamefont {Patterson}, \citenamefont {Sheekey}, \citenamefont {Arp}, \citenamefont {Holleis}, \citenamefont {Koh}, \citenamefont {Choi}, \citenamefont {Xie}, \citenamefont {Xu}, \citenamefont {Redekop}, \citenamefont {Babikyan}, \citenamefont {Zhou}, \citenamefont {Cheng}, \citenamefont {Taniguchi}, \citenamefont {Watanabe}, \citenamefont {Jin}, \citenamefont {Lantagne-Hurtubise}, \citenamefont {Alicea},\ and\ \citenamefont {Young}}]{Caitlin2024}%
  \BibitemOpen
  \bibfield  {author} {\bibinfo {author} {\bibfnamefont {C.~L.}\ \bibnamefont {Patterson}}, \bibinfo {author} {\bibfnamefont {O.~I.}\ \bibnamefont {Sheekey}}, \bibinfo {author} {\bibfnamefont {T.~B.}\ \bibnamefont {Arp}}, \bibinfo {author} {\bibfnamefont {L.~F.~W.}\ \bibnamefont {Holleis}}, \bibinfo {author} {\bibfnamefont {J.~M.}\ \bibnamefont {Koh}}, \bibinfo {author} {\bibfnamefont {Y.}~\bibnamefont {Choi}}, \bibinfo {author} {\bibfnamefont {T.}~\bibnamefont {Xie}}, \bibinfo {author} {\bibfnamefont {S.}~\bibnamefont {Xu}}, \bibinfo {author} {\bibfnamefont {E.}~\bibnamefont {Redekop}}, \bibinfo {author} {\bibfnamefont {G.}~\bibnamefont {Babikyan}}, \bibinfo {author} {\bibfnamefont {H.}~\bibnamefont {Zhou}}, \bibinfo {author} {\bibfnamefont {X.}~\bibnamefont {Cheng}}, \bibinfo {author} {\bibfnamefont {T.}~\bibnamefont {Taniguchi}}, \bibinfo {author} {\bibfnamefont {K.}~\bibnamefont {Watanabe}}, \bibinfo {author} {\bibfnamefont {C.}~\bibnamefont {Jin}}, \bibinfo {author} {\bibfnamefont {E.}~\bibnamefont
  {Lantagne-Hurtubise}}, \bibinfo {author} {\bibfnamefont {J.}~\bibnamefont {Alicea}},\ and\ \bibinfo {author} {\bibfnamefont {A.~F.}\ \bibnamefont {Young}},\ }\href {https://arxiv.org/abs/2408.10190} {\bibinfo {title} {Superconductivity and spin canting in spin-orbit proximitized rhombohedral trilayer graphene}} (\bibinfo {year} {2024}),\ \Eprint {https://arxiv.org/abs/2408.10190} {arXiv:2408.10190 [cond-mat.mes-hall]} \BibitemShut {NoStop}%
\bibitem [{\citenamefont {Raines}\ \emph {et~al.}(2024)\citenamefont {Raines}, \citenamefont {Glazman},\ and\ \citenamefont {Chubukov}}]{Raines2024unconventional}%
  \BibitemOpen
  \bibfield  {author} {\bibinfo {author} {\bibfnamefont {Z.~M.}\ \bibnamefont {Raines}}, \bibinfo {author} {\bibfnamefont {L.~I.}\ \bibnamefont {Glazman}},\ and\ \bibinfo {author} {\bibfnamefont {A.~V.}\ \bibnamefont {Chubukov}},\ }\bibfield  {title} {\bibinfo {title} {Unconventional discontinuous transitions in a two-dimensional system with spin and valley degrees of freedom},\ }\href {https://doi.org/10.1103/PhysRevB.110.155402} {\bibfield  {journal} {\bibinfo  {journal} {Phys. Rev. B}\ }\textbf {\bibinfo {volume} {110}},\ \bibinfo {pages} {155402} (\bibinfo {year} {2024})}\BibitemShut {NoStop}%
\bibitem [{\citenamefont {Coleman}(2015)}]{Coleman2015}%
  \BibitemOpen
  \bibfield  {author} {\bibinfo {author} {\bibfnamefont {P.}~\bibnamefont {Coleman}},\ }\href {https://doi.org/10.1017/cbo9781139020916} {\emph {\bibinfo {title} {Introduction to Many-Body Physics}}}\ (\bibinfo  {publisher} {Cambridge University Press},\ \bibinfo {year} {2015})\ pp.\ \bibinfo {pages} {469--474}\BibitemShut {NoStop}%
\bibitem [{\citenamefont {Saito}\ \emph {et~al.}(2020)\citenamefont {Saito}, \citenamefont {Ge}, \citenamefont {Watanabe}, \citenamefont {Taniguchi},\ and\ \citenamefont {Young}}]{Saito2020}%
  \BibitemOpen
  \bibfield  {author} {\bibinfo {author} {\bibfnamefont {Y.}~\bibnamefont {Saito}}, \bibinfo {author} {\bibfnamefont {J.}~\bibnamefont {Ge}}, \bibinfo {author} {\bibfnamefont {K.}~\bibnamefont {Watanabe}}, \bibinfo {author} {\bibfnamefont {T.}~\bibnamefont {Taniguchi}},\ and\ \bibinfo {author} {\bibfnamefont {A.~F.}\ \bibnamefont {Young}},\ }\bibfield  {title} {\bibinfo {title} {Independent superconductors and correlated insulators in twisted bilayer graphene},\ }\href {https://doi.org/10.1038/s41567-020-0928-3} {\bibfield  {journal} {\bibinfo  {journal} {Nat. Phys.}\ }\textbf {\bibinfo {volume} {16}},\ \bibinfo {pages} {926–930} (\bibinfo {year} {2020})}\BibitemShut {NoStop}%
\bibitem [{\citenamefont {Stepanov}\ \emph {et~al.}(2020)\citenamefont {Stepanov}, \citenamefont {Das}, \citenamefont {Lu}, \citenamefont {Fahimniya}, \citenamefont {Watanabe}, \citenamefont {Taniguchi}, \citenamefont {Koppens}, \citenamefont {Lischner}, \citenamefont {Levitov},\ and\ \citenamefont {Efetov}}]{Stepanov2020}%
  \BibitemOpen
  \bibfield  {author} {\bibinfo {author} {\bibfnamefont {P.}~\bibnamefont {Stepanov}}, \bibinfo {author} {\bibfnamefont {I.}~\bibnamefont {Das}}, \bibinfo {author} {\bibfnamefont {X.}~\bibnamefont {Lu}}, \bibinfo {author} {\bibfnamefont {A.}~\bibnamefont {Fahimniya}}, \bibinfo {author} {\bibfnamefont {K.}~\bibnamefont {Watanabe}}, \bibinfo {author} {\bibfnamefont {T.}~\bibnamefont {Taniguchi}}, \bibinfo {author} {\bibfnamefont {F.~H.~L.}\ \bibnamefont {Koppens}}, \bibinfo {author} {\bibfnamefont {J.}~\bibnamefont {Lischner}}, \bibinfo {author} {\bibfnamefont {L.}~\bibnamefont {Levitov}},\ and\ \bibinfo {author} {\bibfnamefont {D.~K.}\ \bibnamefont {Efetov}},\ }\bibfield  {title} {\bibinfo {title} {Untying the insulating and superconducting orders in magic-angle graphene},\ }\href {https://doi.org/10.1038/s41586-020-2459-6} {\bibfield  {journal} {\bibinfo  {journal} {Nature}\ }\textbf {\bibinfo {volume} {583}},\ \bibinfo {pages} {375–378} (\bibinfo {year} {2020})}\BibitemShut {NoStop}%
\bibitem [{\citenamefont {Veyrat}\ \emph {et~al.}(2020)\citenamefont {Veyrat}, \citenamefont {Déprez}, \citenamefont {Coissard}, \citenamefont {Li}, \citenamefont {Gay}, \citenamefont {Watanabe}, \citenamefont {Taniguchi}, \citenamefont {Han}, \citenamefont {Piot}, \citenamefont {Sellier},\ and\ \citenamefont {Sacépé}}]{Veyrat2020}%
  \BibitemOpen
  \bibfield  {author} {\bibinfo {author} {\bibfnamefont {L.}~\bibnamefont {Veyrat}}, \bibinfo {author} {\bibfnamefont {C.}~\bibnamefont {Déprez}}, \bibinfo {author} {\bibfnamefont {A.}~\bibnamefont {Coissard}}, \bibinfo {author} {\bibfnamefont {X.}~\bibnamefont {Li}}, \bibinfo {author} {\bibfnamefont {F.}~\bibnamefont {Gay}}, \bibinfo {author} {\bibfnamefont {K.}~\bibnamefont {Watanabe}}, \bibinfo {author} {\bibfnamefont {T.}~\bibnamefont {Taniguchi}}, \bibinfo {author} {\bibfnamefont {Z.}~\bibnamefont {Han}}, \bibinfo {author} {\bibfnamefont {B.~A.}\ \bibnamefont {Piot}}, \bibinfo {author} {\bibfnamefont {H.}~\bibnamefont {Sellier}},\ and\ \bibinfo {author} {\bibfnamefont {B.}~\bibnamefont {Sacépé}},\ }\bibfield  {title} {\bibinfo {title} {Helical quantum {H}all phase in graphene on srtio 3},\ }\href {https://doi.org/10.1126/science.aax8201} {\bibfield  {journal} {\bibinfo  {journal} {Science}\ }\textbf {\bibinfo {volume} {367}},\ \bibinfo {pages} {781–786} (\bibinfo {year} {2020})}\BibitemShut
  {NoStop}%
\bibitem [{\citenamefont {Coissard}\ \emph {et~al.}(2022)\citenamefont {Coissard}, \citenamefont {Wander}, \citenamefont {Vignaud}, \citenamefont {Grushin}, \citenamefont {Repellin}, \citenamefont {Watanabe}, \citenamefont {Taniguchi}, \citenamefont {Gay}, \citenamefont {Winkelmann}, \citenamefont {Courtois}, \citenamefont {Sellier},\ and\ \citenamefont {Sac{\'{e}}p{\'{e}}}}]{Coissard2022}%
  \BibitemOpen
  \bibfield  {author} {\bibinfo {author} {\bibfnamefont {A.}~\bibnamefont {Coissard}}, \bibinfo {author} {\bibfnamefont {D.}~\bibnamefont {Wander}}, \bibinfo {author} {\bibfnamefont {H.}~\bibnamefont {Vignaud}}, \bibinfo {author} {\bibfnamefont {A.~G.}\ \bibnamefont {Grushin}}, \bibinfo {author} {\bibfnamefont {C.}~\bibnamefont {Repellin}}, \bibinfo {author} {\bibfnamefont {K.}~\bibnamefont {Watanabe}}, \bibinfo {author} {\bibfnamefont {T.}~\bibnamefont {Taniguchi}}, \bibinfo {author} {\bibfnamefont {F.}~\bibnamefont {Gay}}, \bibinfo {author} {\bibfnamefont {C.~B.}\ \bibnamefont {Winkelmann}}, \bibinfo {author} {\bibfnamefont {H.}~\bibnamefont {Courtois}}, \bibinfo {author} {\bibfnamefont {H.}~\bibnamefont {Sellier}},\ and\ \bibinfo {author} {\bibfnamefont {B.}~\bibnamefont {Sac{\'{e}}p{\'{e}}}},\ }\bibfield  {title} {\bibinfo {title} {Imaging tunable quantum {H}all broken-symmetry orders in graphene},\ }\href {https://doi.org/10.1038/s41586-022-04513-7} {\bibfield  {journal} {\bibinfo  {journal} {Nature}\
  }\textbf {\bibinfo {volume} {605}},\ \bibinfo {pages} {51} (\bibinfo {year} {2022})}\BibitemShut {NoStop}%
\bibitem [{\citenamefont {Liu}\ \emph {et~al.}(2021)\citenamefont {Liu}, \citenamefont {Wang}, \citenamefont {Watanabe}, \citenamefont {Taniguchi}, \citenamefont {Vafek},\ and\ \citenamefont {Li}}]{Liu2021}%
  \BibitemOpen
  \bibfield  {author} {\bibinfo {author} {\bibfnamefont {X.}~\bibnamefont {Liu}}, \bibinfo {author} {\bibfnamefont {Z.}~\bibnamefont {Wang}}, \bibinfo {author} {\bibfnamefont {K.}~\bibnamefont {Watanabe}}, \bibinfo {author} {\bibfnamefont {T.}~\bibnamefont {Taniguchi}}, \bibinfo {author} {\bibfnamefont {O.}~\bibnamefont {Vafek}},\ and\ \bibinfo {author} {\bibfnamefont {J.~I.~A.}\ \bibnamefont {Li}},\ }\bibfield  {title} {\bibinfo {title} {Tuning electron correlation in magic-angle twisted bilayer graphene using {C}oulomb screening},\ }\href {https://doi.org/10.1126/science.abb8754} {\bibfield  {journal} {\bibinfo  {journal} {Science}\ }\textbf {\bibinfo {volume} {371}},\ \bibinfo {pages} {1261–1265} (\bibinfo {year} {2021})}\BibitemShut {NoStop}%
\bibitem [{\citenamefont {Liu}\ \emph {et~al.}(2022{\natexlab{a}})\citenamefont {Liu}, \citenamefont {Zhang}, \citenamefont {Watanabe}, \citenamefont {Taniguchi},\ and\ \citenamefont {Li}}]{Liu2022b}%
  \BibitemOpen
  \bibfield  {author} {\bibinfo {author} {\bibfnamefont {X.}~\bibnamefont {Liu}}, \bibinfo {author} {\bibfnamefont {N.~J.}\ \bibnamefont {Zhang}}, \bibinfo {author} {\bibfnamefont {K.}~\bibnamefont {Watanabe}}, \bibinfo {author} {\bibfnamefont {T.}~\bibnamefont {Taniguchi}},\ and\ \bibinfo {author} {\bibfnamefont {J.~I.~A.}\ \bibnamefont {Li}},\ }\bibfield  {title} {\bibinfo {title} {Isospin order in superconducting magic-angle twisted trilayer graphene},\ }\href {https://doi.org/10.1038/s41567-022-01515-0} {\bibfield  {journal} {\bibinfo  {journal} {Nat. Phys.}\ }\textbf {\bibinfo {volume} {18}},\ \bibinfo {pages} {522–527} (\bibinfo {year} {2022}{\natexlab{a}})}\BibitemShut {NoStop}%
\bibitem [{\citenamefont {You}\ and\ \citenamefont {Vishwanath}(2022)}]{You2022}%
  \BibitemOpen
  \bibfield  {author} {\bibinfo {author} {\bibfnamefont {Y.-Z.}\ \bibnamefont {You}}\ and\ \bibinfo {author} {\bibfnamefont {A.}~\bibnamefont {Vishwanath}},\ }\bibfield  {title} {\bibinfo {title} {{K}ohn-{L}uttinger superconductivity and intervalley coherence in rhombohedral trilayer graphene},\ }\href {https://doi.org/10.1103/PhysRevB.105.134524} {\bibfield  {journal} {\bibinfo  {journal} {Phys. Rev. B}\ }\textbf {\bibinfo {volume} {105}},\ \bibinfo {pages} {134524} (\bibinfo {year} {2022})}\BibitemShut {NoStop}%
\bibitem [{\citenamefont {Han}\ \emph {et~al.}(2024)\citenamefont {Han}, \citenamefont {Lu}, \citenamefont {Yao}, \citenamefont {Shi}, \citenamefont {Yang}, \citenamefont {Seo}, \citenamefont {Ye}, \citenamefont {Wu}, \citenamefont {Zhou}, \citenamefont {Liu}, \citenamefont {Shi}, \citenamefont {Hua}, \citenamefont {Watanabe}, \citenamefont {Taniguchi}, \citenamefont {Xiong}, \citenamefont {Fu},\ and\ \citenamefont {Ju}}]{han2024signatureschiralsuperconductivityrhombohedral}%
  \BibitemOpen
  \bibfield  {author} {\bibinfo {author} {\bibfnamefont {T.}~\bibnamefont {Han}}, \bibinfo {author} {\bibfnamefont {Z.}~\bibnamefont {Lu}}, \bibinfo {author} {\bibfnamefont {Y.}~\bibnamefont {Yao}}, \bibinfo {author} {\bibfnamefont {L.}~\bibnamefont {Shi}}, \bibinfo {author} {\bibfnamefont {J.}~\bibnamefont {Yang}}, \bibinfo {author} {\bibfnamefont {J.}~\bibnamefont {Seo}}, \bibinfo {author} {\bibfnamefont {S.}~\bibnamefont {Ye}}, \bibinfo {author} {\bibfnamefont {Z.}~\bibnamefont {Wu}}, \bibinfo {author} {\bibfnamefont {M.}~\bibnamefont {Zhou}}, \bibinfo {author} {\bibfnamefont {H.}~\bibnamefont {Liu}}, \bibinfo {author} {\bibfnamefont {G.}~\bibnamefont {Shi}}, \bibinfo {author} {\bibfnamefont {Z.}~\bibnamefont {Hua}}, \bibinfo {author} {\bibfnamefont {K.}~\bibnamefont {Watanabe}}, \bibinfo {author} {\bibfnamefont {T.}~\bibnamefont {Taniguchi}}, \bibinfo {author} {\bibfnamefont {P.}~\bibnamefont {Xiong}}, \bibinfo {author} {\bibfnamefont {L.}~\bibnamefont {Fu}},\ and\ \bibinfo {author} {\bibfnamefont
  {L.}~\bibnamefont {Ju}},\ }\href {https://arxiv.org/abs/2408.15233} {\bibinfo {title} {Signatures of chiral superconductivity in rhombohedral graphene}} (\bibinfo {year} {2024}),\ \Eprint {https://arxiv.org/abs/2408.15233} {arXiv:2408.15233 [cond-mat.mes-hall]} \BibitemShut {NoStop}%
\bibitem [{\citenamefont {Vasyukov}\ \emph {et~al.}(2013)\citenamefont {Vasyukov}, \citenamefont {Anahory}, \citenamefont {Embon}, \citenamefont {Halbertal}, \citenamefont {Cuppens}, \citenamefont {Neeman}, \citenamefont {Finkler}, \citenamefont {Segev}, \citenamefont {Myasoedov}, \citenamefont {Rappaport}, \citenamefont {Huber},\ and\ \citenamefont {Zeldov}}]{Vasyukov2013}%
  \BibitemOpen
  \bibfield  {author} {\bibinfo {author} {\bibfnamefont {D.}~\bibnamefont {Vasyukov}}, \bibinfo {author} {\bibfnamefont {Y.}~\bibnamefont {Anahory}}, \bibinfo {author} {\bibfnamefont {L.}~\bibnamefont {Embon}}, \bibinfo {author} {\bibfnamefont {D.}~\bibnamefont {Halbertal}}, \bibinfo {author} {\bibfnamefont {J.}~\bibnamefont {Cuppens}}, \bibinfo {author} {\bibfnamefont {L.}~\bibnamefont {Neeman}}, \bibinfo {author} {\bibfnamefont {A.}~\bibnamefont {Finkler}}, \bibinfo {author} {\bibfnamefont {Y.}~\bibnamefont {Segev}}, \bibinfo {author} {\bibfnamefont {Y.}~\bibnamefont {Myasoedov}}, \bibinfo {author} {\bibfnamefont {M.~L.}\ \bibnamefont {Rappaport}}, \bibinfo {author} {\bibfnamefont {M.~E.}\ \bibnamefont {Huber}},\ and\ \bibinfo {author} {\bibfnamefont {E.}~\bibnamefont {Zeldov}},\ }\bibfield  {title} {\bibinfo {title} {A scanning superconducting quantum interference device with single electron spin sensitivity},\ }\href {https://doi.org/10.1038/nnano.2013.169} {\bibfield  {journal} {\bibinfo  {journal} {Nat.
  Nanotechnology}\ }\textbf {\bibinfo {volume} {8}},\ \bibinfo {pages} {639–644} (\bibinfo {year} {2013})}\BibitemShut {NoStop}%
\bibitem [{\citenamefont {Liu}\ \emph {et~al.}(2022{\natexlab{b}})\citenamefont {Liu}, \citenamefont {Farahi}, \citenamefont {Chiu}, \citenamefont {Papic}, \citenamefont {Watanabe}, \citenamefont {Taniguchi}, \citenamefont {Zaletel},\ and\ \citenamefont {Yazdani}}]{Liu2022}%
  \BibitemOpen
  \bibfield  {author} {\bibinfo {author} {\bibfnamefont {X.}~\bibnamefont {Liu}}, \bibinfo {author} {\bibfnamefont {G.}~\bibnamefont {Farahi}}, \bibinfo {author} {\bibfnamefont {C.-L.}\ \bibnamefont {Chiu}}, \bibinfo {author} {\bibfnamefont {Z.}~\bibnamefont {Papic}}, \bibinfo {author} {\bibfnamefont {K.}~\bibnamefont {Watanabe}}, \bibinfo {author} {\bibfnamefont {T.}~\bibnamefont {Taniguchi}}, \bibinfo {author} {\bibfnamefont {M.~P.}\ \bibnamefont {Zaletel}},\ and\ \bibinfo {author} {\bibfnamefont {A.}~\bibnamefont {Yazdani}},\ }\bibfield  {title} {\bibinfo {title} {Visualizing broken symmetry and topological defects in a quantum {H}all ferromagnet},\ }\href {https://doi.org/10.1126/science.abm3770} {\bibfield  {journal} {\bibinfo  {journal} {Science}\ }\textbf {\bibinfo {volume} {375}},\ \bibinfo {pages} {321} (\bibinfo {year} {2022}{\natexlab{b}})}\BibitemShut {NoStop}%
\bibitem [{\citenamefont {Nuckolls}\ \emph {et~al.}(2023)\citenamefont {Nuckolls}, \citenamefont {Lee}, \citenamefont {Oh}, \citenamefont {Wong}, \citenamefont {Soejima}, \citenamefont {Hong}, \citenamefont {Călugăru}, \citenamefont {Herzog-Arbeitman}, \citenamefont {Bernevig}, \citenamefont {Watanabe}, \citenamefont {Taniguchi}, \citenamefont {Regnault}, \citenamefont {Zaletel},\ and\ \citenamefont {Yazdani}}]{Nuckolls2023}%
  \BibitemOpen
  \bibfield  {author} {\bibinfo {author} {\bibfnamefont {K.~P.}\ \bibnamefont {Nuckolls}}, \bibinfo {author} {\bibfnamefont {R.~L.}\ \bibnamefont {Lee}}, \bibinfo {author} {\bibfnamefont {M.}~\bibnamefont {Oh}}, \bibinfo {author} {\bibfnamefont {D.}~\bibnamefont {Wong}}, \bibinfo {author} {\bibfnamefont {T.}~\bibnamefont {Soejima}}, \bibinfo {author} {\bibfnamefont {J.~P.}\ \bibnamefont {Hong}}, \bibinfo {author} {\bibfnamefont {D.}~\bibnamefont {Călugăru}}, \bibinfo {author} {\bibfnamefont {J.}~\bibnamefont {Herzog-Arbeitman}}, \bibinfo {author} {\bibfnamefont {B.~A.}\ \bibnamefont {Bernevig}}, \bibinfo {author} {\bibfnamefont {K.}~\bibnamefont {Watanabe}}, \bibinfo {author} {\bibfnamefont {T.}~\bibnamefont {Taniguchi}}, \bibinfo {author} {\bibfnamefont {N.}~\bibnamefont {Regnault}}, \bibinfo {author} {\bibfnamefont {M.~P.}\ \bibnamefont {Zaletel}},\ and\ \bibinfo {author} {\bibfnamefont {A.}~\bibnamefont {Yazdani}},\ }\bibfield  {title} {\bibinfo {title} {Quantum textures of the many-body wavefunctions in
  magic-angle graphene},\ }\href {https://doi.org/10.1038/s41586-023-06226-x} {\bibfield  {journal} {\bibinfo  {journal} {Nature}\ }\textbf {\bibinfo {volume} {620}},\ \bibinfo {pages} {525–532} (\bibinfo {year} {2023})}\BibitemShut {NoStop}%
\bibitem [{\citenamefont {Kim}\ \emph {et~al.}(2023)\citenamefont {Kim}, \citenamefont {Choi}, \citenamefont {Lantagne-Hurtubise}, \citenamefont {Lewandowski}, \citenamefont {Thomson}, \citenamefont {Kong}, \citenamefont {Zhou}, \citenamefont {Baum}, \citenamefont {Zhang}, \citenamefont {Holleis}, \citenamefont {Watanabe}, \citenamefont {Taniguchi}, \citenamefont {Young}, \citenamefont {Alicea},\ and\ \citenamefont {Nadj-Perge}}]{Kim2023}%
  \BibitemOpen
  \bibfield  {author} {\bibinfo {author} {\bibfnamefont {H.}~\bibnamefont {Kim}}, \bibinfo {author} {\bibfnamefont {Y.}~\bibnamefont {Choi}}, \bibinfo {author} {\bibfnamefont {{\'E}.}~\bibnamefont {Lantagne-Hurtubise}}, \bibinfo {author} {\bibfnamefont {C.}~\bibnamefont {Lewandowski}}, \bibinfo {author} {\bibfnamefont {A.}~\bibnamefont {Thomson}}, \bibinfo {author} {\bibfnamefont {L.}~\bibnamefont {Kong}}, \bibinfo {author} {\bibfnamefont {H.}~\bibnamefont {Zhou}}, \bibinfo {author} {\bibfnamefont {E.}~\bibnamefont {Baum}}, \bibinfo {author} {\bibfnamefont {Y.}~\bibnamefont {Zhang}}, \bibinfo {author} {\bibfnamefont {L.}~\bibnamefont {Holleis}}, \bibinfo {author} {\bibfnamefont {K.}~\bibnamefont {Watanabe}}, \bibinfo {author} {\bibfnamefont {T.}~\bibnamefont {Taniguchi}}, \bibinfo {author} {\bibfnamefont {A.~F.}\ \bibnamefont {Young}}, \bibinfo {author} {\bibfnamefont {J.}~\bibnamefont {Alicea}},\ and\ \bibinfo {author} {\bibfnamefont {S.}~\bibnamefont {Nadj-Perge}},\ }\bibfield  {title} {\bibinfo {title}
  {Imaging inter-valley coherent order in magic-angle twisted trilayer graphene},\ }\href {https://doi.org/10.1038/s41586-023-06663-8} {\bibfield  {journal} {\bibinfo  {journal} {Nature}\ }\textbf {\bibinfo {volume} {623}},\ \bibinfo {pages} {942} (\bibinfo {year} {2023})}\BibitemShut {NoStop}%
\bibitem [{\citenamefont {Thomson}\ \emph {et~al.}(2022)\citenamefont {Thomson}, \citenamefont {Sorensen}, \citenamefont {Nadj-Perge},\ and\ \citenamefont {Alicea}}]{Thomson2022}%
  \BibitemOpen
  \bibfield  {author} {\bibinfo {author} {\bibfnamefont {A.}~\bibnamefont {Thomson}}, \bibinfo {author} {\bibfnamefont {I.~M.}\ \bibnamefont {Sorensen}}, \bibinfo {author} {\bibfnamefont {S.}~\bibnamefont {Nadj-Perge}},\ and\ \bibinfo {author} {\bibfnamefont {J.}~\bibnamefont {Alicea}},\ }\bibfield  {title} {\bibinfo {title} {Gate-defined wires in twisted bilayer graphene: From electrical detection of intervalley coherence to internally engineered {M}ajorana modes},\ }\href {https://doi.org/10.1103/PhysRevB.105.L081405} {\bibfield  {journal} {\bibinfo  {journal} {Phys. Rev. B}\ }\textbf {\bibinfo {volume} {105}},\ \bibinfo {pages} {L081405} (\bibinfo {year} {2022})}\BibitemShut {NoStop}%
\bibitem [{\citenamefont {Xie}\ \emph {et~al.}(2023)\citenamefont {Xie}, \citenamefont {Lantagne-Hurtubise}, \citenamefont {Young}, \citenamefont {Nadj-Perge},\ and\ \citenamefont {Alicea}}]{Xie2023}%
  \BibitemOpen
  \bibfield  {author} {\bibinfo {author} {\bibfnamefont {Y.-M.}\ \bibnamefont {Xie}}, \bibinfo {author} {\bibfnamefont {E.}~\bibnamefont {Lantagne-Hurtubise}}, \bibinfo {author} {\bibfnamefont {A.~F.}\ \bibnamefont {Young}}, \bibinfo {author} {\bibfnamefont {S.}~\bibnamefont {Nadj-Perge}},\ and\ \bibinfo {author} {\bibfnamefont {J.}~\bibnamefont {Alicea}},\ }\bibfield  {title} {\bibinfo {title} {Gate-defined topological josephson junctions in {B}ernal bilayer graphene},\ }\href {https://doi.org/10.1103/PhysRevLett.131.146601} {\bibfield  {journal} {\bibinfo  {journal} {Phys. Rev. Lett.}\ }\textbf {\bibinfo {volume} {131}},\ \bibinfo {pages} {146601} (\bibinfo {year} {2023})}\BibitemShut {NoStop}%
\bibitem [{\citenamefont {Wei}\ \emph {et~al.}(2023)\citenamefont {Wei}, \citenamefont {Zeng},\ and\ \citenamefont {MacDonald}}]{Wei2023weaklocalizationprobeintervalley}%
  \BibitemOpen
  \bibfield  {author} {\bibinfo {author} {\bibfnamefont {N.}~\bibnamefont {Wei}}, \bibinfo {author} {\bibfnamefont {Y.}~\bibnamefont {Zeng}},\ and\ \bibinfo {author} {\bibfnamefont {A.~H.}\ \bibnamefont {MacDonald}},\ }\href {https://arxiv.org/abs/2312.11259} {\bibinfo {title} {Weak localization as a probe of intervalley coherence in graphene multilayers}} (\bibinfo {year} {2023}),\ \Eprint {https://arxiv.org/abs/2312.11259} {arXiv:2312.11259 [cond-mat.mes-hall]} \BibitemShut {NoStop}%
\bibitem [{\citenamefont {Bena}\ and\ \citenamefont {Montambaux}(2009)}]{bena2009remarks}%
  \BibitemOpen
  \bibfield  {author} {\bibinfo {author} {\bibfnamefont {C.}~\bibnamefont {Bena}}\ and\ \bibinfo {author} {\bibfnamefont {G.}~\bibnamefont {Montambaux}},\ }\bibfield  {title} {\bibinfo {title} {Remarks on the tight-binding model of graphene},\ }\href {https://doi.org/10.1088/1367-2630/11/9/095002} {\bibfield  {journal} {\bibinfo  {journal} {New J. Phys.}\ }\textbf {\bibinfo {volume} {11}},\ \bibinfo {pages} {095003} (\bibinfo {year} {2009})}\BibitemShut {NoStop}%
\end{thebibliography}%

\clearpage

\appendix
\onecolumngrid

\section{Model Hamiltonian and Hartree-Fock procedure}
\label{app-sec:hamiltonian}

The minimal tight-binding Hamiltonian describing Bernal bilayer graphene, expanded near the two valleys labeled by $\tau \in \{\pm 1\}$, reads~\cite{mccann2013electronic, Jung2014}
\begin{equation}\begin{split}
    \hat{H}_{0} = \sum_{\vb{k}} \sum_{\tau s \sigma \sigma'} h(\vb{K}^\tau + \vb{k})_{\sigma \sigma'} 
        c_{\tau s \sigma \vb{k}}^\dag c_{\tau s \sigma' \vb{k}}.
\end{split}\end{equation}
Here the operator $c_{\tau s \sigma \vb{k}}$ annihilates an electron at momentum $\vb{k}$ measured from the Brillouin zone corner $\vb{K}^\tau = \tau (4 \pi / 3 a, 0)$ with lattice constant $a = \SI{2.46}{\angstrom}$, spin index $s \in \{\uparrow, \downarrow\}$ and sublattice index $\sigma \in (A_1, B_1, A_2, B_2)$. In this basis the Hamiltonian matrix reads
\begin{equation}\begin{split}
    h(\vb{q})_{\sigma \sigma'} = \mqty[
        u / 2 & -\gamma_0 f_{\vb{q}} & \gamma_4 f_{\vb{q}} & \gamma_3 f_{\vb{q}}^\dag \\
        -\gamma_0 f_{\vb{q}}^\dag & \Delta + u / 2 & \gamma_1 & \gamma_4 f_{\vb{q}} \\
        \gamma_4 f_{\vb{q}}^\dag & \gamma_1 & \Delta - u / 2 & -\gamma_0 f_{\vb{q}} \\
        \gamma_3 f_{\vb{q}} & \gamma_4 f_{\vb{q}}^\dag & -\gamma_0 f_{\vb{q}}^\dag & -u / 2
    ]_{\sigma \sigma'},
    \label{eq:tight_binding_h}
\end{split}\end{equation}
where the function $f_{\vb{q}} = \smash{e^{i q_y a / \sqrt{3}} + 2 e^{-i q_y a / 2 \sqrt{3}} \cos{\left(q_x a / 2\right)}}$ describes in-plane, nearest-neighbor hopping in the sublattice-centered convention~\cite{mccann2013electronic, bena2009remarks} and $u$ denotes the interlayer potential difference induced by the perpendicular displacement field $D$ through \cref{eq:definition-u}. Experimentally accessible values of this potential difference range up to $u \sim \SI{120}{\milli\electronvolt}$ for applied $D \leq \SI{1.6}{\volt\per\nano\meter}$~\cite{Zhang2023, Yiran2024, Holleis2023, li2024tunable}. The other parameters appearing in \cref{eq:tight_binding_h}, \ie~the on-site potential $\Delta$ and the hopping terms $\gamma_0$, $\gamma_1$, $\gamma_3$ and $\gamma_4$, are fixed by fitting against ab-initio calculations~\cite{Jung2014} and are listed in \cref{tab:parameters}.

Proximity to a TMD, such as a WSe$_2$ monolayer, induces both Ising- and Rashba-type SOC terms to the closest graphene layer (we consider the top layer for concreteness),
\begin{equation}\begin{split}
    \hat{H}_{\mathrm{I}} = \frac{\lambda_{\mathrm{I}}}{2} \sum_{\vb{k}} 
        \vb{c}^\dag_{\vb{k}} \left( \tau^z s^z \mathbb{P}_2 \right) \vb{c}_{\vb{k}}, \qquad
    \hat{H}_{\mathrm{R}} = \frac{\lambda_{\mathrm{R}}}{2} \sum_{\vb{k}} 
        \vb{c}_{\vb{k}}^\dag \left( \tau^z s^y \sigma^x - s^x \sigma^y \right) \mathbb{P}_2 \vb{c}_{\vb{k}},
\end{split}\end{equation}
where $\trans{\vb{c}}_{\vb{k}} = \mqty[ c_{+ \uparrow A_1 \vb{k}} & \ldots & c_{- \downarrow B_2 \vb{k}} ]$ and $\lambda_{\mathrm{I}}$ and $\lambda_{\mathrm{R}}$ respectively denote the strength of Ising and Rashba SOC. The operator $\mathbb{P}_2$ projects onto the top layer of BLG and the Pauli matrices $\tau^\mu$, $s^\mu$ and $\sigma^\mu$ act on the valley, spin and sublattice degrees of freedom respectively. As explained in the main text, due to the suppression of Rashba SOC effects at large $D$ fields~\cite{Zaletel2019} we focus on Ising-type SOC throughout this work.

We then consider Coulomb repulsion between electrons, which can be conveniently separated into its long-range $\hat{H}_{\mathrm{C}}$ and short-range $\hat{H}_{\mathrm{V}}$ parts as
\begin{equation}\begin{split}
    \hat{H}_{\mathrm{C}} &= \frac{1}{2A} \sum_{\vb{q}} V_{\mathrm{C}}(\vb{q}) \normord{\rho(\vb{q}) \rho(-\vb{q})}, \\
    \hat{H}_{\mathrm{V}} &= \frac{J_{\mathrm{H}}}{2A} \sum_{\vb{k} \vb{k}' \vb{q}} \sum_{\tau ss' \sigma \sigma'}
        \eta(\vb{q})_{\tau \sigma \sigma'}
        \normord{c_{(-\tau) s \sigma \vb{k}}^\dag c_{\tau s \sigma (\vb{k} + \vb{q})}
            c_{\tau s' \sigma' \vb{k}'}^\dag c_{(-\tau) s' \sigma' (\vb{k}' - \vb{q})} },
            \label{eq:Coulomb}
\end{split}\end{equation}
where $A$ is the sample area, $J_{\mathrm{H}}$ controls the strength of short-range interactions, and $\normord{}$ denotes normal ordering. The short-range Hamiltonian $\hat{H}_{\mathrm{V}}$ is also referred to in the literature as the Hund's coupling, as it can be approximately recast in terms of a coupling between spin operators in the two valleys~\cite{Chatterjee2022}. The long-range part $\hat{H}_{\mathrm{C}}$ couples to the long-wavelength component of the electron density, $\smash{\rho(\vb{q}) = \sum_{\vb{k} \alpha} c_{\alpha \vb{k}}^\dag c_{\alpha (\vb{k} + \vb{q})}}$, which involves only intravalley scattering, where $\alpha = (\tau, s, \sigma)$ encompasses valley, spin and sublattice indices. We consider an interaction potential $V_{\mathrm{C}}(\vb{q})$ that incorporates screening from the two gates situated at a distance $d$ on both sides of the BLG device,
\begin{equation}
    V_{\mathrm{C}}(\vb{q}) = \frac{q_{\text{e}}^2}{2 \epsilon_{\mathrm{r}} \epsilon_0 q} 
        \tanh{(q d)},
\end{equation}
with $q_{\text{e}}$ the electron charge, $\epsilon_{\mathrm{r}}$ the relative permittivity, and $\epsilon_0$ the vacuum permittivity. In this work we set $d \approx \SI{30}{\nano\meter}$ and model additional screening arising from electrons in BLG by treating $\epsilon_{\mathrm{r}}$ as a free parameter.
Generically, short-range components of Coulomb repulsion yield a \emph{ferromagnetic} Hund's coupling with $J_{\mathrm{H}} > 0$, which favors aligning spins across valleys. In principle, other lattice-scale effects, including electron-phonon interactions, may generate additional contributions. In this work we nevertheless fix the sign to be ferromagnetic, $J_{\mathrm{H}} > 0$, in order to reproduce experimental observations.

\begin{table}
    \centering
    \begin{tabular}{c c c c c c}
        \toprule 
        $\gamma_0$ & $\gamma_1$ & $\gamma_3$ & $\gamma_4$ & $\Delta$ \\
        \midrule 
        $\quad \SI{2.61}{\electronvolt} \quad$ &
        $\quad \SI{361}{\milli\electronvolt} \quad$ &
        $\quad \SI{283}{\milli\electronvolt} \quad$ &
        $\quad \SI{138}{\milli\electronvolt} \quad$ &
        $\quad \SI{15}{\milli\electronvolt} \quad$ \\
        \bottomrule
    \end{tabular}
    \caption{Parameters used in the non-interacting Hamiltonian matrix, \cref{eq:tight_binding_h}, and their numerical values obtained from fitting to \emph{ab initio} calculations in Ref.~\onlinecite{Jung2014}.
    \label{tab:parameters}
    }
\end{table}

Finally, the phase factors $\eta(\vb{q})_{\tau \sigma \sigma'}$ in the Hund's coupling Hamiltonian $\smash{\hat{H}_{\mathrm{V}}}$ must be chosen to preserve $\text{C}_3$ rotation symmetry. As explained in Appendix B of Ref.~\onlinecite{Koh2024}, we can choose a gauge and adopt the following form:
\begin{equation}\begin{split}
    \eta(\vb{q})_{\tau \sigma \sigma'} = 
        \begin{dcases}
            e^{2 i \tau (\lambda_\sigma - \lambda_{\sigma'}) \theta_{\vb{q}}} & \vb{q} \neq \vb{0} \\
            \delta_{\sigma \sigma'} & \vb{q} = \vb{0},
        \end{dcases}
\end{split}\end{equation}
where we introduced the sublattice-valued index $\lambda_\sigma = 0, +1, -1$ for the sublattices $B_2$, $A_1$ and $A_2/B_1$, respectively, and $\theta_{\vb{q}} = \arg{(q_x + i q_y)}$.

Our starting point for the Hartree-Fock algorithm is the spin-orbit proximitized, non-interacting Hamiltonian $\smash{\hat{H}_{0}} + \smash{\hat{H}_{\mathrm{I}}}$ at a fixed electronic density $n_e$ and displacement field $D$. We then decouple the interaction terms $\hat{H}_{\mathrm{C}}$ and $\hat{H}_{\mathrm{V}}$ by introducing mean-field order parameters with different broken symmetries. In each symmetry sector, the best Slater-determinant ground state is obtained self-consistently through a symmetry-restricted Hartree-Fock algorithm. Finally, the optimal instance of each symmetry sector are compared; the mean-field ground state is taken as the state with the lowest energy.  Details of the numerical implementation can be found in Appendix B of Ref.~\onlinecite{Koh2024}.

\section{Analysis in the Hund's coupling free limit}
\label{app:no-Hunds}

In this Appendix we consider the case where Hund's coupling is neglected ($J_{\mathrm{H}} = 0$). First let us consider $\lambda_{\mathrm{I}} = 0$ and discuss the enlarged symmetry group that characterizes this limit. Here $\mathrm{SU}(2)_{\mathrm{s}}$ spin rotations can be enacted separately in each valley due to the lack of intervalley exchange processes. In what follows it will be useful to consider, in addition to global $\mathrm{SU}(2)$ rotations parameterized by $e^{i \theta \vb{s} \cdot \vb{n} / 2}$ with $\theta$ a rotation angle about the $\vb{n}$ unit vector and $\vb{s} = (s^x, s^y, s^z)$ a vector of spin Pauli matrices, the corresponding valley-contrasting spin rotations $e^{i \phi \tau^z \vb{s} \cdot \vb{n} / 2}$. 

\subsection{Ground state degeneracies}

Using this extra symmetry we can show that various ground states listed in \cref{tab:symmetry-orders-legends} become degenerate in the special case of $J_{\mathrm{H}} = 0$. For example, SP and SVL states (with order parameters $\tau^0 s^z$ and $\tau^z s^z$ respectively) can be connected by a $\pi$ spin rotation along (say) the $x$-axis, which applies only to electrons in valley $\vb{K}^-$. The operator enacting this transformation is $\smash{\mathcal{U} = e^{i \pi (\tau^0 - \tau^z) s^x / 4}}$. Similarly, the two $g = 2$ IVC states (IVC$_0$ and IVC$_{\text{z}}$ with order parameters $\tau^x s^0$ and $\tau^x s^z$ respectively) are related by a $\pi$ spin rotation around the $z$ axis for one valley only, implemented by the operator $\smash{\mathcal{U}' = e^{i \pi (\tau^0 - \tau^z) s^z / 4}}$. Note that the operator $\mathcal{U}'$ is also contained in the $\mathrm{U}(1) \times \mathrm{U}(1)$ symmetry group that characterizes the situation \emph{with} Ising SOC but \emph{without} Hund's coupling, which we explore numerically below to connect with the results of Ref.~\onlinecite{Ming2023}. The degeneracy between the IVC$_0$ and IVC$_{\text{z}}$ states (and their spin-orbit-proximitized cousins SVL+IVC$_0$ and SVL+IVC$_{\text{z}}$) therefore survives in this case.

Two of the $g = 1$ IVC states discussed in this work (SP-IVC and SVL-IVC, where intervalley coherence respectively sets in within spin-polarized and spin-valley-locked Fermi surfaces) are also degenerate in the $J_{\mathrm{H}} = 0$ and $\lambda_{\mathrm{I}} = 0$ limit. To see this, note that their order parameters are given by a combination of $\tau^x s^0$ and $\tau^0 s^z$ (SP-IVC); or a combination of $\tau^x s^x$ and $\tau^z s^z$ (SVL-IVC). We have seen above that their respective valley-diagonal components $\tau^0 s^z$ and $\tau^z s^z$ are related by the operator $\mathcal{U}$. One can similarly show that (up to a global $\mathrm{U}(1)$ valley rotation) their intervalley coherent order parameters $\tau^x s^0$ and $\tau^x s^x$ are mapped to each other under the action of $\mathcal{U}$.

Finally, the valley-polarized (VP) solution discussed in the main text is a peculiar case that needs to be treated separately. Indeed, because of its vanishing spin polarization this state cannot be mapped to other Stoner half metals, such as SP and SVL states, using independent spin rotations in the two valleys. To understand why it is nevertheless degenerate with SP and SVL states, we return to the long-range part of Coulomb interactions, in the first line of \cref{eq:Coulomb}. Such a term has a larger SU(4) symmetry group that acts within the spin and valley subspace; thus its contribution  to the ground state energy is identical for VP, SP and SVL states---which all host 2 out of 4 spin-valley flavors predominantly occupied. The kinetic energy part in \cref{eq:tight_binding_h} breaks SU(4), but nevertheless time-reversal symmetry enforces that the non-interacting band structures in the two valleys are mirror copies of each other: the kinetic energy (for a particular filling) is thus identical regardless of whether two valley flavors (VP), two spin flavors (SP), or the two flavors preferred by Ising SOC (SVL) are occupied.

\subsection{Hartree-Fock phase diagrams}

\begin{figure*}
    \centering
    \includegraphics[width = 1\linewidth]{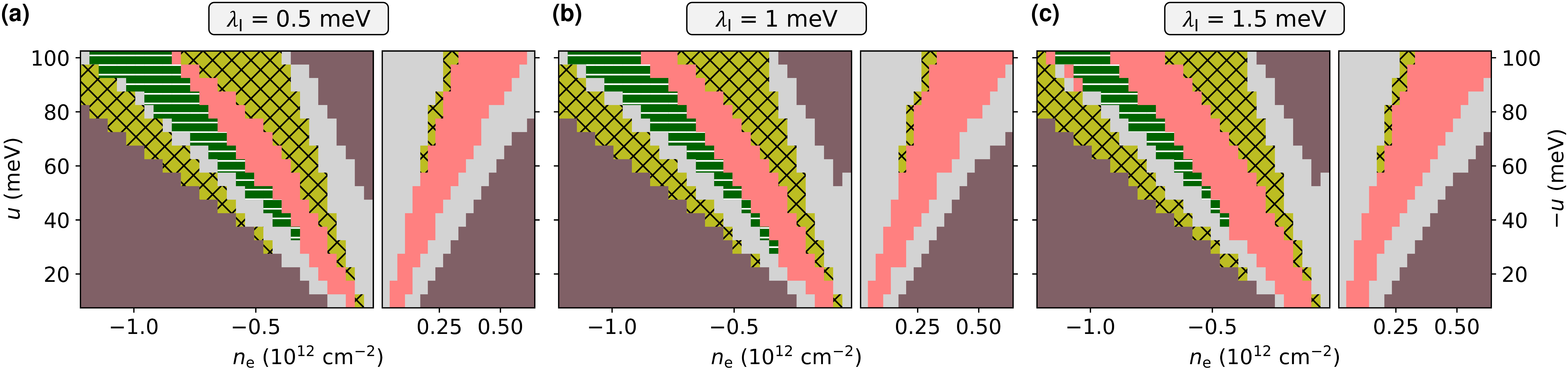}
    \phantomsubfloat{\label{fig:results-no-hunds-lambdaI-05}}
    \phantomsubfloat{\label{fig:results-no-hunds-lambdaI-10}}
    \phantomsubfloat{\label{fig:results-no-hunds-lambdaI-15}}
    \vspace{-1.6\baselineskip}
    \caption{\textbf{Phase diagrams without Hund's coupling.} Hole- and electron-doped phase diagrams of BLG as a function of charge density $n_{\text{e}}$ and interlayer potential $u$ in the absence of Hund's coupling ($J_{\mathrm{H}} = 0$). We consider Ising SOC strengths \textbf{(a)} $\lambda_{\mathrm{I}} = \SI{0.5}{\milli\electronvolt}$, \textbf{(b)} $\lambda_{\mathrm{I}} = \SI{1}{\milli\electronvolt}$, and \textbf{(c)} $\lambda_{\mathrm{I}} = \SI{1.5}{\milli\electronvolt}$. The case without Ising SOC ($\lambda_{\mathrm{I}} = 0$) is presented in \cref{fig:results-wo-soc-Jh-00}.}
    \label{fig:results-no-hunds}
\end{figure*}

In \cref{fig:results-no-hunds} we consider the effects of Ising SOC on the phase diagram of BLG when Hund's coupling is neglected---this corresponds to the situation investigated in Ref.~\onlinecite{Ming2023}. Apart from quantitative differences arising from our choice of a weaker interaction strength, we reproduce the main features observed in this work. In particular, the location of Stoner ferromagnets (the SVL half metal and the SVP quarter and three-quarter metals) roughly coincide. The $g = 1$ and $g = 2$ IVC states are also reproduced, although Ref.~\onlinecite{Ming2023} does not see the leftmost, generalized three-quarter-metal $g = 1$ IVC state obtained in our \cref{fig:results-no-hunds}---presumably because of their larger interaction strength 
($\epsilon_{\mathrm{r}} = 10$) that favors Stoner ferromagnets over IVC states (see also \cref{fig:results-interaction-strength}). Similarly to the conclusions reached in Ref.~\onlinecite{Ming2023}, increasing Ising SOC leads to the expansion of the SVL phase at the expense of the neighboring symmetry-broken $g = 2$ order: the SVL+IVC$_0$ state (which is degenerate with SVL+IVC$_z$ as explained above).

Two key differences arise between results without Hund's coupling, presented here and in Ref.~\onlinecite{Ming2023}, and those in the main text. First, setting $J_{\mathrm{H}} = 0$ removes the energetic preference towards developing spin polarization, which drives the formation of the spin-canted (C) and the spin-canted IVC (C-IVC) states. Second, the non-generic $\mathrm{U}(1) \times \mathrm{U}(1)$ symmetry in this limit makes the $g = 2$ IVC region degenerate, while introducing $J_{\mathrm{H}}>0$ stabilizes the SVL+IVC$_{\text{z}}$ order in \cref{fig:results-w-soc}, at the expense of the SVL+IVC$_0$ state. These two orders are similar but nevertheless differ by the symmetries they preserve: whereas the spin-singlet variant SVL+IVC$_0$ preserves microscopic time reversal $\mathcal{T}$, its spin-triplet cousin SVL+IVC$_{\text{z}}$ preserves the effective antiunitary $\mathcal{T}_{\mathrm{v}}$.

\end{document}